\documentclass[aps,amsmath,amssymb,superscriptaddress,nofootinbib]{revtex4}

\usepackage{graphicx}
\usepackage{hyperref}
\usepackage{color}
\usepackage{epsf}
\usepackage{bm}
 \usepackage{slashed}
\usepackage{hyperref}
\usepackage{enumitem}
\usepackage{lipsum}
\usepackage{bbold}
\usepackage{sidecap}
\usepackage{subfigure}
\usepackage{color}

\newcommand{\bea}{\begin{eqnarray}}
\newcommand{\eea}{\end{eqnarray}}
\newcommand{\la}{\label}
\newcommand{\be}{\begin{equation}}
\newcommand{\ee}{\end{equation}}

\def\12{\frac{1}{2}}

\newcommand{\bb}{\underline{\beta}}

\def\XXint#1#2#3{{\setbox0=\hbox{$#1{#2#3}{\int}$}
     \vcenter{\hbox{$#2#3$}}\kern-.5\wd0}}

\numberwithin{equation}{section}

\begin{document}

\title{Streched String with Self-Interaction at the Hagedorn Point: \\Spatial Sizes and Black Hole}

\author{Yachao Qian and Ismail Zahed}
\address{Department of Physics and Astronomy,
Stony Brook University,  Stony Brook, NY 11794-3800.}

\begin{abstract}
We  analyze the length, mass and spatial distribution of a discretized  transverse string in $D_\perp$ dimensions with fixed end-points 
near its Hagedorn temperature. We suggest that such a string may dominate the (holographic) 
Pomeron kinematics for dipole-dipole scattering at
intermediate and small impact parameters. Attractive self-string interactions cause the transverse string size to contract away from
its diffusive size, a mechanism reminiscent of the string-black-hole transmutation. The string shows sizable asymmetries in the transverse plane that  translate to primordial azimuthal asymmetries in the  stringy particle production in the Pomeron kinematics for current pp
and pA collisions at collider energies.
\end{abstract}

\date{\today}

\maketitle

\section{\label{sec:introduction}introduction}

Recent high multiplicity events in pA and pp collisions at the LHC~\cite{CMS:2012qk,Abelev:2012ola,Aad:2012gla} have revealed some similarities with
heavy ion collisions at ultra-relativistic energies:  1)  a prompt and large entropy deposition;  2)  a large radial and elliptic flow. The results have renewed the interest in the formation of a fireball and the relevance of hydrodynamics
in small hadronic collisions at high energy~\cite{Shuryak:2013ke,Bozek:2013uha,
deSouza:2015ena,Kalaydzhyan:2014zqa,Kalaydzhyan:2014tfa}. The purpose of this paper is to explore this idea further by using 
self-interacting strings close to their Hagedorn point.  This study complements our recent investigation  of cold
and self-interacting strings at low-x~\cite{Qian:2014rda}.


Hadron collisions at high energies are dominated by soft Pomeron and Reggeon exchanges. The Pomeron is 
an effective $0^{++}$ exchange corresponding to the highest Regge trajectory. Its small intercept 
$\alpha_P(0)-1\approx 0.08$ is used to explain the small growth of the hadron-hadron cross section at large $\sqrt{s}$
with the rapidity interval $\chi={\rm ln}(s/s_0)$ in the context of Reggeon calculus~\cite{Grisaru:1973vw,Grisaru:1974cf,Lipatov:1976zz,Kuraev:1976ge}. First principle 
perturbative  QCD calculations describes a harder Pomeron by re-summing the soft collinear Bremshtralung 
with a larger intercept and zero slope~\cite{Kuraev:1977fs,Balitsky:1978ic}.

Non-perturbative arguments based on duality suggests that the soft Pomeron involves a closed string
exchange in the t-channel. The string world-sheet could be thought as  a fishnet of planar gluon diagrams
in QCD in the large number of colors limit. The quantum theory of planar diagrams in the double limit of strong
coupling and large number of colors is tractable in supersymmetric theories using the holographic
principle~\cite{Maldacena:1997re}.   Many descriptions of the soft Pomeron in holographic duals to QCD have been suggested 
 without supersymmetry, reproducing a number of results in both DIS and diffractive scattering~\cite{Qian:2014rda,Qian:2014jna,Shuryak:2013sra,Shuryak:2013ke,Zahed:2012sg,Stoffers:2012mn,Stoffers:2012ai,Stoffers:2012zw,Basar:2012jb}.
Recently, we have suggested that highly resolved and cold string with a number of quanta $N\equiv 1/x$ can 
be used to account for low-x non-perturbative physics~\cite{Qian:2014rda}. Perturbative low-x physics based on the color
glass condensate has been discussed by many~\cite{McLerran:2001sr,TapiaTakaki:2010zz,Iancu:2003xm,Iancu:2002xk,Navelet:2002zz,Iancu:2001yq,Levin:2001eq,McLerran:1993ka,Gelis:2010nm,Marquet:2005hu,Iancu:2003uh,Iancu:2009nd,Gelis:2012ri}.

One of  the most remarkable feature of free strings is the exponential growth with their mass of the degeneracy 
in their spectra, which translates to a constant entropy to mass ratio~\cite{Fubini:1969wp,Fubini:1969qb}.  Excited strings offer a very
efficient way to scramble information and create entropy. A competing mechanism for scrambling information
appears in the opposite realm of the physical spectrum in the form of  black-holes. Bekenstein noticed that the 
black-hole entropy  grows in proportion to its area therefore to its mass to a power larger than 1 in any dimension~\cite{Bekenstein:1973ur,Bekenstein:1972tm,Bekenstein:1974ax}. 
This has led Susskind 
and others~\cite{Polchinski:2001ju,Susskind:1994hb,Susskind:1994vu,Damour:1999aw,Horowitz:1996nw} to suggest that fundamental interacting single strings reduce to black-holes 
at sufficiently strong self-coupling.

Recently Shuryak and one of us have suggested that the transmutation of strings to black-holes under
self-interaction maybe revealed in hadron-hadron collisions at high energy when probing small impact parameters. 
The idea is that the standard Pomeron as a string exchange in pp  collisions dominates the cross section
for typical impact parameters ${\bf b}\approx 1.5$ fm. However, at smaller impact parameters, the string gets 
highly excited with a rapid build up of entropy. This translates to a high multiplicity event possibly at the origin
of the ridge observed recently at the LHC~\cite{CMS:2012qk,Abelev:2012ola,Aad:2012gla}. 

In this paper,  we will consider the Pomeron as a closed string exchange in the bottom-up approach to 
holographic QCD in AdS$_5$ with a wall to account for the finite QCD string tension~\cite{Rho:1999jm,Janik:2000aj,Janik:2000pp}. For typically large 
impact parameters, the string lies mostly on the wall for which the AdS$_5$ metric is nearly flat. So, in
leading order we will ignore the effects of curvature and consider a string in flat $2+D_\perp=5$ dimensions.
The effects of curvature on our analysis will be treated through the use of an effective 
transverse dimension $2<D_\perp<3$. 
We will fix the entropy of the string and study its transmutation to a black hole through self-interaction 
by following the process  in real space using the scalar Polyakov action.

In section 2 we detail the discretized version of the transverse scalar string in flat $D_\perp$ dimensions. 
In section 3 we introduce the concept of an effective temperature in a micro-canonical description of
a single non-interacting string. In section 4 we investigate the effects of attractive self-interactions between the
string bits using the Feynman variational principle both in flat and effectively curved $D_\perp$. In section 5
we detail numerically the geometrical and angular deformations of the string for single and multiple but interacting
string exchanges. Our conclusions are in section 6.



\section{\label{}Discretized Free Transverse String}

Scattering of dipoles in the pomeron kinematics with a large rapidity interval $\chi={\ln}(s/s_0)$ and fixed impact parameter $b$ 
is dominated by a closed t-channel string exchange. In leading order in $\chi$, the exchange amplitude can be shown to be that 
of a free transverse string at fixed Unruh temperature $T=a/2\pi$ with the mean world-sheet acceleration $a=\chi/{\bf b}$~\cite{Basar:2012jb, Stoffers:2012zw, Stoffers:2012ai, Stoffers:2013tla, Qian:2014jna}. The free  transverse string with fixed end-points in $D_\perp$ dimensions is characterized by the scalar Polyakov action

\be
S_\perp = \frac{\sigma_T}{2} \int d \tau \int_0^\pi d \sigma \ \ \left[  \left( \dot{x}_\perp  \right)^2  + \left( {x'}_\perp \right)^2    \right]
\label{XS1}
\ee
with the end-point condition 
\be
x_\perp^i (\sigma=0, \tau) = 0 \ \ \ \ \ \ x_\perp^i (\sigma=\pi, \tau) = \bold{b}^i
\ee
The string tension is $\sigma_T=1/(2\pi \alpha^\prime)$ with $\alpha^\prime=l_s^2$. For simplicity, we will set $2l_s\equiv1$ throughout
and restore it by inspection when needed. The purpose of the present work is to show how a a fixed end-point string with a high multiplicity
content behaves both in flat and effectively curved space-time. In particular its spatial size and deformation near its Hagedorn point as a way
to model dipole-dipole collisions in the hot Pomeron regime. Initial geometrical string deformations maybe the source
of large prompt azimuthal deformations in the inelastic channels and for high multiplicity events. 

The transverse free string (\ref{XS1})  can be thought as a collection of $N$  string bits connected by identical
strings~\cite{Karliner:1988hd,Bergman:1997ki}. The transverse Hamiltonian follows from (\ref{XS1}) canonically.
Using the mode decompostion for the amplitudes  $x_\perp^i$ 

\be
x_\perp^i (k , \tau) = \bold{b}^i \frac{k}{N }  +   \sum_{n=1}^{N-1}  X_{n}^i (\tau) \sin \left( \frac{n k}{N } \pi\right)   \ \ \ \ \ \ \ \ \    
(k=0,1, \cdots, N)
\label{N1}
\ee
the canonical momenta are $P_n^i(\tau)={\dot{X}}_n^i$. The Hamiltonian is then

\bea
\mathcal{H}_\perp  
=  
\frac{1}{2 } \sum_{n=1}^{N-1}    \left(         P _{n}^i (\tau) P_{n }^i (\tau)   + 
\Omega_n^2  {X}_{n}^i (\tau)  {X}_{n }^i (\tau) \right)  + \frac{  b^2}{  \pi^2 } 
\label{HAR}
\eea
with $\Omega_n = \frac{2 N}{\pi} \sin  \left(  \frac{n \pi}{2 N} \right)$.
Each oscillator in (\ref{HAR}) carries a string bit mass $m_N=2/N$ and a large compressibility $k_N=4/(\pi^2m_N)$.
The ground state of this dangling  N-string bit Hamiltonian  is a product of Gaussians~\cite{Karliner:1988hd}

\be
\Psi[X]=\prod_{n, i} \Psi (X_n^i) = \prod_{n, i} \left( \frac{\Omega_n}{\pi}  \right)^{\frac{1}{4}} \exp\left[  - \frac{\Omega_n}{2} (X_n^i)^2   \right]
\rightarrow {\cal N}\left(0, \frac 1{\Omega_n}\right)
\ee
In its ground state, each of the discretized string bit coordinates $X_n^i$ is normally distributed with probability $|\Psi(X_n^i)|^2$.
This gives rise to a random walk of the string bits along the chain in the transverse direction with fixed end-points. This is also true
for the continuum with $\Omega_n\rightarrow n$. The ground state energy is

\be
M_\perp^2 = \frac{1}{2} \left< \mathcal{H}_\perp \right>   =  \frac{D_\perp}{4} \sum_{n=1}^{N-1}    \Omega_n     + \frac{  b^2}{  2 \pi^2 } 
\rightarrow \frac{ D_\perp}{\pi^2  }  N^2 + \frac{  b^2}{  2\pi^2 } 
\label{H0}
\ee
and the string transverse squared size is

\be 
R_\perp^2  =  \frac{1}{N} \sum_{k= 0}^N  \left< \left(  x_k^i  - \bold{b}^i \frac{k}{N} \right)^2 \right> 
= \frac{D_\perp}{4}\sum_{n=1}^{N-1} \frac{1}{\Omega_n}\rightarrow  \frac{D_\perp}{4} \ln (N)
\label{RT0}
\ee 

In a recent analysis we have identified the string resolution as $N\equiv 1/x$ with $x$ the fraction of wee-parton momentum carried
by the string bits. We have further suggested that the holographic string  with self-interactions provide a mechanism of saturation that borrows on the mechanism of black-hope formation in fundamental strings as advocated by Susskind and others~\cite{Polchinski:2001ju,Susskind:1994hb,Susskind:1994vu}. We refer to these strings as cold strings with quantum or zero-point effects as dominant. The classical or Hagedorn regime
of the string is the one we would like to develop here as a microscopic mechanism for the high multiplicity events in dipole-dipole
scattering and ultimately in pp and pA scatterings.


 \section{Free String at Finite Temperature}

To describe the string close to its classical or Hagedorn point, we 
introduce an effective temperature $1/\underline{\beta}$ 
which is conjugate of the squared and normal ordered mass operator~\cite{Susskind:1994hb,Susskind:1994vu,Damour:1999aw}, 

 \be
:\mathcal{H}_\perp:  =    \sum_{n=1}^{N-1}  \Omega_n    (a_n^i)^\dagger a_n^i   + \frac{  b^2}{  \pi^2 } 
\ee
with the standard commutators $\left[ a_n^i ,  (a_{n'}^{i'})^\dagger  \right] = \delta_{n n^\prime} \delta_{i i^\prime}$. 
All expectation values at finite $1/\underline{\beta}$ will be carried using the density matrix 
$e^{ -\bb :\mathcal{H}_\perp:} /\mathcal{Z}_\perp $, with the transverse partition function

 \be
\mathcal{Z}_\perp  =\left<   e^{ -\bb :\mathcal{H}_\perp:}  \right>
= \exp \left( - \bb \frac{  b^2}{  \pi^2 }  \right) \prod_{i=1}^{D_\perp} \prod_{n=1}^{N-1} \frac{1}{ 1 -  e^{ -   \bb \Omega_n  }}
\ee
This is a micro-canonical description of a  single thermal string.
In practical terms, it corresponds to string bits $X_n^i$ normally distributed with 

\be
X_n^i   \sim  \mathcal{N} \left(0  ,  \frac{1}{ \Omega_n \left( e^{\bb \Omega_n} - 1  \right)  } \right)  
\ee
The squared mass is then $2M_\perp^2\equiv-\partial{\cal Z}_\perp/\partial \bb$ or

\be
\label{TS1}
2M_\perp^2 
=    \left<\hspace{-1 mm} \left< \hspace{1 mm} : \mathcal{H}_\perp :   \hspace{1 mm} \right>\hspace{-1 mm}\right>
=  D_\perp  \sum_{n=1}^{N-1}  \frac{ \Omega_n }{e^{\bb \Omega_n} - 1} + \frac{b^2}{  \pi^2}  \rightarrow
\frac{D_\perp}{ \bb^2}  \int_{\bb}^\infty d x \,\,  \frac{ x }{e^{x} - 1} + \frac{b^2}{   \pi^2}   
\approx \frac{D_\perp}{6} \frac{\pi^2}{  \bb^2} +   \frac{b^2}{  \pi^2}  
\ee
and its squared transverse size is

\be 
\label{TS2}
 R_\perp^2 
\equiv  \frac{1}{N} \sum_{k= 0}^N  \left<\hspace{-1 mm} \left< \hspace{1 mm}:\left(  x_k^i  - \bold{b}^i \frac{k}{N} \right)^2 : \hspace{1 mm} \right>\hspace{-1 mm}\right>
= \frac{D_\perp}{2} \sum_{n=1}^{N-1}  \frac{1}{\Omega_n} \frac{1}{e^{\bb \Omega_n} -1}\rightarrow
  \frac{D_\perp}{2} \int_{\bb}^\infty   \frac{d x }{x} \frac{1}{e^{x} - 1}\approx \frac{D_\perp}{2 \bb}
\ee 
where $ \left<\hspace{-0.7 mm} \left< \hspace{0.7 mm} \cdots \hspace{0.7 mm} \right>\hspace{-0.7 mm}\right> $ is the expectation 
value carried using the density matrix. The effective entropy is

\bea
\label{TS3}
S_\perp =&& -\bb  \frac{\partial \ln \mathcal{Z}_\perp}{\partial \bb}  + \ln \mathcal{Z}_\perp 
= D_\perp \sum_{n=1}^{N-1} \left[  \frac{\bb \Omega_n }{e^{\bb \Omega_n} - 1} -\ln \left(   1 -  e^{ -   \bb \Omega_n  } \right) \right]
\nonumber\\
\rightarrow&& \frac{D_\perp}{\bb}     \int_{\bb}^\infty d x \,\,  \left[  \frac{x }{e^{x} - 1} -\ln \left(   1 -  e^{ -  x } \right) \right] \approx 
\frac{D_\perp}{3} \frac{  \pi^2 }{  \bb} 
\eea
which can be recasted using the Hagedorn temperature $1/\beta_H$

\be
\label{TS4}
S_\perp \approx  2 \pi \sqrt{\frac{D_\perp}{6}} M_\perp  \rightarrow 2 \pi \sqrt{\frac{D_\perp \alpha'}{6}} {M_\perp}\equiv  \beta_H {M_\perp}
\ee
after re-instating the string unit with $M_\perp/l_s\rightarrow M_\perp$. 
Below the value of $1/\underline{\beta}$ will be fixed by fixing the mass or the entropy of the thermal 
string. We note that for large $1/\underline{\beta}$ the string behaves classically with dwarfed quantum 
or zero point contributions. Hence the normal ordering. To make contact with physical observables, we
identify  $S_\perp$ with the prompt multiplicity and approximate it 
with the final charge multiplicity $N_{\rm ch}$ (upper bound). For a single string exchange

\be
S_\perp\approx \frac{D_\perp}{3} \frac{  \pi^2 }{  \bb} \approx 7.5\,N_{\rm ch}
\label{CHARGE}
\ee

\begin{figure}[!htb]
\minipage{0.48\textwidth}
\includegraphics[height=50mm]{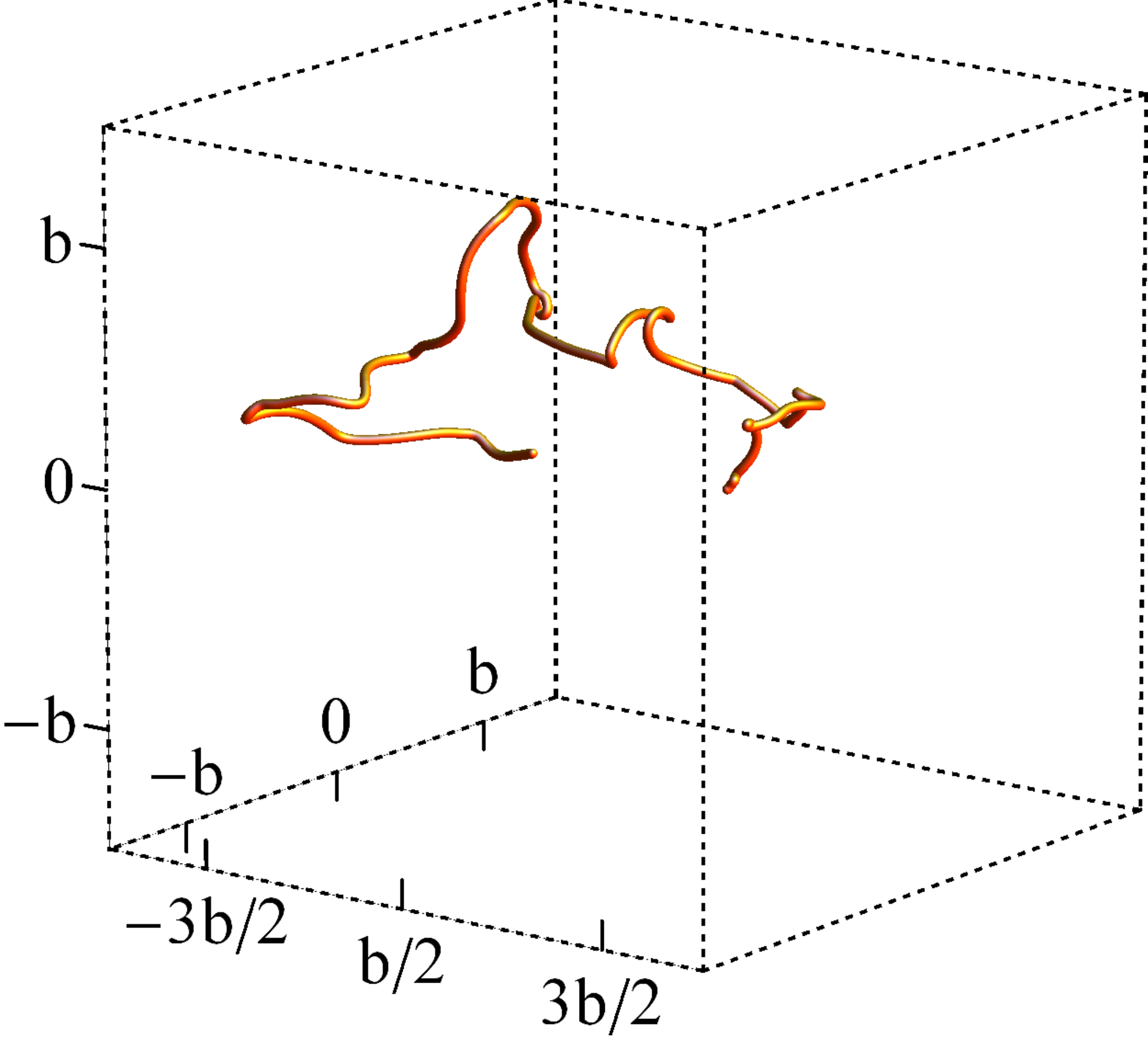}
\endminipage\hfill
\minipage{0.48\textwidth}
\includegraphics[height=50mm]{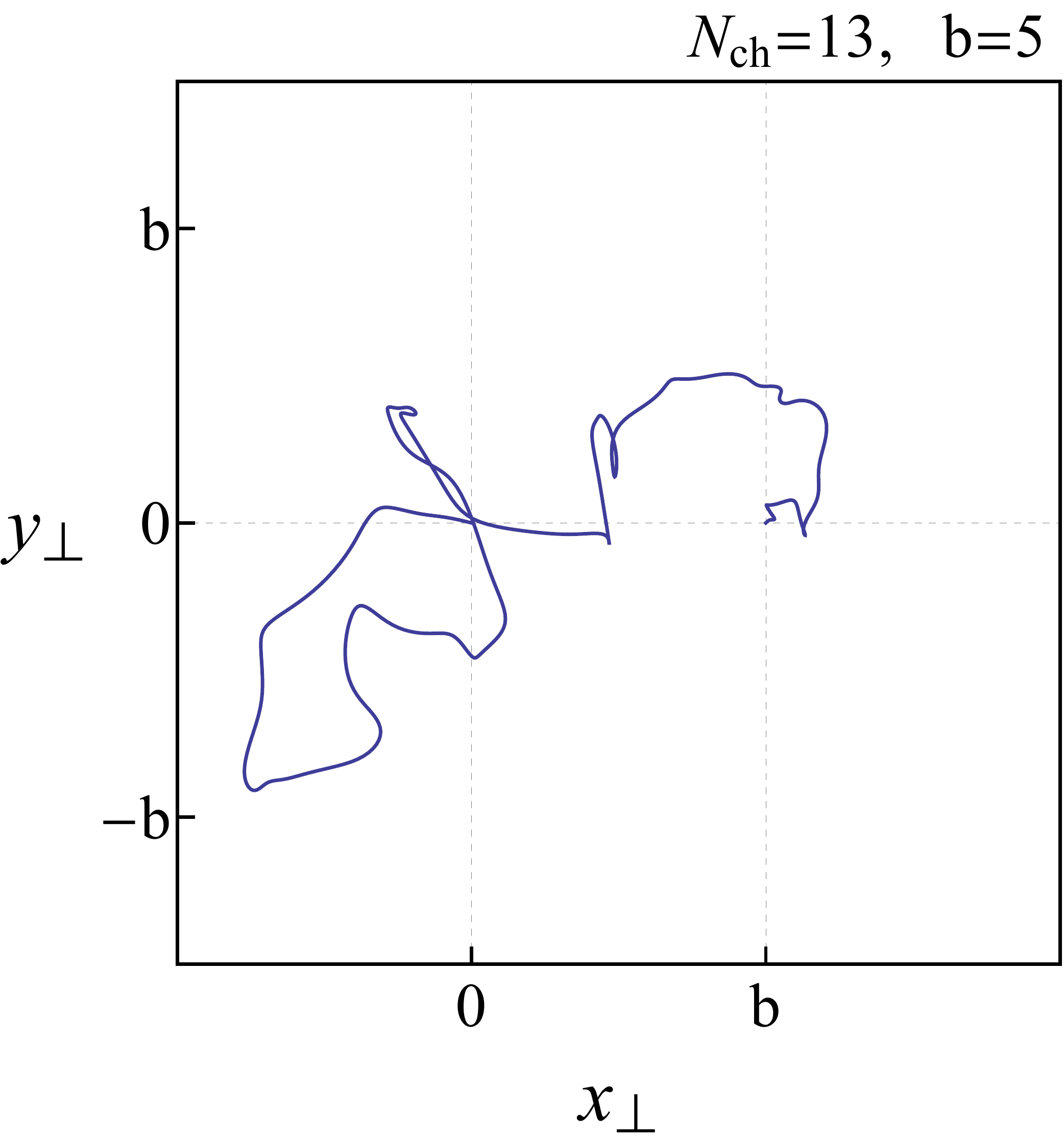}
\endminipage
  \caption{Streched string with fixed ${\bf b}=5=10l_s$ and multiplicity $N_{ch}=13$ in the holographic $D_\perp=3$
  (left) and projected onto the spatial 2-dimensional transverse space (right). }\label{XDStringNch13}
\end{figure}

\begin{figure}[!htb]
\minipage{0.48\textwidth}
\includegraphics[height=50mm]{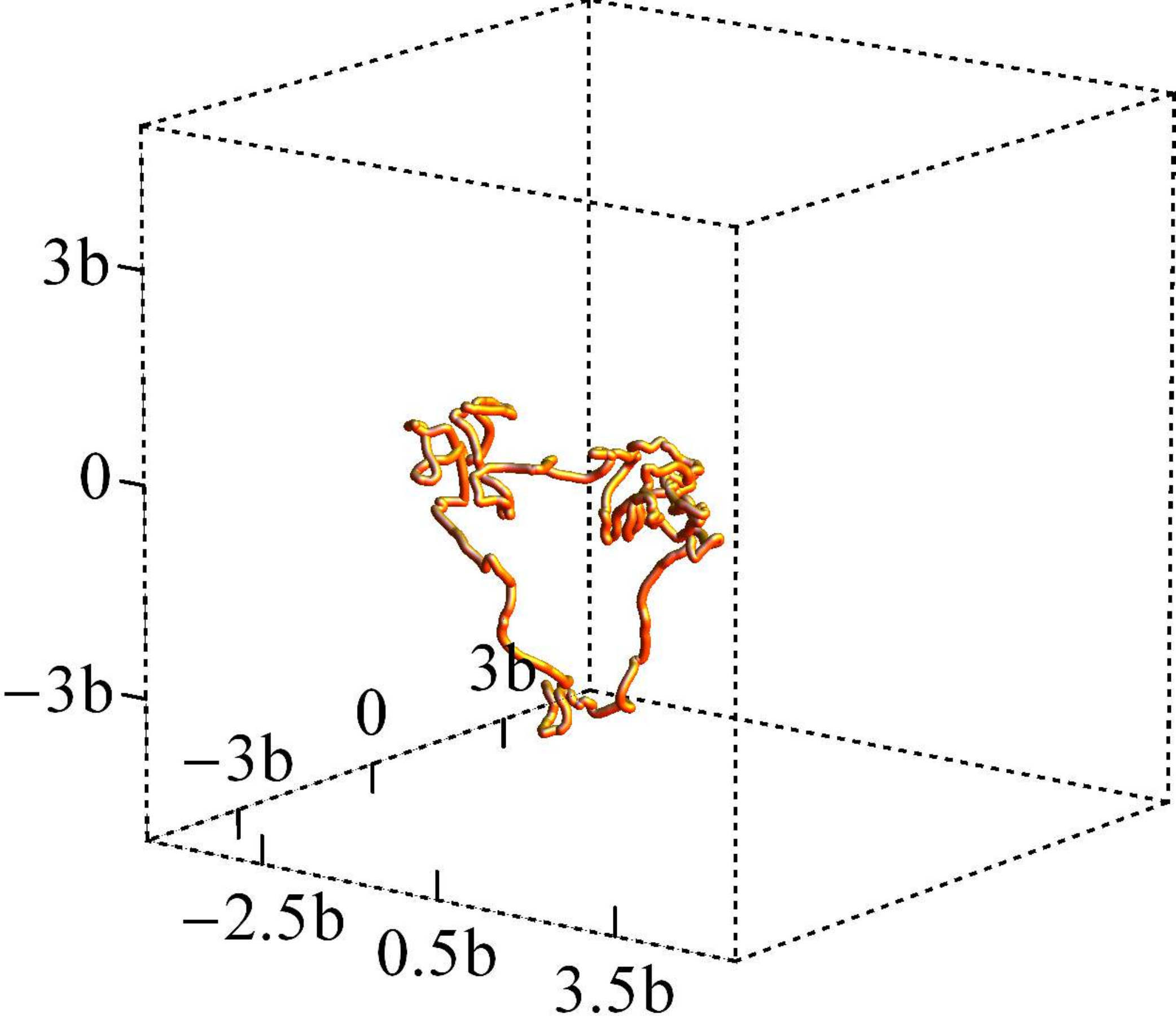}
\endminipage\hfill
\minipage{0.48\textwidth}
\includegraphics[height=50mm]{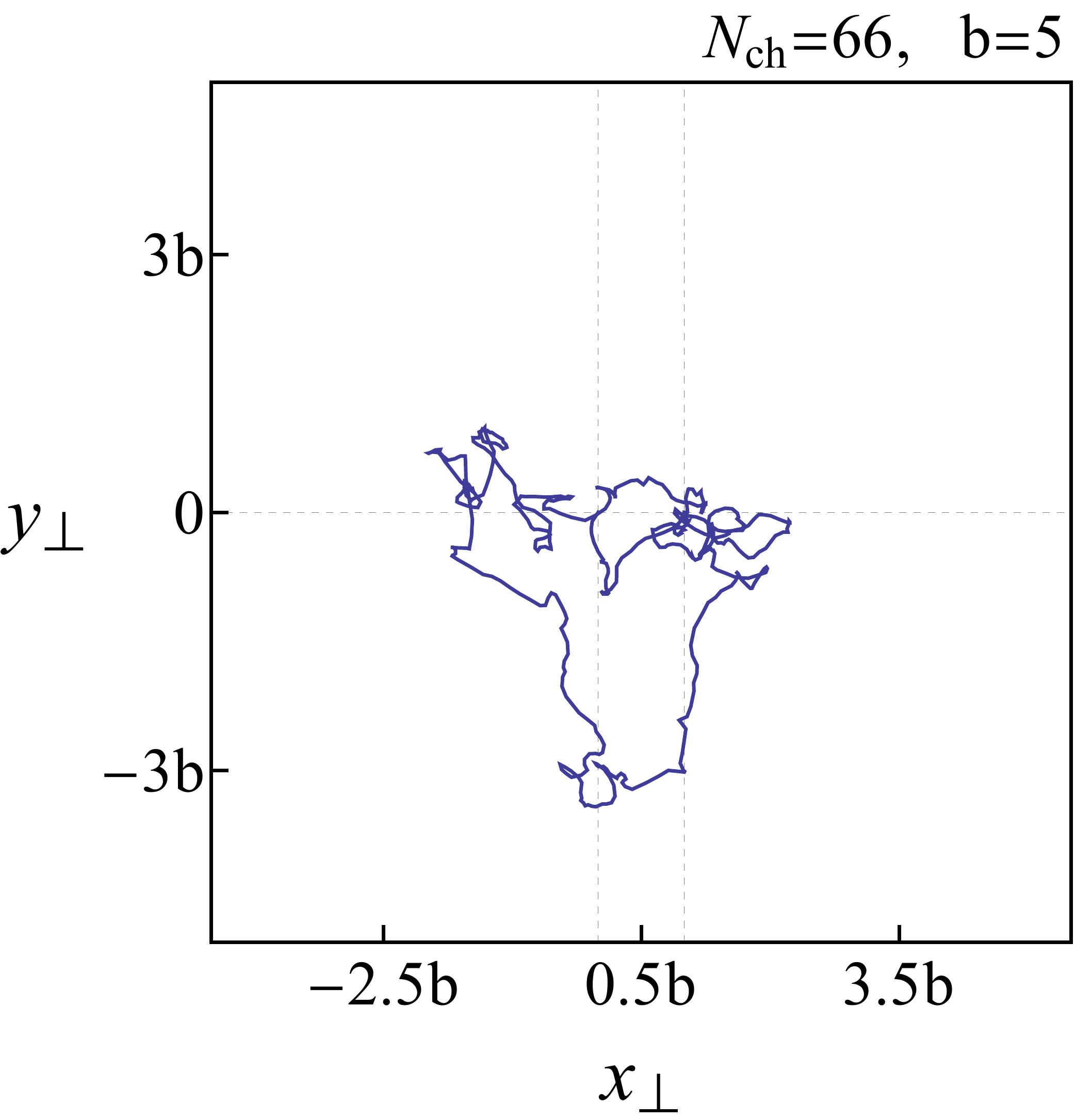}
\endminipage
  \caption{ Streched string with fixed ${\bf b}=5=10l_s$ and multiplicity $N_{ch}=66$ in the holographic $D_\perp=3$
  (left) and projected onto the spatial 2-dimensional transverse space (right).}\label{XDStringNch66}
\end{figure}

\begin{figure}[!htb]
\minipage{0.48\textwidth}
\includegraphics[height=50mm]{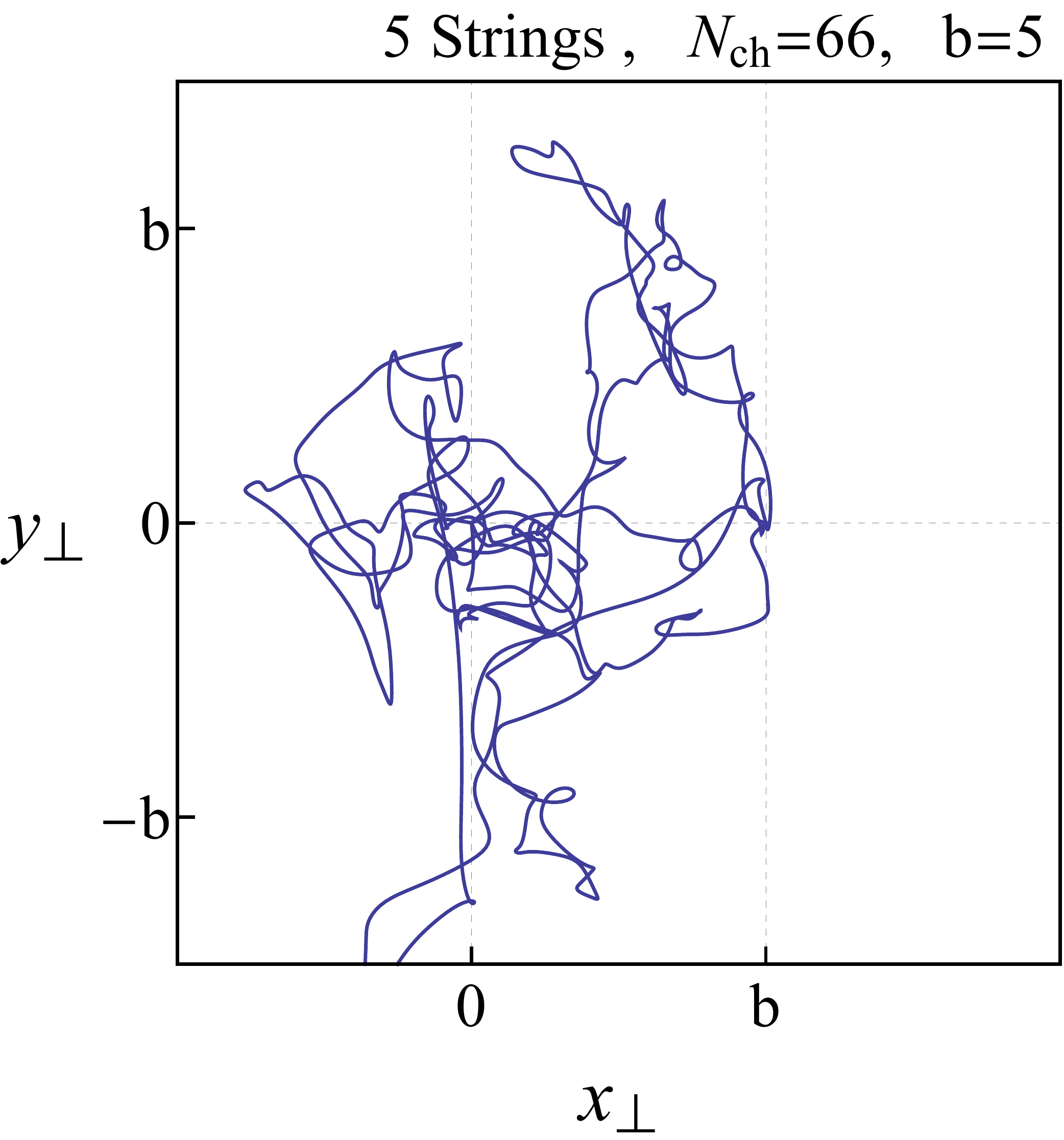}
\endminipage\hfill
\minipage{0.48\textwidth}
\includegraphics[height=50mm]{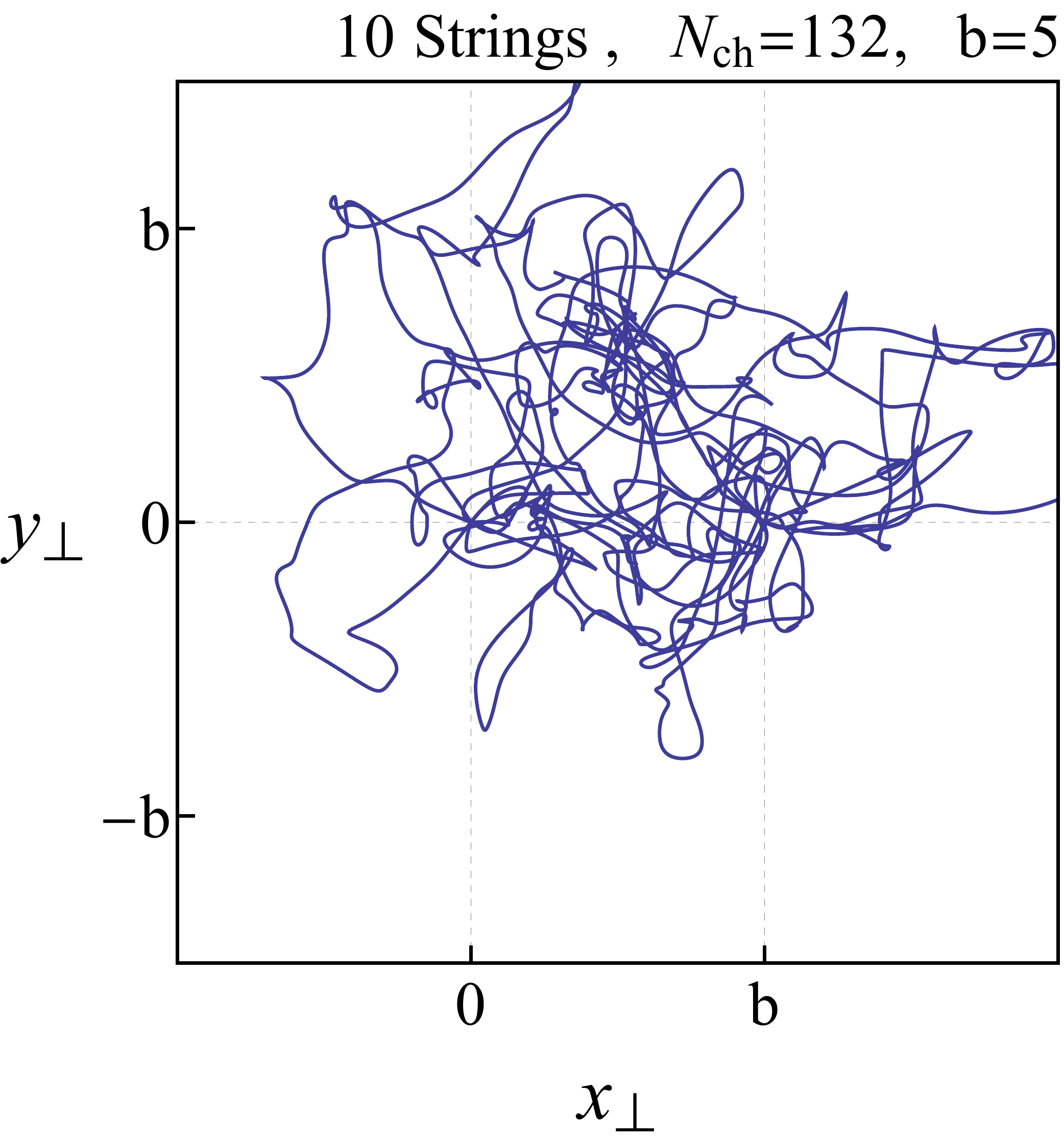}
\endminipage
  \caption{ 5 string shapes (left) with a total multiplicity $N_{\rm ch}=66$, and 
  10 string shapes (right) with a total multiplicity $N_{\rm ch}=132$.
  The string end-points are fixed at ${\bf b}=5=10\,l_s$}\label{2DStringsL}
\end{figure}

\begin{figure}[!htb]
\minipage{0.48\textwidth}
\includegraphics[height=50mm]{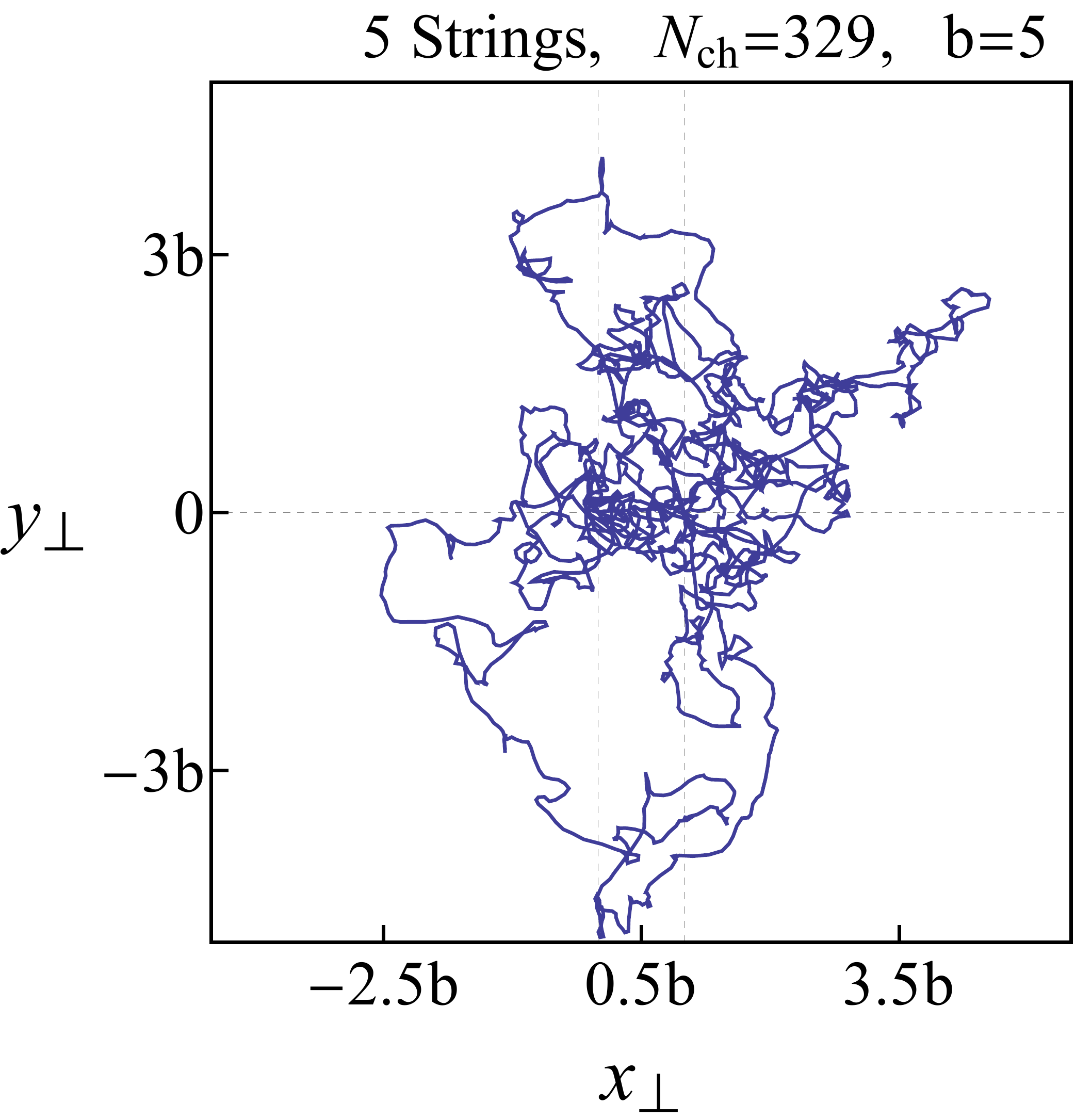}
\endminipage\hfill
\minipage{0.48\textwidth}
\includegraphics[height=50mm]{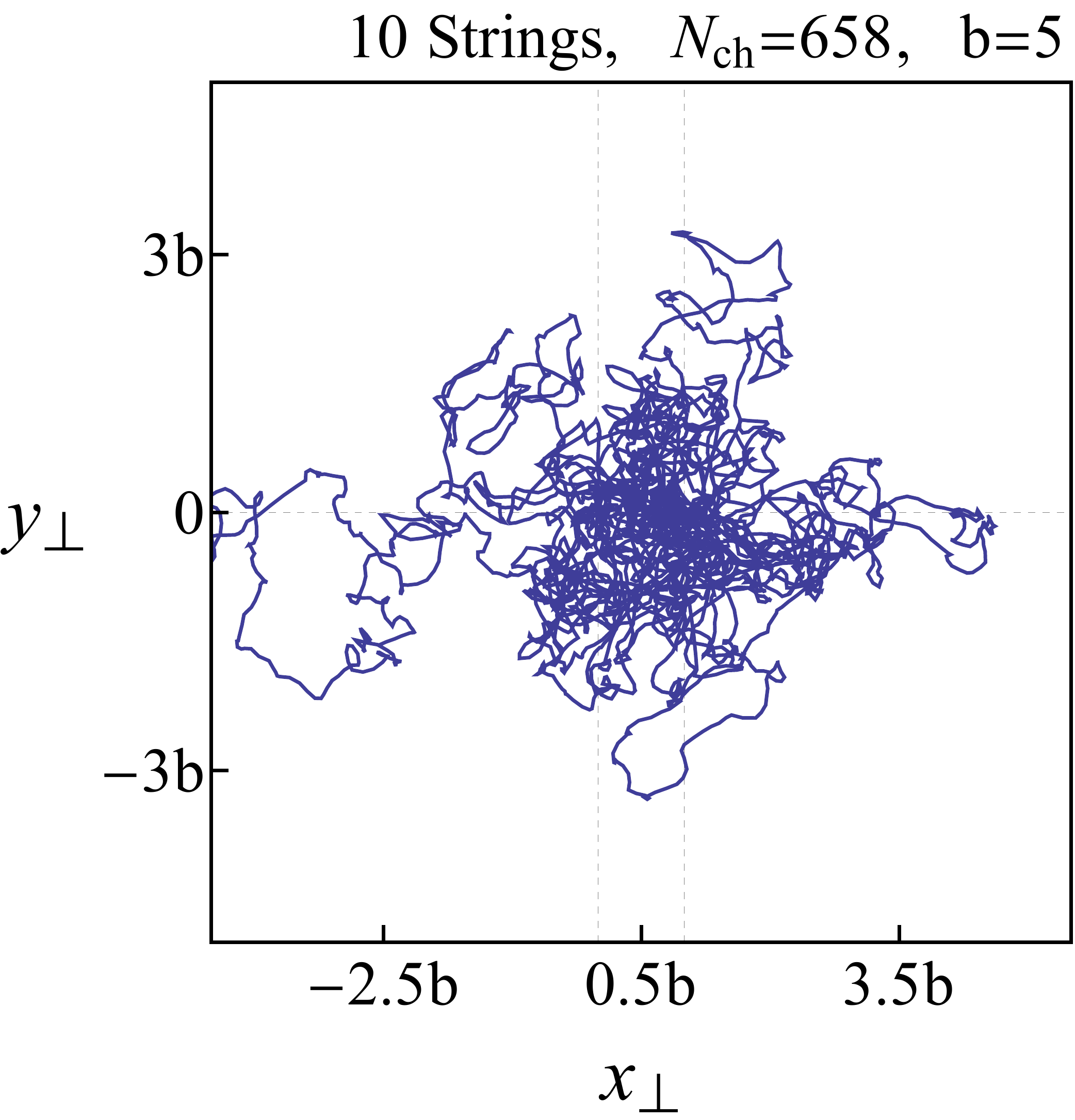}
\endminipage
 \caption{ 5 string shapes (left) with a total multiplicity $N_{\rm ch}=329$, and 
  10 string shapes (right) with a total multiplicity $N_{\rm ch}=658$.
  The string end-points are fixed at ${\bf b}=5\equiv 10\,l_s$}\label{2DStringsH}
\end{figure}

Throughout, high temperature means $  \bb\ll 1$
and classical means $N\bb\gg 1$. Analytically, we will take $N \rightarrow \infty$ and fix $\bb\ll 1$.  
Numerically, the best we can do is set $N=500$ which fixes the range  $\bb \approx (0.1, 0.02)$, since
$\bb\leq 0.1$  is small and $N\bb \geq 10$  is large. For a single string this translates to a charge multiplicity $N_{ch}$ in the range
$(13, 66)$. In Fig.~\ref{XDStringNch13} and Fig.~\ref{XDStringNch66} we show a single string for a fixed distance 
$\bold{b} = 5\equiv 10l_s\approx 1\,{\rm fm}$ with charge multiplicity $N_{ch} =13$ and $N_{ch} =66$ respectively. The left figure is the string projected in the transverse spatial plane, while the right figure is the string in the holographic but flat $D_\perp =3$ dimensions.  
The effects of the warping in the holographic direction will be discussed below.

pp and pA scattering in the holographic context may involve more than a single string exchange~\cite{Shuryak:2013sra,Stoffers:2012mn,Stoffers:2012ai}.
Multiple string exchanges involve colder strings in their diffusive regime with a higher multiplicity.
For 5 and 10 multiple string exchanges,  the charge multiplicity $N_{ch}$ is in the range 
$(66, 329)$ and $(132 , 658)$ respectively. In  Fig.~\ref{2DStringsL} (left) we display 5 strings with 
$\bb = 0.1$  or a charge multiplicity of $N_{ch} = 66$. In  Fig.~\ref{2DStringsL} (right) we display 10 strings with 
$\bb = 0.1$  or a charge multiplicity of $N_{ch} = 132$. Fig.~\ref{2DStringsH} is the same as Fig.~\ref{2DStringsL}
but with $\bb = 0.02$  or $N_{ch} = 329$ for 5 strings and $N_{ch} = 658$ for 10 strings.

 \section{Thermal String with Self-interactions}

We now follow our recent analysis for the cold string in~\cite{Qian:2014rda} and explicit the string self-interaction by assuming it to be 
dominated by the two-body string bits interactions mediated by a static exchange in $D_\perp+2$ dimensions. Specifically,

\be\la{interaction}
V =-\frac{1}{2}  g^2 \sum_{k \neq k'} \int \frac{d^{D_\perp + 1} p}{(2 \pi)^{D_\perp + 1}}  \frac{  M(\vec{x}_k )   M(\vec{x}_{k'} )}{p^2 + m^2} \exp \left( i \vec{p} \cdot (\vec{x}_k - \vec{x}_{k'})  \right)
\ee
where $M(\vec{x}_k)$ is the mass of the discrete point at $\vec{x}_k$.  The exchange is generic and is parameterzed with an
attractive coupling $g$ and a mass $m$. In holographic QCD, the exchanged mass is that of the lowest scalar~\cite{Kalaydzhyan:2014tfa,Liu:2014fda,Liu:2014qrt}.
Thoughout, we will set $m=0$ as any finite $m$ can be re-absorbed into a re-definition of the coupling and use $g$ as a parameter.

The interacting Hamiltonian is now
$\mathcal{H}_\perp   \rightarrow \mathcal{H}_\perp^0  + 2 M_\perp V$. The partition function for the interacting string 
is now formally given by

\be
\label{ZZ1}
\mathcal{Z} =  \left<   e^{-\bb \mathcal{H}_\perp}  \right> =  \left<   e^{-\bb \left( \mathcal{H}_\perp^0 + 2 M_\perp V \right)}  \right>
\ee
where the averaging is carried using a complete set of free harmonic oscillators with trial frequencies $\omega_n$ instead of
the free frequencies $\Omega_n$. This corresponds to the interacting string bits $X_n^i$  normally distributed with

\be
X_n^i   \sim  \mathcal{N} \left(0  ,  \frac{1}{\omega_n \left[e^{\frac{\bb}{2} \left( \omega_n + \frac{\Omega_n^2}{\omega_n}   \right) } - 1\right] }\right)  
\ee

\subsection{Variational Analysis}

To estimate (\ref{ZZ1}) we will use the Feynman variational principle~\cite{Diakonov:1983hh,Feynman:1955zz}

\be
\label{FZZ}
\mathcal{Z} \geq \mathcal{Z}_0 \exp \left( - 2  \bb   M_\perp \left<\hspace{-1 mm} \left< \hspace{1 mm}   V \hspace{1 mm} \right>\hspace{-1 mm}\right>  \right)
\ee
with

\be
\mathcal{Z}_0 =  \left< e^{- \bb \mathcal{H}_\perp^0}   \right>   =\exp \left( - \bb \frac{  b^2}{  \pi^2 }  \right) \prod_{i=1}^{D_\perp} \prod_{n=1}^{N-1} \frac{1}{ 1 - \exp \left[  - \frac{\bb}{2} \left( \omega_n + \frac{\Omega_n^2 }{\omega_n } \right) \right]}
\ee
 
In leading order in the interaction the squared mass and size of the interacting string are given by

\be\la{leadingMperp}
2M_\perp^2 
=     \left<\hspace{-1 mm} \left< \hspace{1 mm} \mathcal{H}_\perp^0  \hspace{1 mm} \right>\hspace{-1 mm}\right>
= \frac{D_\perp}{2}  \sum_{n=1}^{N-1}  \frac{   \omega_n + \frac{\Omega_n^2}{\omega_n}  }{\exp \left[ \frac{\bb}{2} \left( \omega_n + \frac{\Omega_n^2}{\omega_n}   \right) \right] - 1} + \frac{b^2}{   \pi^2}  
\ee
\be \la{leadingRperp}
   R_\perp^2 \hspace{1 mm} 
=  \frac{1}{N} \sum_{k= 0}^N    \left<\hspace{-1 mm} \left< \hspace{1 mm} \left(  x_k^i  - \bold{b}^i \frac{k}{N} \right)^2 \hspace{1 mm} \right>\hspace{-1 mm}\right>
= \frac{D_\perp}{2} \sum_{n=1}^{N-1}  \frac{1}{\omega_n} \frac{ 1 }{\exp \left[ \frac{\bb}{2} \left( \omega_n + \frac{\Omega_n^2}{\omega_n}   \right) \right] - 1}
\ee 
The discretized string mass distribution $M(\vec{x}_k) \longrightarrow {M_\perp}/{N + 1}$ so that the averaged pair-interaction reads

\bea\la{Vaverage}
 \left<\hspace{-1 mm} \left< \hspace{1 mm} V  \hspace{1 mm} \right>\hspace{-1 mm}\right>
&\approx& -\frac{1}{2}  g^2 \frac{M_\perp^2}{N^2} \sum_{k \neq k'} \int\frac{d^{D_\perp + 1} p}{(2 \pi)^{D_\perp + 1}}   \frac{  e^{     i \vec{p} \cdot \vec{b} \frac{(k- k')}{N}  } }{p^2 + m^2 } \exp \left(  - \frac{p^2}{2 D_\perp}  \left<\hspace{-1.8 mm} \left< \hspace{1.8 mm}  \left(\vec{x}_k - \bold{b}^i \frac{k}{N} - \vec{x}_{k'} +  \bold{b}^i \frac{k'}{N}\right)^2   \hspace{1.8 mm} \right>\hspace{-1.8 mm}\right> \right) \nonumber\\
&\approx& -\frac{1}{2}  g^2 \frac{M_\perp^2}{N^2} \sum_{k \neq k'}\int\frac{d^{D_\perp + 1} p}{(2 \pi)^{D_\perp + 1}}   \frac{   e^{     i \vec{p} \cdot \vec{b} \frac{(k- k')}{N}  }}{p^2 + m^2 } \exp \left(  -   \frac{p^2}{2 D_\perp}   \left<\hspace{-1.8 mm} \left< \hspace{1.8 mm}  \left( \vec{x}_k  - \bold{b}^i \frac{k}{N} \right)^2 + \left( \vec{x}_{k'} - \bold{b}^i \frac{k'}{N}  \right)^2   \hspace{1.8 mm} \right>\hspace{-1.8 mm}\right>   \right)  \nonumber\\
\eea
where we have exponentiated the averaging and then used the quadratic nature of the distributions.
Since the position of the string bits are normally distributed, we can carry the averaging in the exponent
explicitly. The result is

\bea
\label{VVaverage}
 \left<\hspace{-1 mm} \left< \hspace{1 mm} V  \hspace{1 mm} \right>\hspace{-1 mm}\right>
&\approx&- \frac{1}{2}  g^2 \frac{M_\perp^2}{N^2} \sum_{k \neq k'} \int\frac{d^{D_\perp + 1} p}{(2 \pi)^{D_\perp + 1}}\frac{  e^{     i \vec{p} \cdot \vec{b} \frac{(k- k')}{N}  } }{p^2 + m^2} \exp \left(  -  \frac{ p^2}{2}  \sum_{n=1}^{N-1}  \frac{  \left[  \sin^2  \left(  \frac{n k}{N} \pi  \right) +  \sin^2  \left(  \frac{n k'}{N} \pi  \right) \right] }{  \omega_n \left( e^{  \frac{\bb}{2} \left( \omega_n + \frac{\Omega_n^2}{\omega_n}   \right)}- 1  \right)} \right )  \nonumber\\
&\approx&- \frac{1}{2}  g^2 M_\perp^2 \int\frac{d^{D_\perp + 1} p}{(2 \pi)^{D_\perp + 1}}   \frac{  1 }{p^2 + m^2 }    \frac{4 \sin^2 \left(  \frac{\vec{p} \cdot \vec{b}}{2} \right)}{(\vec{p} \cdot \vec{b})^2}   \exp \left(  -   p^2  \frac{R_\perp^2}{D_\perp}  \right )  
\eea
We note that~(\ref{Vaverage}) is overall similar to the result established in~\cite{Qian:2014rda}, except that now both $M_\perp, R_\perp$ 
are implicit functions of the effective temperature $1/\bb$. Inserting (\ref{VVaverage}) back into (\ref{FZZ}) shows that for the interacting
string the free energy is bounded from below

 
\be\la{freeenergy}
F \geq -    g^2 M_\perp^3  \int\frac{d^{D_\perp + 1} p}{(2 \pi)^{D_\perp + 1}}   \frac{  1 }{p^2 + m^2 } \frac{4 \sin^2 \left(  \frac{\vec{p} \cdot \vec{b}}{2} \right)}{(\vec{p} \cdot \vec{b})^2}   \exp \left(  -   p^2  \frac{R_\perp^2}{D_\perp}  \right )  +  \frac{D_\perp}{\bb}  \sum_{n=1}^{N-1} \ln \left( 1 -  \exp  \left[ -  \frac{\bb}{2} \left( \omega_n + \frac{\Omega_n^2 }{\omega_n } \right)  \right] \right) + \frac{b^2 }{ \pi^2}
\ee
The bound in~\ref{freeenergy} is parametrized by the set of frequencies $\omega_n$ which are fixed variationally through

\bea
\label{LOWER}
\frac{\delta F}{\delta \omega_n} &\geq&  \frac{D_\perp}{2} \left( 1 - \frac{\Omega_n^2}{\omega_n^2} \right) \frac{1}{  \exp  \left[  \frac{\bb}{2} \left( \omega_n + \frac{\Omega_n^2 }{\omega_n } \right)  \right] - 1}  \nonumber\\
&-& \frac{\delta M_\perp^2}{\delta \omega_n} \times  \frac{3 g^2 M_\perp}{2}     \int\frac{d^{D_\perp + 1} p}{(2 \pi)^{D_\perp + 1}}   \frac{  1 }{p^2 + m^2 } \frac{4 \sin^2 \left(  \frac{\vec{p} \cdot \vec{b}}{2} \right)}{(\vec{p} \cdot \vec{b})^2}   \exp \left(  -   p^2  \frac{R_\perp^2}{D_\perp}  \right )  \nonumber\\
&-& \left( - \frac{1}{D_\perp} \frac{\delta R_\perp^2}{\delta \omega_n}  \right)\times  g^2  M_\perp^3     \int\frac{d^{D_\perp + 1} p}{(2 \pi)^{D_\perp + 1}}   \frac{ p^2  }{p^2 + m^2 } \frac{4 \sin^2 \left(  \frac{\vec{p} \cdot \vec{b}}{2} \right)}{(\vec{p} \cdot \vec{b})^2}   \exp \left(  -   p^2  \frac{R_\perp^2}{D_\perp}  \right )  =0
\eea

\subsection{\label{ } High Temperature Limit}
 
 To find the lower bound in (\ref{LOWER}) is in general involved. However, at high temperature the contributions simplify

\bea
 \frac{\delta M_\perp^2}{\delta \omega_n} = \left( 1 - \frac{\Omega_n^2}{\omega_n^2}  \right) \frac{D_\perp}{2} \frac{\exp \left[ \frac{\bb}{2} \left( \omega_n + \frac{\Omega_n^2}{\omega_n}   \right) \right] - 1 -  \frac{\bb}{2} \left( \omega_n + \frac{\Omega_n^2}{\omega_n}   \right)  \exp \left[ \frac{\bb}{2} \left( \omega_n + \frac{\Omega_n^2}{\omega_n}   \right) \right]  }{\left( \exp \left[ \frac{\bb}{2} \left( \omega_n + \frac{\Omega_n^2}{\omega_n}   \right) \right] - 1  \right)^2} 
\approx 0 \left( \frac{1}{\bb} \right)
\eea
 
\bea
 - \frac{1}{D_\perp} \frac{\delta R_\perp^2}{\delta \omega_n} &=&   \frac{1}{2\omega_n^2}   \frac{ 1 }{\exp \left[ \frac{\bb}{2} \left( \omega_n + \frac{\Omega_n^2}{\omega_n}   \right) \right] - 1} + \frac{\bb}{4\omega_n}  \left( 1 - \frac{\Omega_n^2}{\omega_n^2}  \right) \frac{  \exp \left[ \frac{\bb}{2} \left( \omega_n + \frac{\Omega_n^2}{\omega_n}   \right) \right]   }{\left( \exp \left[ \frac{\bb}{2} \left( \omega_n + \frac{\Omega_n^2}{\omega_n}   \right) \right] - 1 \right)^2} 
\approx \frac{2 \omega_n}{\bb \left(  \omega_n^2 + \Omega_n^2 \right)^2}\nonumber\\
\eea
So in leading order in $1/\bb$ or close to the Hagedorn temperature, the lower bound in (\ref{LOWER}) is reduced to finding $\omega_n$ which are solutions to

 \bea
\frac{\delta F}{\delta \omega_n} &\approx&    \frac{ D_\perp}{\bb \omega_n}  \frac{\omega_n^2 - \Omega_n^2}{\omega_n^2 + \Omega_n^2}   -   \frac{2 \omega_n}{\bb \left(  \omega_n^2 + \Omega_n^2 \right)^2} \times  g^2  M_\perp^3     \int\frac{d^{D_\perp + 1} p}{(2 \pi)^{D_\perp + 1}}   \frac{ p^2  }{p^2 + m^2 } \frac{4 \sin^2 \left(  \frac{\vec{p} \cdot \vec{b}}{2} \right)}{(\vec{p} \cdot \vec{b})^2}   \exp \left(  -   p^2  \frac{R_\perp^2}{D_\perp}  \right )    = 0
\eea
Thus $\omega_n^2 =  {\eta}^2 +  \sqrt{\eta^4 + \Omega_n^4}$ with

\be
\label{II}
\eta^2= \frac{g^2 M_\perp^3}{  D_\perp} \int\frac{d^{D_\perp + 1} p}{(2 \pi)^{D_\perp + 1}}   \frac{ p^2 }{p^2 + m^2 } \frac{4 \sin^2 \left(  \frac{\vec{p} \cdot \vec{b}}{2} \right)}{(\vec{p} \cdot \vec{b})^2}   \exp \left(  -   p^2  \frac{R_\perp^2}{D_\perp}  \right )
\ee
Since $M_\perp, R_\perp$ in (\ref{II}) involve $\omega_n$ implicitly, the evaluation of $\eta$ follows iteratively using numerical analysis.

\subsection{  Numerical Results: $D_\perp=3$}

\begin{figure}[!htb]
\minipage{0.48\textwidth}
\includegraphics[height=50mm]{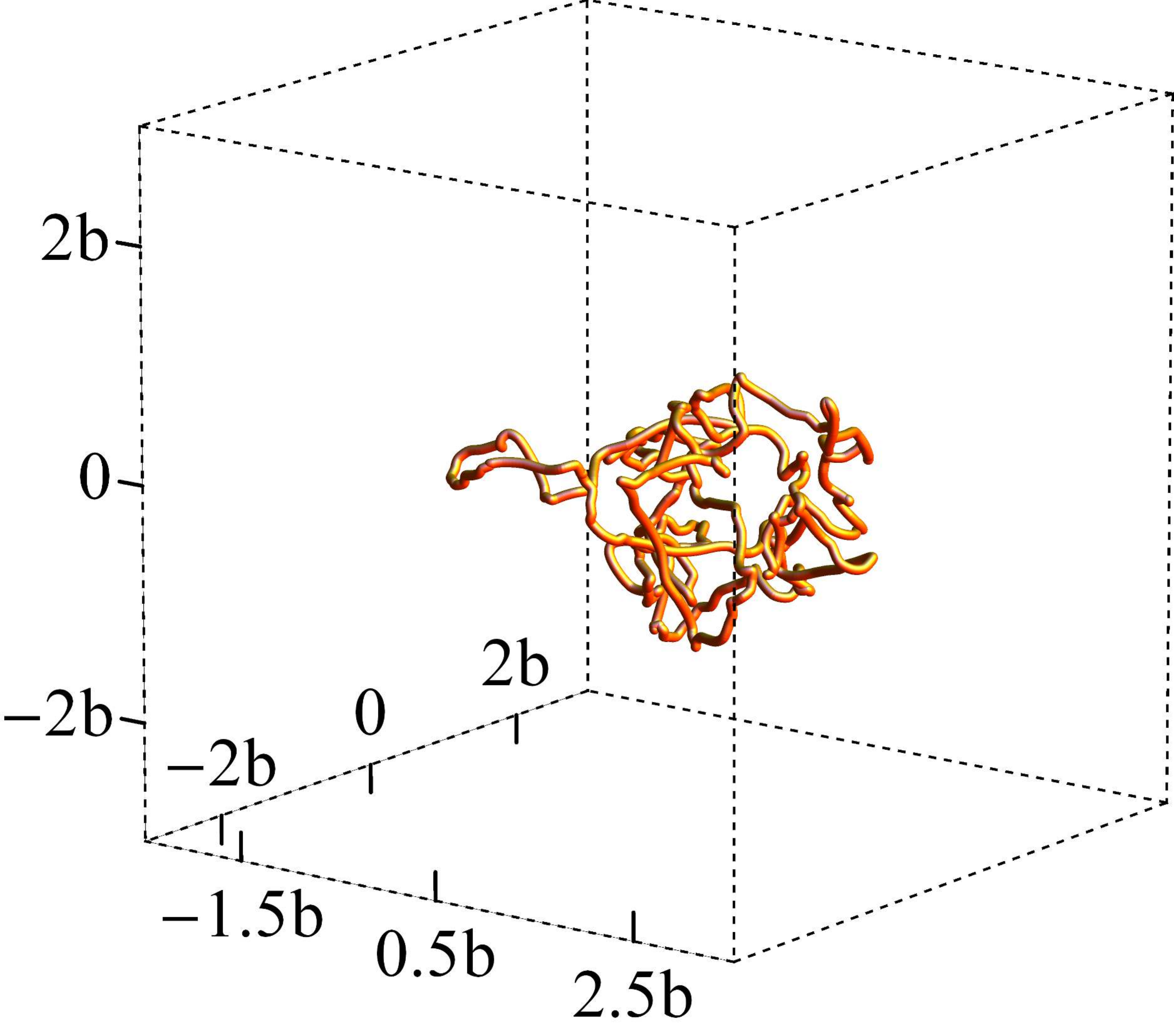}
\endminipage\hfill
\minipage{0.48\textwidth}
\includegraphics[height=50mm]{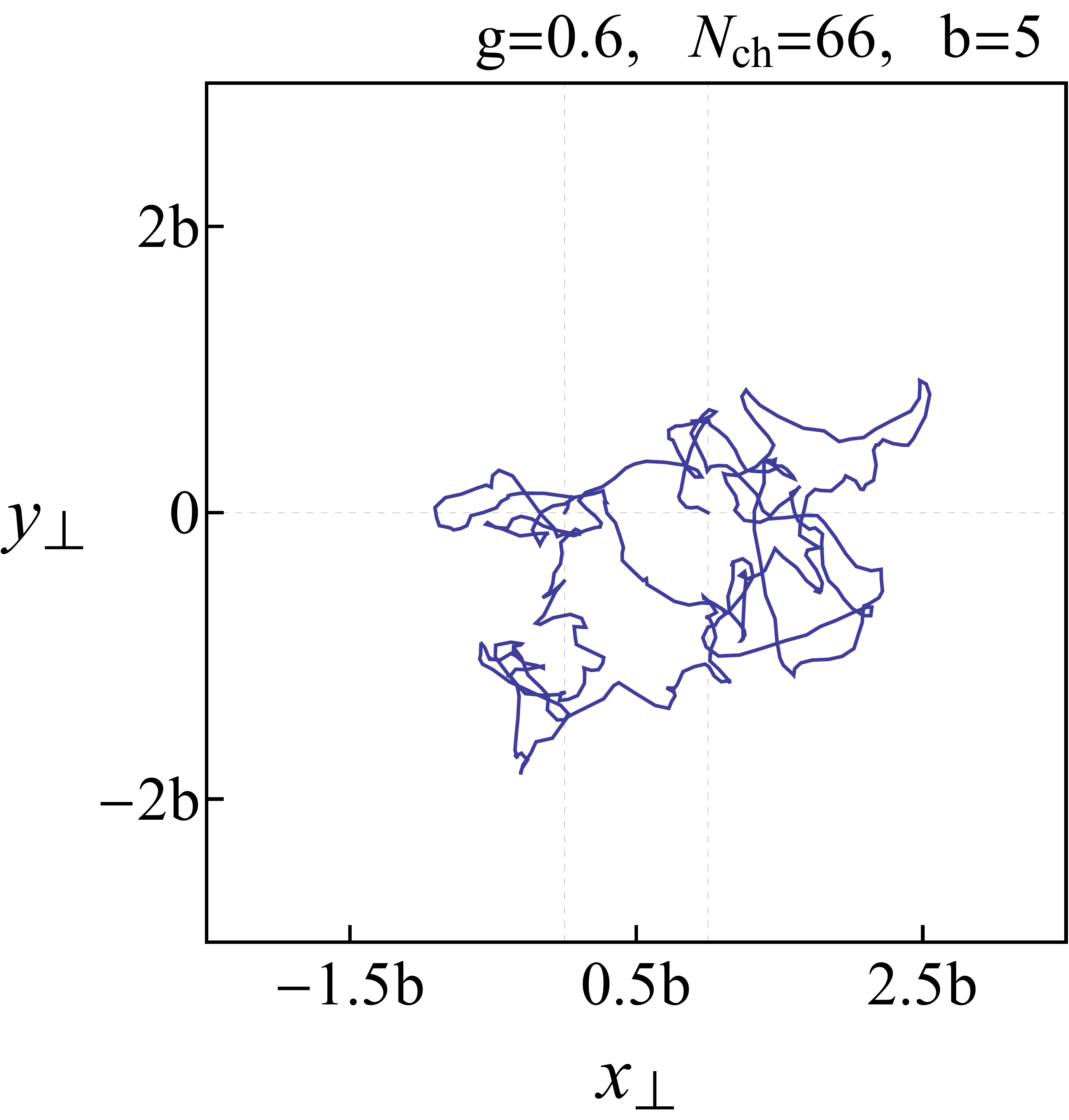}
\endminipage
  \caption{Interacting string with $g=0.6$ and $N_{ch}=66$ for a separation of $b=5=10l_s$ in $D_\perp=3$
  (left) and projected onto the 2-spatial dimensions (right).}\label{XDStringNch66g06}
\end{figure}

\begin{figure}[!htb]
\minipage{0.48\textwidth}
\includegraphics[height=50mm]{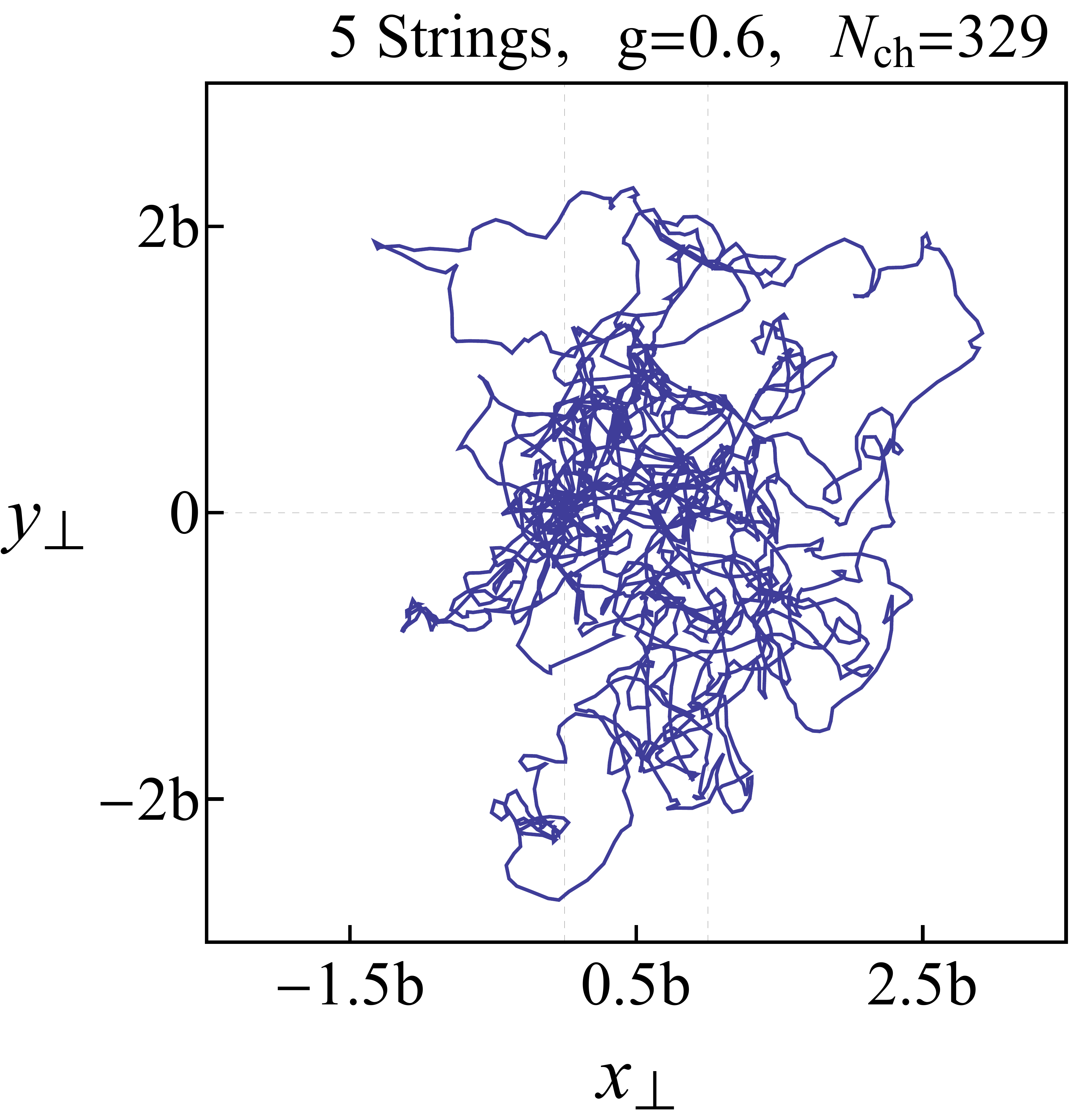}
\endminipage\hfill
\minipage{0.48\textwidth}
\includegraphics[height=50mm]{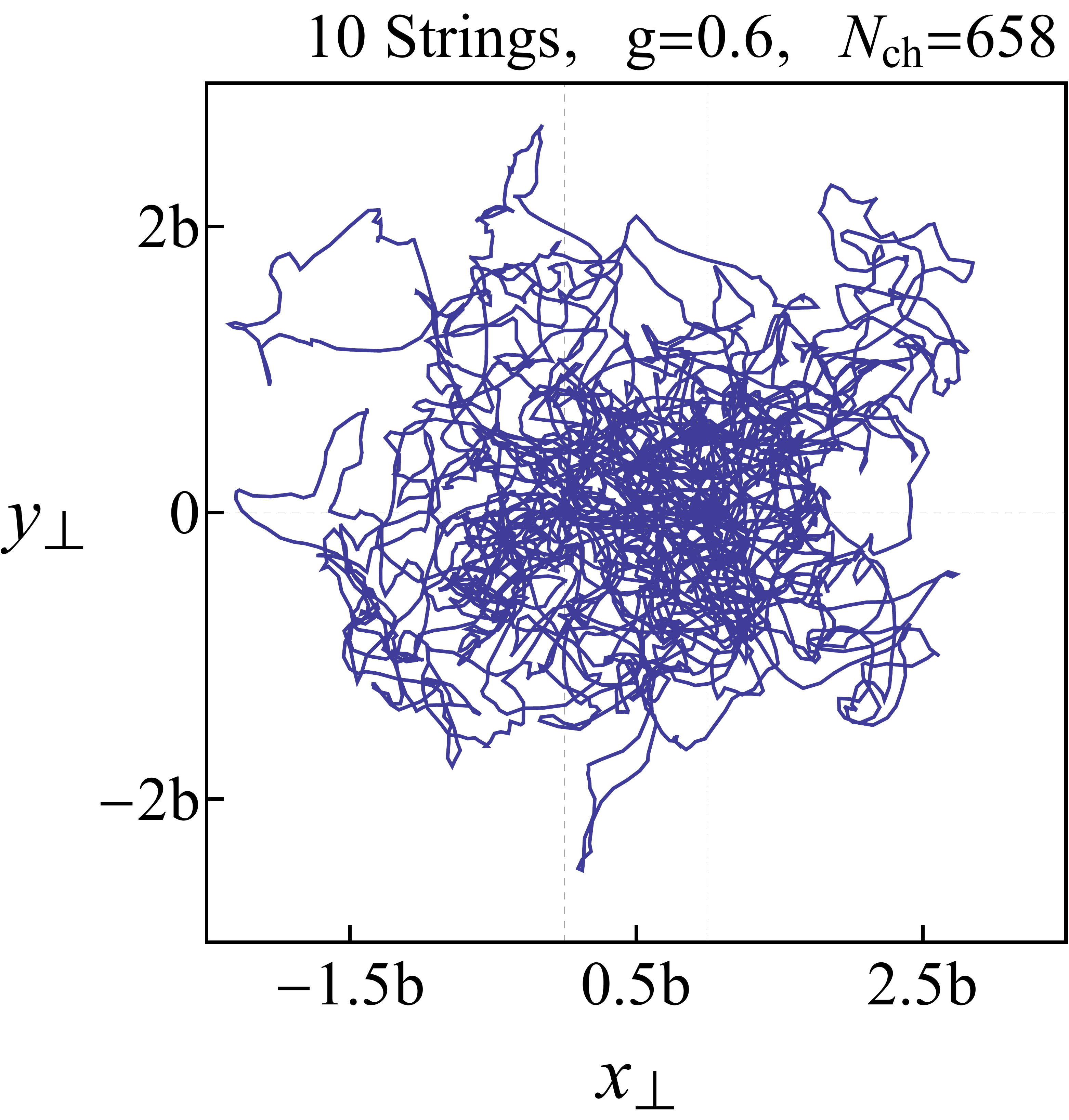}
\endminipage
  \caption{5 interacting strings with $g=0.6$ and $N_{ch}=329$ for a separation of $b=5=10l_s$ in $D_\perp=3$
  (left) and  the same for 10 interacting strings  and $N_{ch}=658$ (right).}
  \label{2DStringsg06}
\end{figure}

In  Fig.~\ref{XDStringNch66g06} we show the string shape for an interacting string 
with fixed end-points $\bold{b} = 5=10l_s$,  an effective temperature parameter $1/\bb =1/0.1$ or a charge multiplicity
of $N_{ch}=66$, and a coupling $g =0.6$. 
On the right the string is displaced in $D_\perp=3$ dimensions with a flat holographic direction. On the left, we show the same string
projected on the 2 transverse spatial directions only. 
Fig.~\ref{2DStringsg06} is the same as Fig.~\ref{XDStringNch66g06} with  the exchange of 5 strings and 10 strings with $g=0.6$.

\begin{figure}[!htb]
\minipage{0.48\textwidth}
\includegraphics[height=45mm]{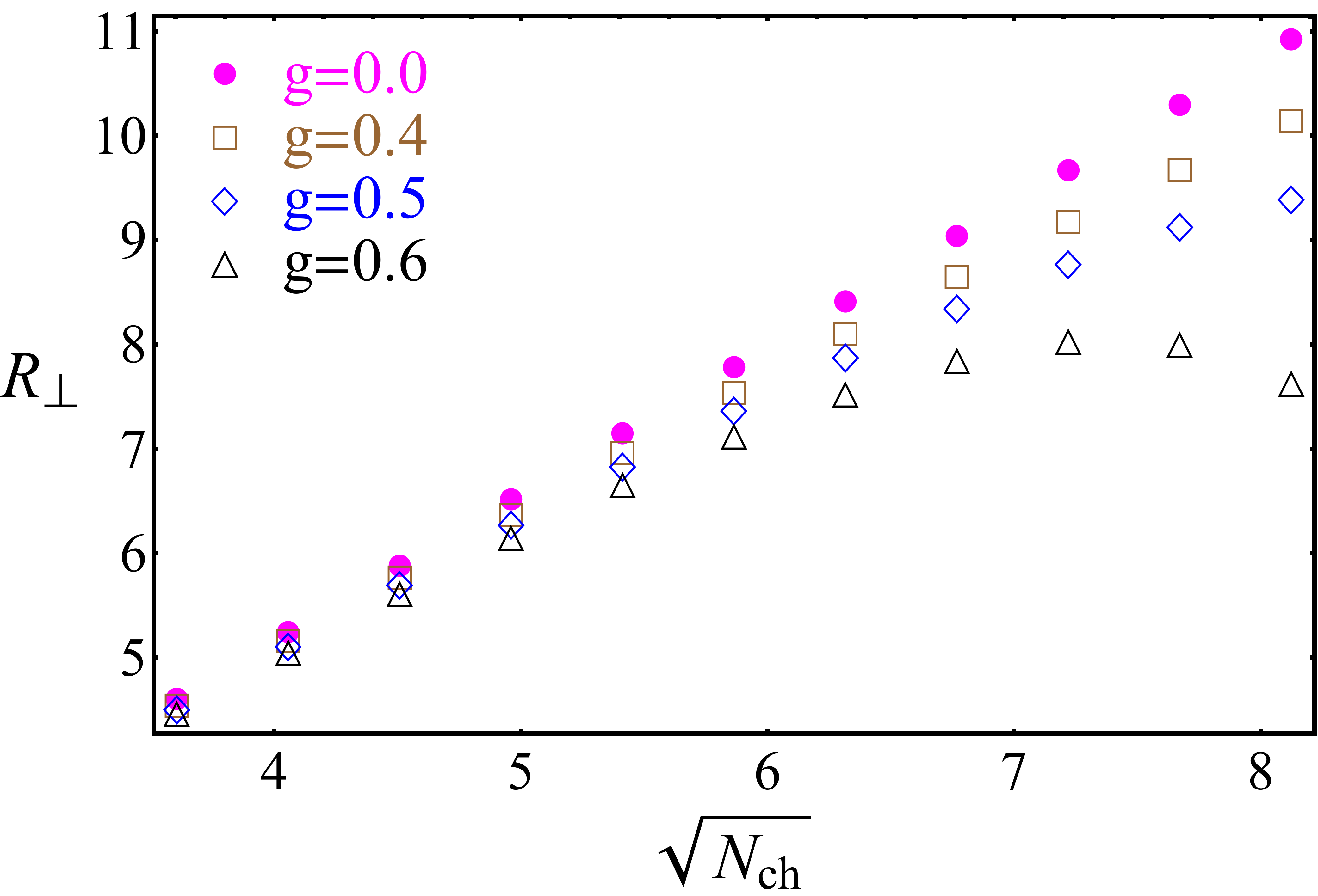}
\endminipage\hfill
\minipage{0.48\textwidth}
\includegraphics[height=45mm]{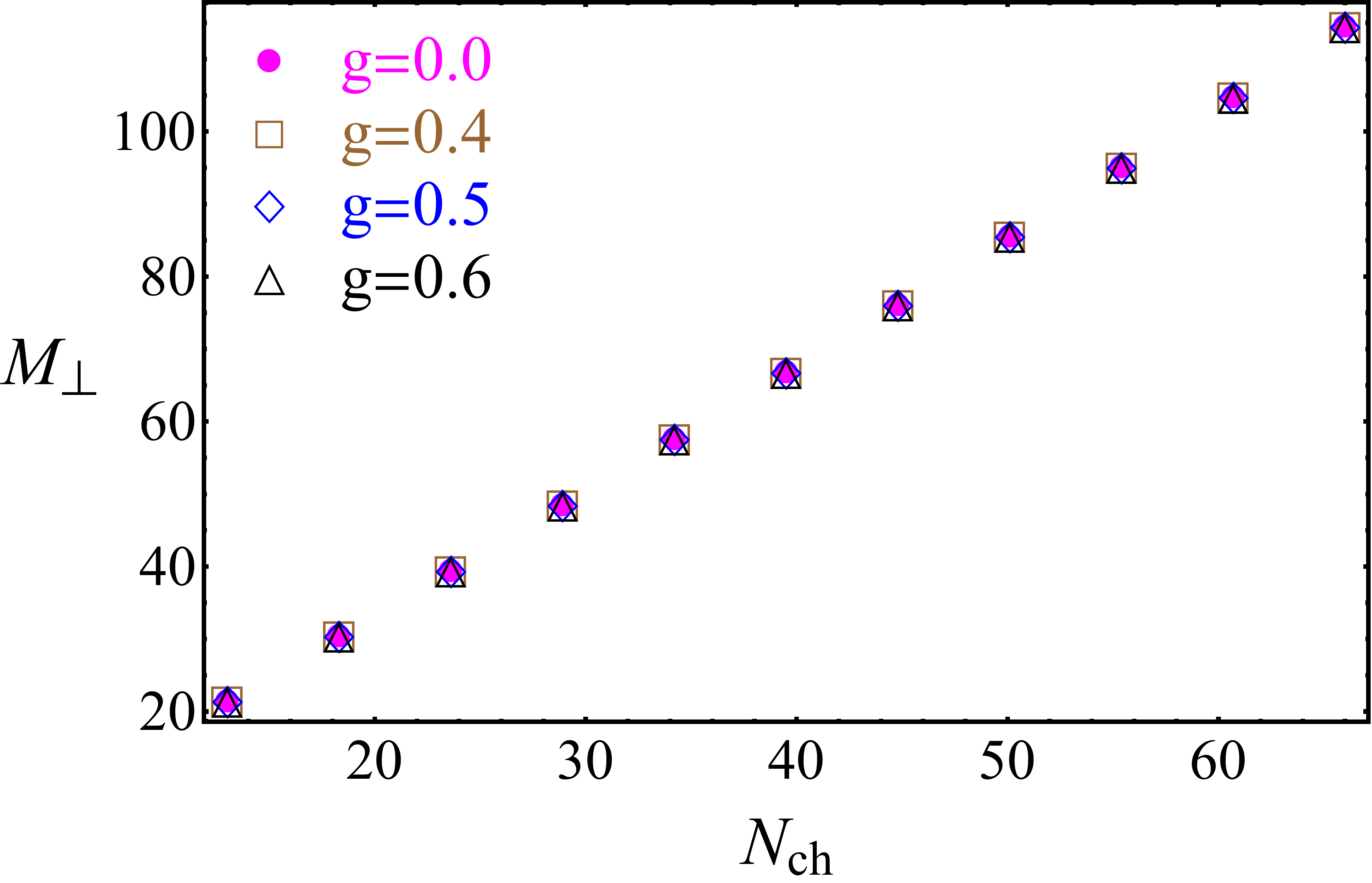}
\endminipage
  \caption{ Transverse size (left) and mass (right) of the string versus its multiplicity for $N=500$ string bits
  and different attractive self-couplings $g$. }
\label{XDiffg}
\end{figure}

\begin{figure}[!htb]
\minipage{0.48\textwidth}
\includegraphics[height=45mm]{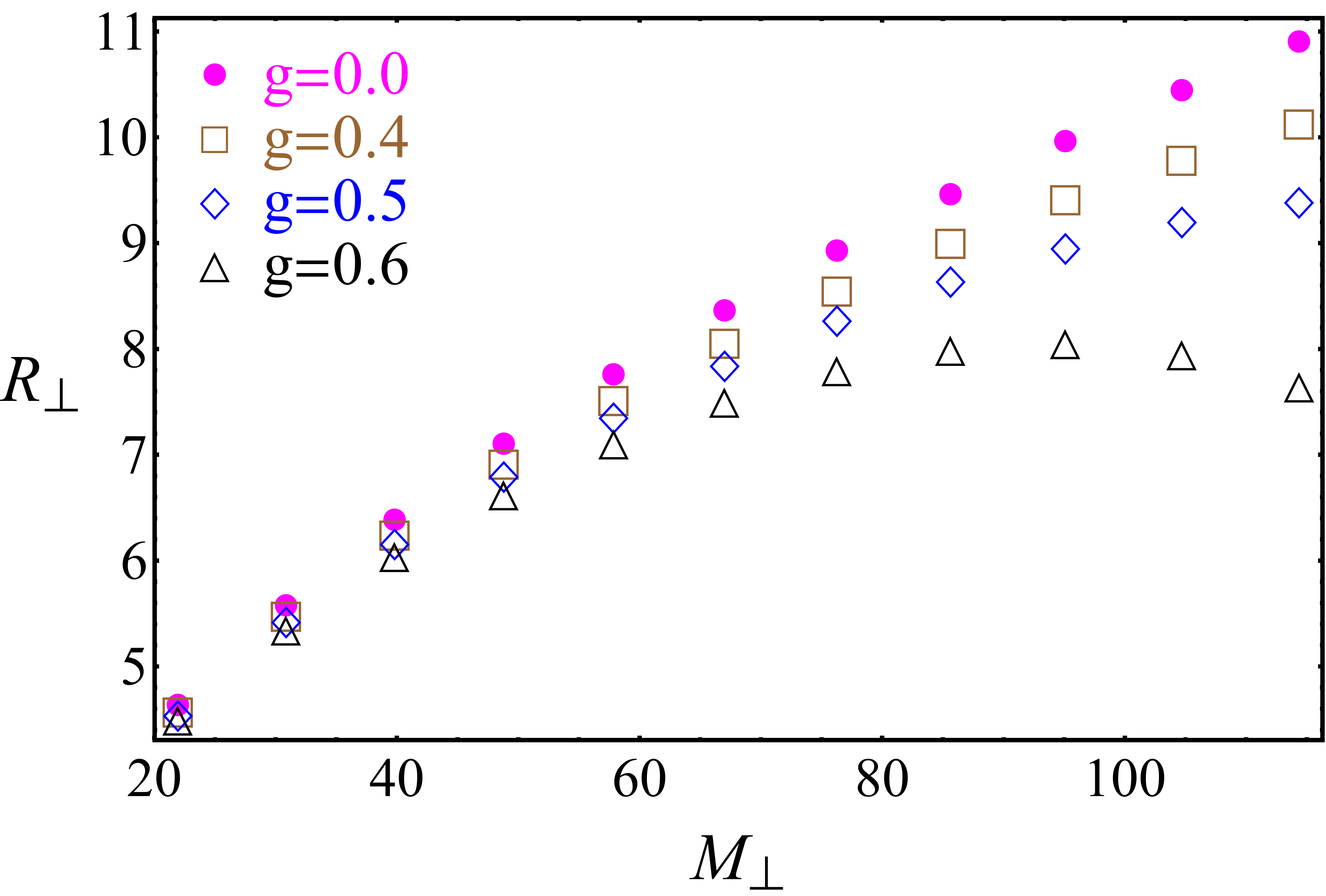}
\endminipage\hfill
\minipage{0.48\textwidth}
\includegraphics[height=45mm]{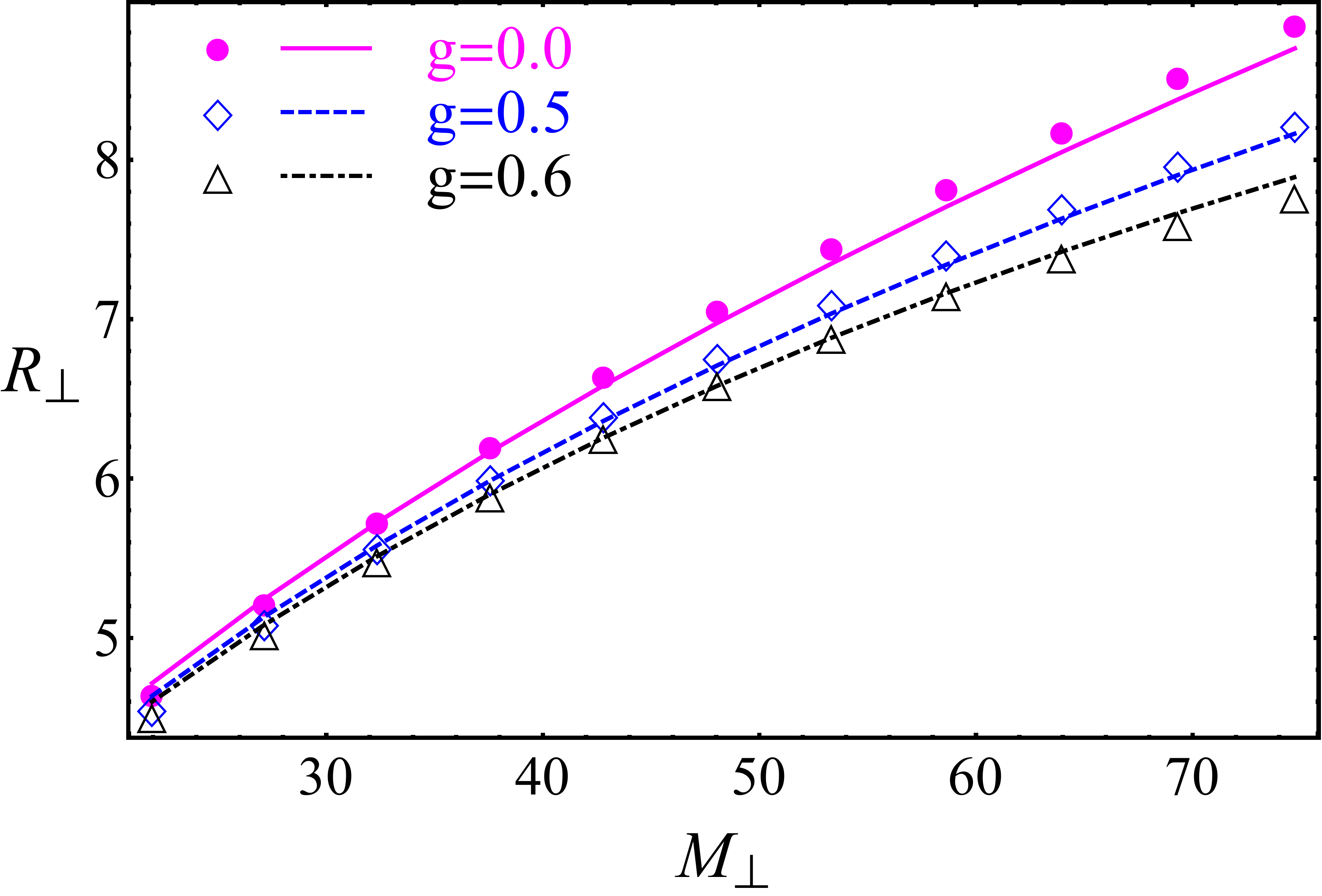}
\endminipage
\caption{ Transverse size of the interacting string for $N=500$ string bits and attractive self-coupling coupling $g$ 
  versus its mass (left). The solid curves are analytical results (right).}\label{RMDiffg}
\end{figure}

\begin{figure}[!htb]
\minipage{0.48\textwidth}
\includegraphics[height=45mm]{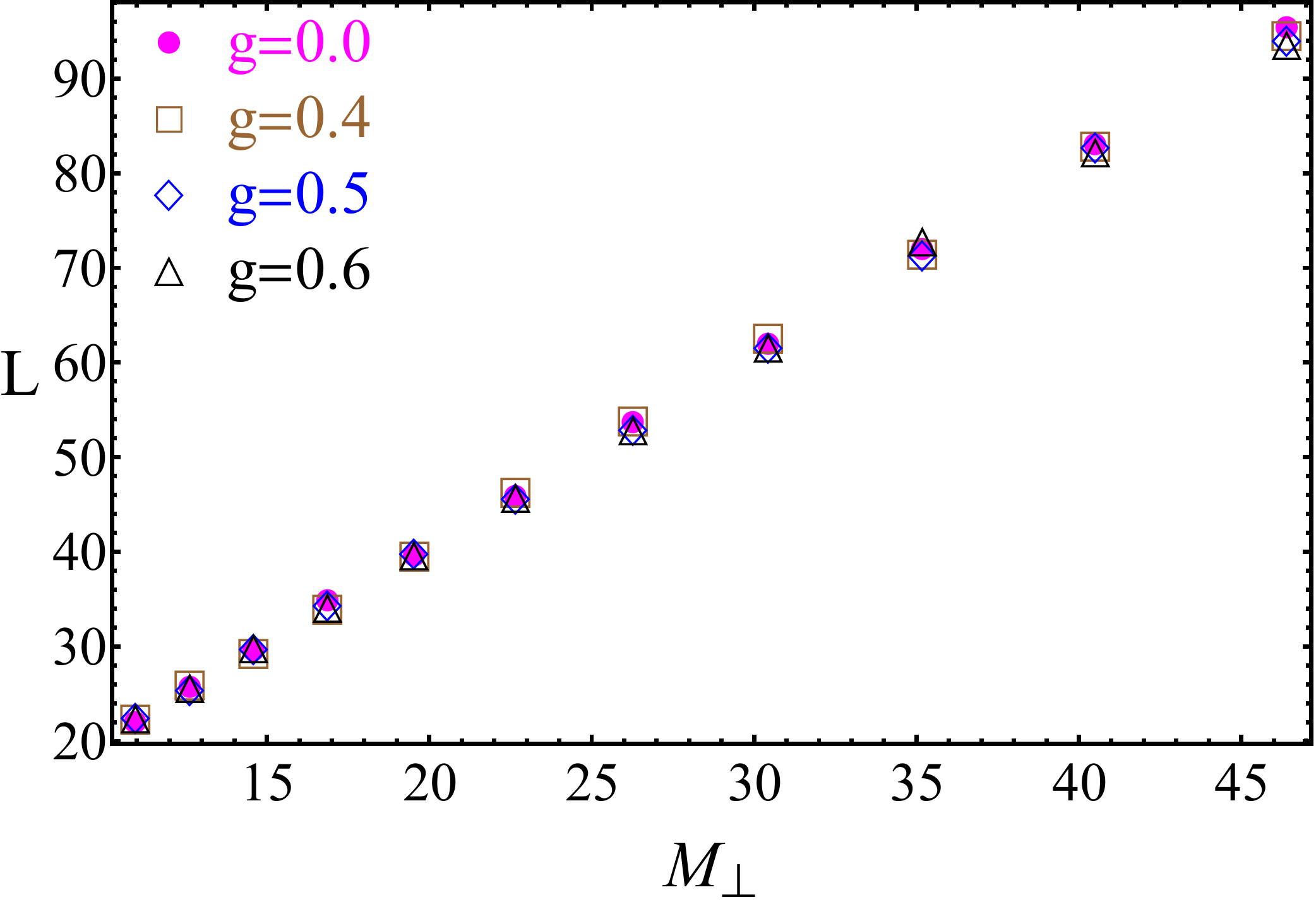}
\endminipage\hfill
\minipage{0.48\textwidth}
\includegraphics[height=45mm]{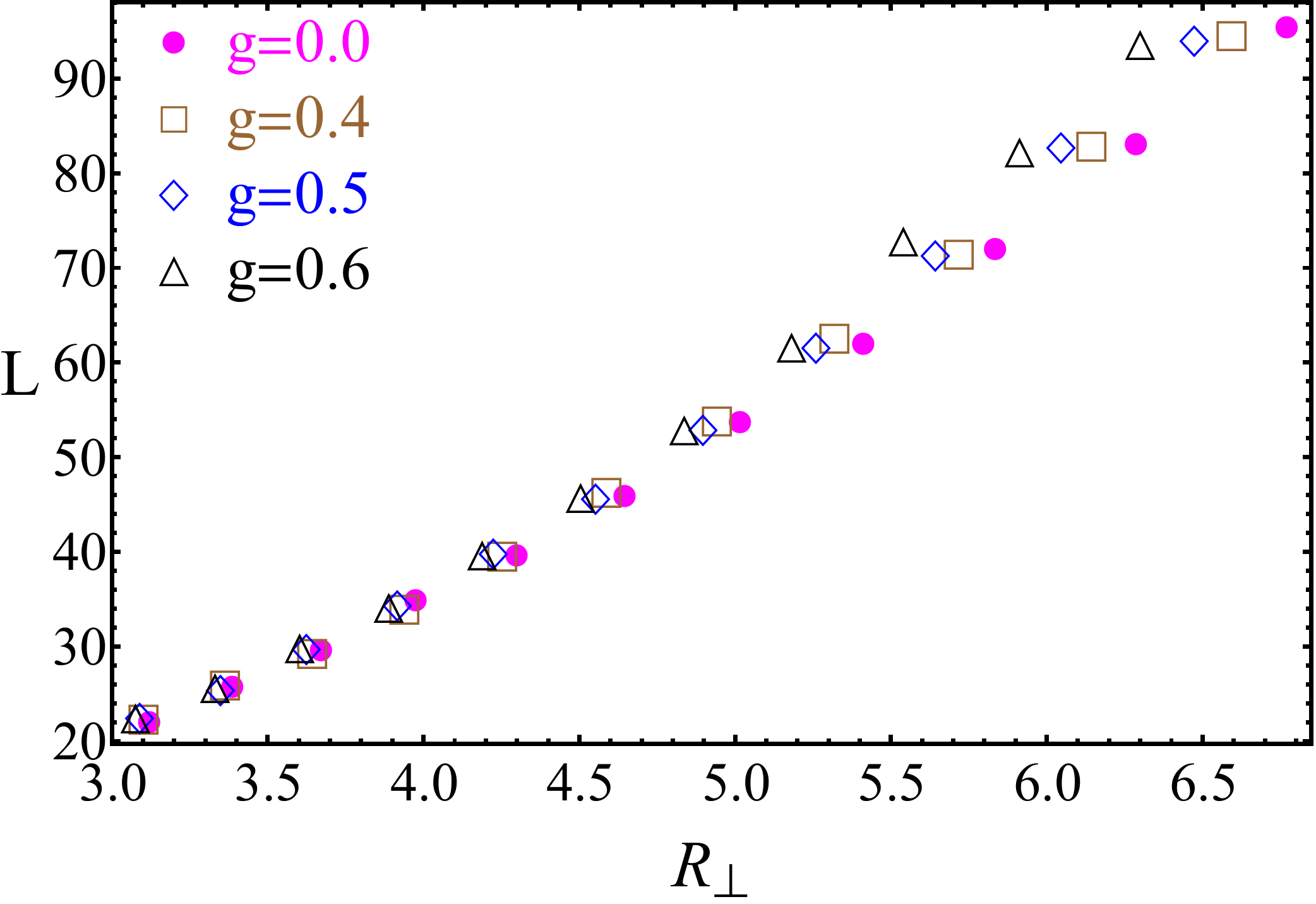}
\endminipage
  \caption{ Total length of the interacting string versus its mass (left) and its transverse size (right) for different attractive 
  self-coupling coupling $g$.The number of string bits is $N=100$. }\label{LMR}
\end{figure}

In Fig.~\ref{XDiffg}~(right) the single string mass versus the charge multiplicity $N_{\rm ch}$
following from  (\ref{leadingMperp}) is shown for a string at high resolution with $N=500$ and different attractive couplings. 
In Fig.~\ref{XDiffg}~(left) the transverse size $R_\perp$ versus $\sqrt{N_{ch}}$ following from~(\ref{leadingRperp}) for the same string parameters.
We note that the attraction does not change the mass or entropy, but does cause the string to contract transversally away from its free
diffusive thermal expansion. In  Fig.~\ref{RMDiffg} we show the transverse size versus the string mass (also entropy) 
or $R_\perp$ as given by~(\ref{leadingRperp})  versus  $M_\perp$ as defined in~(\ref{leadingMperp}). The lines in 
Fig.~\ref{RMDiffg} (right) corresponds to

 \be
 R_\perp^2 \approx  1.5 \sqrt{\frac{3 D_\perp}{2 \pi^2}} \sqrt{1 - 0.012 g^2 M_\perp} \,M_\perp
 \label{FITX}
 \ee
and  in overall agreement with the schematic analysis of the variational result in~(\ref{DIFF3}).  The latter suggests a
first order transmutation to a black-hole for sufficiently strong and attractive self-string interactions. (\ref{FITX}) shows that 
weak coupling but high temperature means $0.012 g^2 M_\perp<1$. Since for most of our analyses we  use
$M_\perp < 100$, this corresponds to $g<1$, hence our choices of $g=0.4, 0.5, 0.6$. For completeness,  
the length of the string ${\rm L}$ defined as

\be
L=
\left<\hspace{-1 mm} \left< \hspace{1 mm}\sum_{k=1}^N\sum_{i=1}^{D_\perp}
\left|x_k^i-x_{k-1}^i\right|\hspace{1 mm} \right>\hspace{-1 mm}\right>
\ee
versus    $M_\perp$~(\ref{leadingMperp}) and  $R_\perp$~(\ref{leadingRperp}) with the resolution $N=100$ for different coupling strengths $g$ are displayed in  Fig.~\ref{LMR}~(left) and Fig.~\ref{LMR}~(right).

\subsection{Schematic Analysis}

An understanding of the self-interacting string in the Hagedorn regime follows from the variational minimization of the free energy 
above. Here we note, that for no self-interaction or $g=0$, the
classical diffusive growth noted in (\ref{TS2}) follows from the fact that the kinetic term in the transverse Hamiltonian 
(\ref{HAR}) scales like $1/R^2$ by the uncertainty principle and does not favor short strings, while the confining 
harmonic term in (\ref{HAR}) does not favor long strings and scales like $R^2$. This trade-off is captured by minimizing
the schematic free energy

\be
{\cal F}_{0\perp}=M_\perp^2\left(\frac 1{R^2}+\frac{R^2}{M_\perp^2}\right)
\label{FREE0}
\ee
$d{\cal F}_{0\perp}/dR=0$ yields (\ref{TS2}). Self-interactions in $2+D_\perp$ are holographically dual to
the exchange of light excitations in bulk. As a result, (\ref{FREE0}) now reads

\be
{\cal F}_{\perp}\equiv {\cal F}_{0\perp}+M_\perp V=M_\perp^2\left(\frac 1{R^2} (1-g_s^2M_\perp) +\frac{R^2}{M_\perp^2}\right)
\label{FREE1}
\ee
after dropping terms of order 1. $d{\cal F}_{\perp}/dR=0$ now occurs for

\be
R^2_\perp\approx  \sqrt{1-g_s^2M_\perp}\,M_\perp
\label{DIFF3}
\ee
which is (\ref{FITX}) for $g^2M_\perp \ll 1$. However, for $g^2M_\perp\approx  1-1/M_\perp^2$ (\ref{DIFF3}) undergoes a first
order change into a fixed size string of few string lengths. The self-interacting string described variationally above 
begins its transmutation   to a black-hole as illustrated by the present schematic analysis.

\subsection{  Numerical Results: $2<D_\perp(\lambda)<3$}

\begin{figure}[!htb]
\minipage{0.48\textwidth}
\includegraphics[height=45mm]{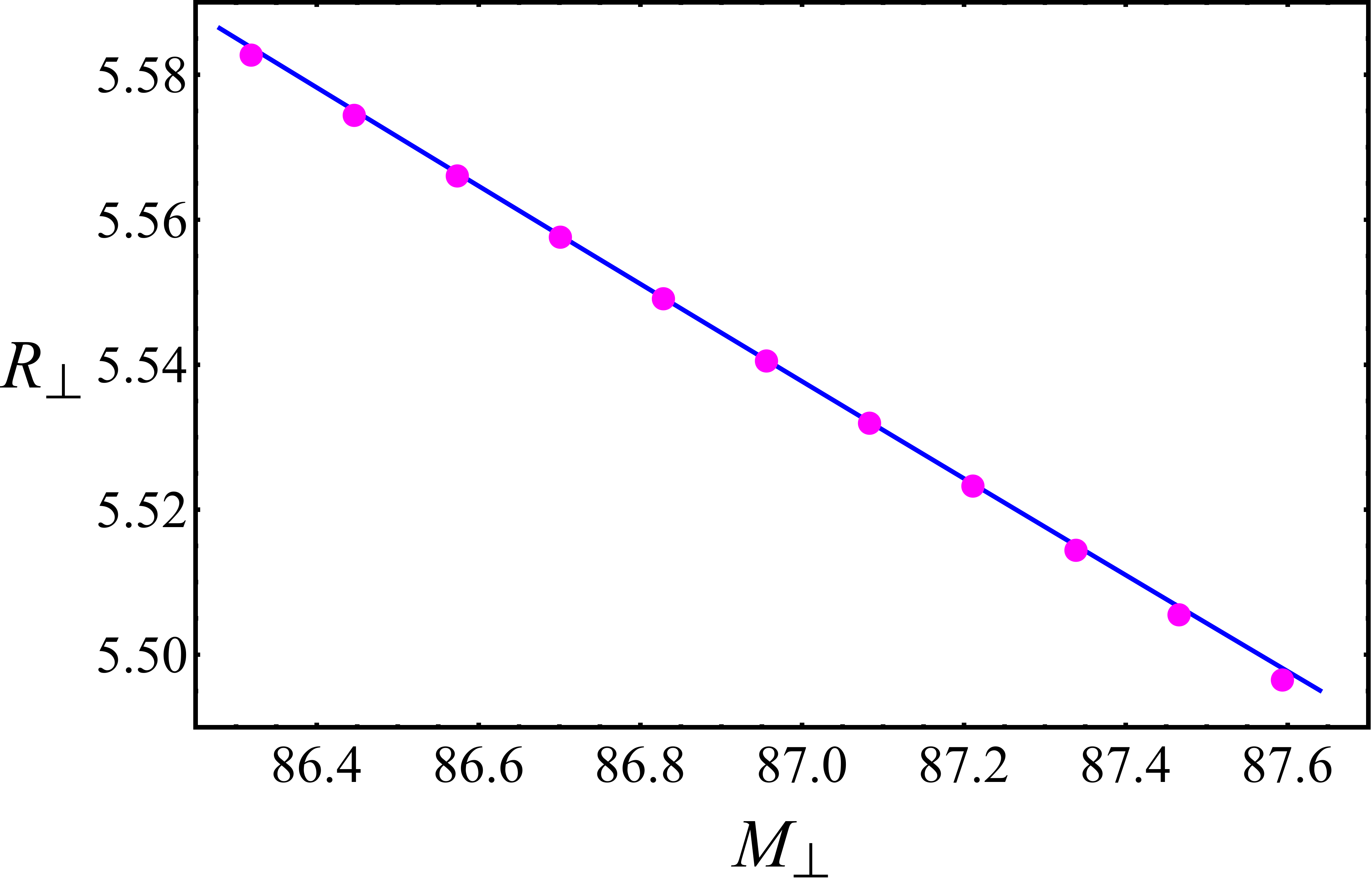}
\endminipage\hfill
\minipage{0.48\textwidth}
\includegraphics[height=50mm]{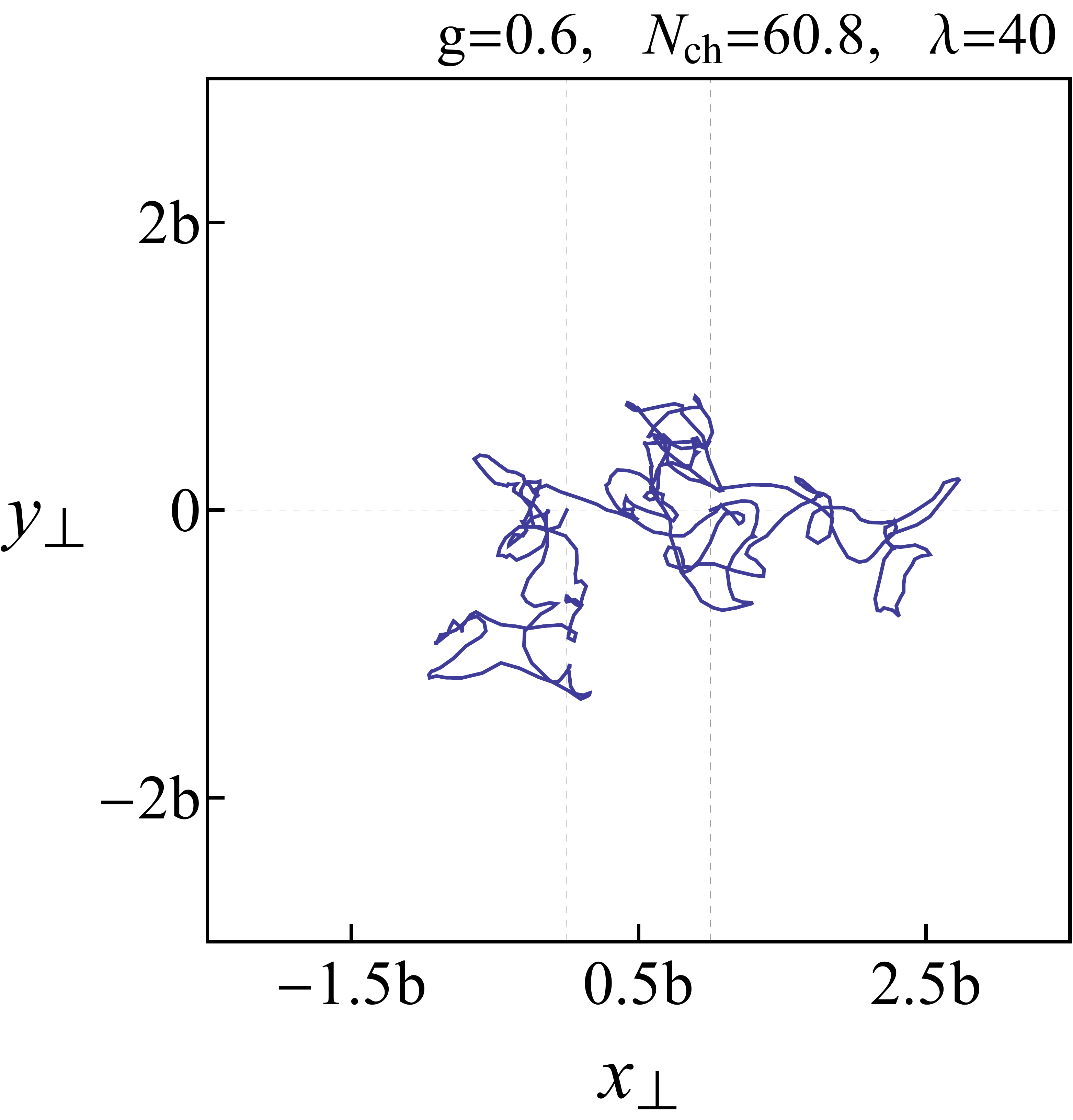}
\endminipage
  \caption{ Left: Solid line is $R_\perp \sim 209 (1/g^2 M_\perp)^{2 \sqrt{\lambda}/3(D_\perp - 1)^2}$ and the dots are numerical results. Right: a typical string with multiplicity $N_{ch} = 60.8$. }\label{Curve}
\end{figure}

 An exact  treatment of the transverse string in curved AdS$_5$ space is beyond the scope of this work. 
In this section we will give simple estimates of the effects of the curvature of AdS$_5$ on some of 
our previous results. One of the main effect of the curved geometry on the Pomeron
is to cause the string transverse degrees of freedom to effectively feel a reduced transverse spatial dimension 
~\cite{Stoffers:2012zw,Stoffers:2012ai,Stoffers:2013tla,Shuryak:2013sra}

\be
D_\perp\rightarrow D_\perp(\lambda)=D_\perp\left(1-\frac{3(D_\perp-1)^2}{2D_\perp\sqrt{\lambda}}+{\cal O}\left(\frac 1\lambda\right)
\right)
\label{DEFFXL}
\ee
with $\lambda=g^2_{YM}N_c$.  (\ref{DEFFXL}) causes the Pomeron intercept to move from $D_\perp/12=0.25$
to $D_\perp(\lambda \approx 40)\approx 0.17$ closer to the empirical  interceptt of $0.08$~\cite{Donnachie:1992ny}.  
A phenomenological way to implement this effect is to
add warping factors on the oscillators in (\ref{XS1}) as we noted in our recent analysis~\cite{Qian:2014jna}.
This will be used 
in our numerical results to follow. A simple estimate follows from the 
substitution (\ref{DEFFXL}) in the schematic analysis. Indeed,
the  estimate in (\ref{FREE1}) shows that the first contribution reflects on the uncertainty principle
which probes short distances and thus is
not sensitive to the curvature of AdS$_5$. The second diffusive contribution is sensitive through $D_\perp$
but will turn out to be
sub-leading as we will show below. The third contribution is long ranged and senses the curvature of AdS$_5$.
Thus

\be
{\cal F}_{0\perp}\rightarrow  M_\perp^2
\left(\frac{1}{R_\perp^2}+\frac{R_\perp^2}{D_\perp(\lambda)\,M_\perp^2}-\frac{g^2M_\perp}{R_\perp^{D_\perp(\lambda)-1}}\right)
\label{EEXL}
\ee
For very small values of $g$ the first two contributions in (\ref{EEXL}) are dominant and the string
transverse size grows diffusively. The minimization of the first two dominant contributions in this regime 
yields $R^2_\perp\approx \sqrt{D_\perp(\lambda)}\,M_\perp$, in agreement with (\ref{TS2}).
However, for

\be
g^2 M_\perp> M_\perp^{\frac{D_\perp(\lambda) - 3}{2}}
\label{CONDX}
\ee
the string size shrinks and the transverse string size follows from balancing the first term 
with the last term due to the interaction. The balance
between the self-interaction and the uncertainty principle, yields a  continuously
decreasing transverse string size

\be
R_\perp\approx \left(\frac 1{g^2 M_\perp}\right)^{\frac{2\sqrt{\lambda}}{3(D_\perp-1)^2}}
\label{RNEW}
\ee
in units of the string length.  A black-hole with a transverse string size emerges for $g^2M_\perp\approx 1$.
In~Fig.~\ref{Curve} (right) we display the interacting string in the effectively curved space for $\lambda=40$, 
$D_\perp = 3$ and $g = 0.6$. In~Fig.~\ref{Curve} (left) we display the transverse size of the interacting
string in the effectively curved space versus $M_\perp$. The dots are from the numerically simulated string,
while the line is a fit to the schematic result (\ref{CONDX}) with 
$R_\perp \approx 209 (1/g^2 M_\perp)^{2 \sqrt{\lambda}/3(D_\perp - 1)^2}$ in a narrow range
of $M_\perp$.

\section{\label{epsilon} Angular Deformations}

 The fluctuating string with fixed end-points exhibits large azimuthal deformations in the transverse plane that can be characterized by the azimuthal moment~\cite{Bzdak:2013rya, Kalaydzhyan:2014tfa}

\be
\epsilon_n = \frac{ \frac{1}{N}\sum_{i}^N  e^{i n \phi_i}  \left( r^\perp_i \right)^n  }{  r_\perp^n }
\ee
with $Nr_\perp^n = \sum_{i}^N     \left( r^\perp_i \right)^n$. Here $\phi$ is the azimuthal angle as measured from the impact parameter line along ${\bf b}$. $r_\perp$ is the averaged size of the string in the transverse plane. For ${\bf b}=0$, we have $ \left< r_\perp^2  \right> /2  = R_\perp^2 / D_\perp$, where $\left< \cdots \right>$ is the average over string ensembles. Specifically,  define $x  \equiv x_\perp^{i=1}$ 
and $y  \equiv x_\perp^{i=2}$ in the transverse plane, where $x $ is parallel to the impact parameter $\bold{b}$ and $y $ perpendicular to it,
\bea
\label{EE1}
x_\perp (k , \tau) =    \sum_{n=1}^{N-1}  X_{n}  (\tau) \sin \left( \frac{n k}{N } \pi\right)+ b   \frac{k}{N }   \qquad \qquad
y_\perp (k , \tau) =   \sum_{n=1}^{N-1}  Y_{n}  (\tau) \sin \left( \frac{n k}{N } \pi\right)    
\eea
where both string bit coordinates $X_n, Y_n$ are normally distributed according to

\be
X_n \sim   \mathcal{N} \left(0  ,  \frac{1}{\omega_n \left[e^{\frac{\bb}{2} \left( \omega_n + \frac{\Omega_n^2}{\omega_n}   \right) } - 1\right] }\right)   \ \ \ \ \ \ \ \ \ \ \ \  Y_n \sim  \mathcal{N} \left(0  ,  \frac{1}{\omega_n \left[e^{\frac{\bb}{2} \left( \omega_n + \frac{\Omega_n^2}{\omega_n}   \right) } - 1\right] }\right)  
\ee
Since (\ref{EE1}) are themselves sum of random walks, they are both normally distributed according to 

\bea
x_\perp (k , \tau)  \sim  \mathcal{N} \left( b   \frac{k}{N } ,  \Sigma_k^2 \right) \ \ \ \ \ \ \ \ \ \ \ \  y_\perp (k , \tau)  \sim   \mathcal{N} \left(0 , \Sigma_k^2  \right)
\eea
with the squared variance

\be
\Sigma_k^2  = \sum_{n=1}^{N-1}  \frac{\sin^2 \left( \frac{n k}{N } \pi\right) }{\omega_n \left[e^{\frac{\bb}{2} \left( \omega_n + \frac{\Omega_n^2}{\omega_n}   \right) } - 1\right] }
\ee

\begin{figure}[!htb]
\minipage{0.48\textwidth}
\includegraphics[height=45mm]{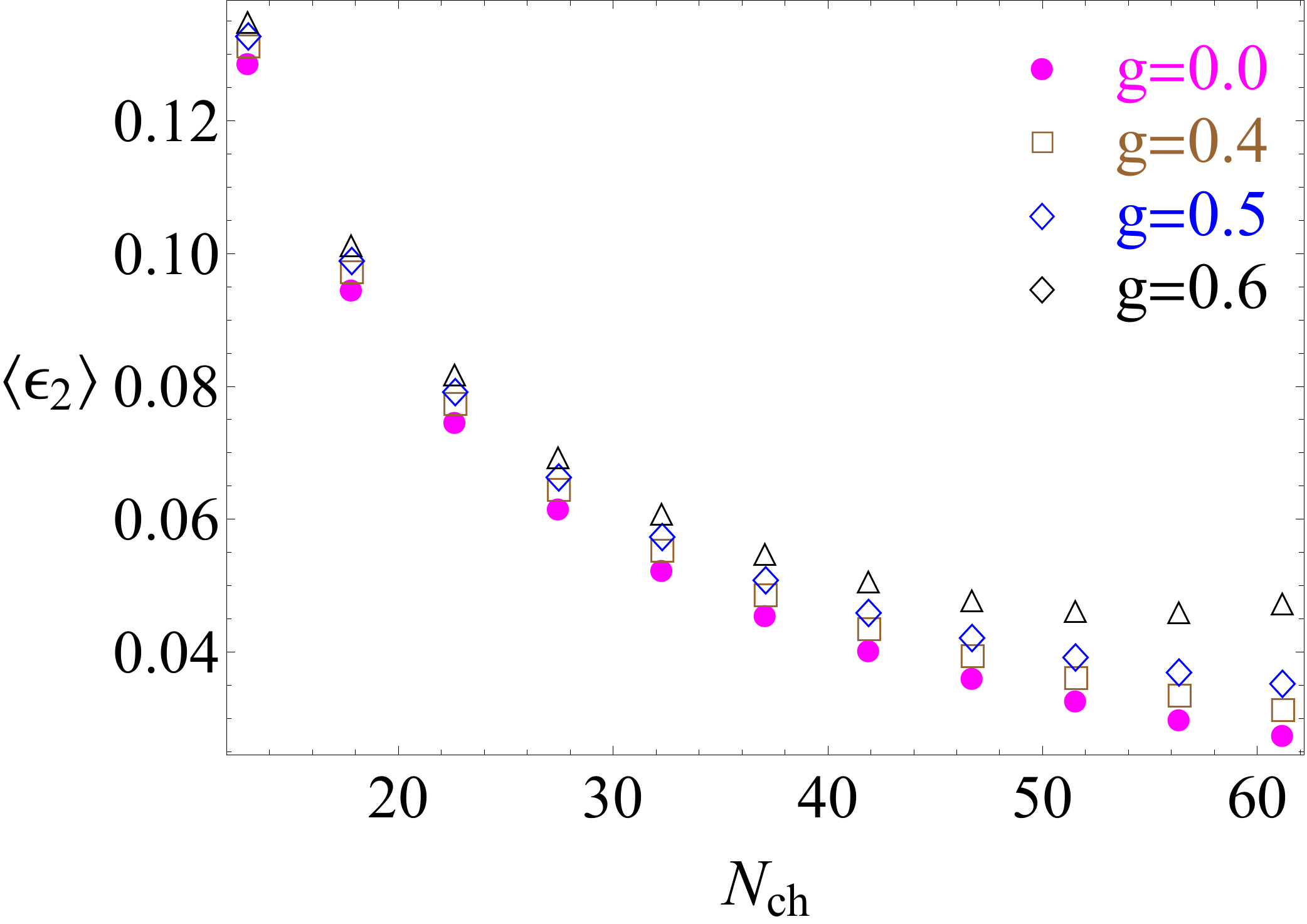}
\endminipage\hfill
\minipage{0.48\textwidth}
\includegraphics[height=45mm]{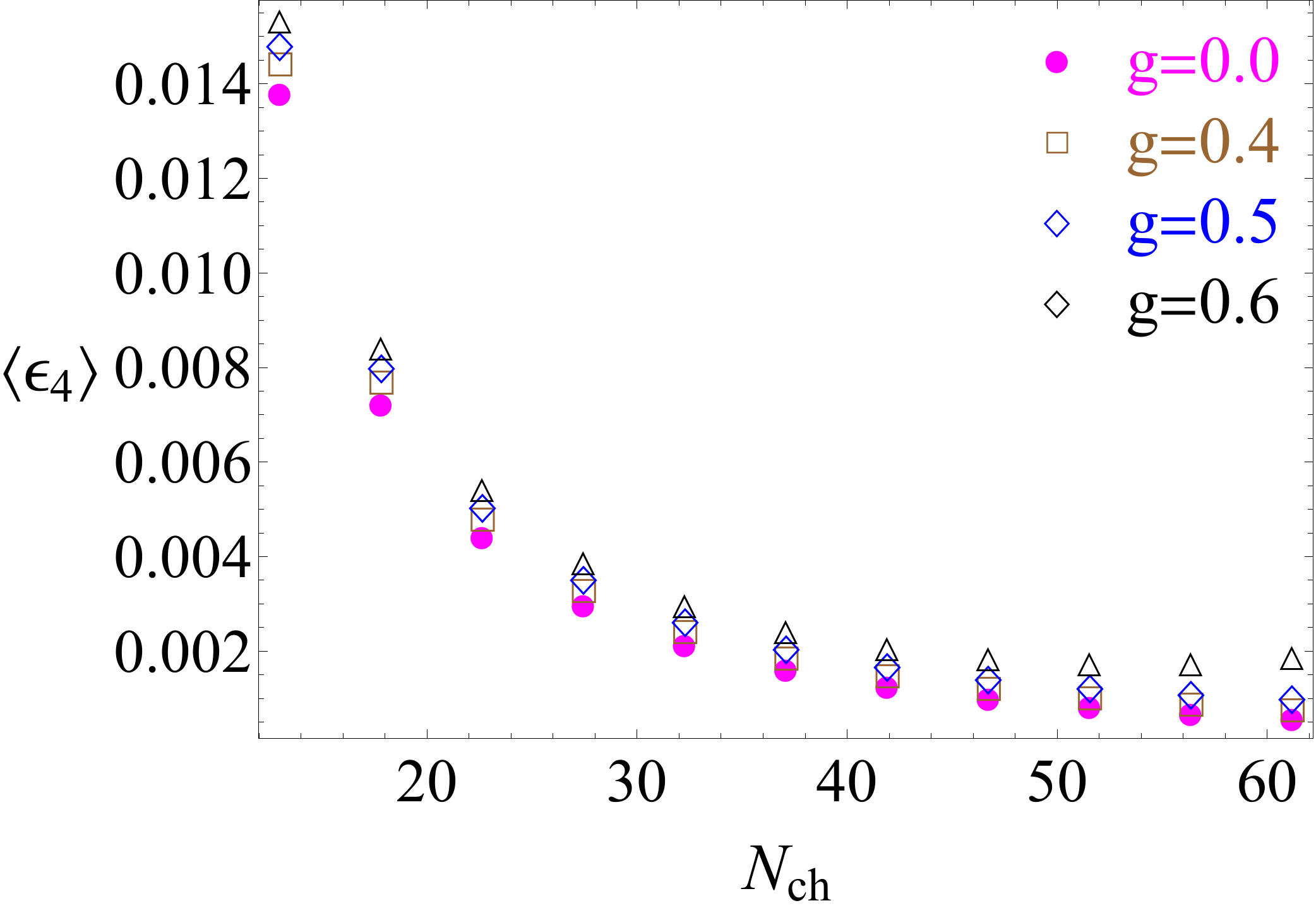}
\endminipage
  \caption{The azimuthal moments $\left<\epsilon_{2,4}\right>$ versus multiplicity for a single string with attractive self-coupling
  $g$. }\label{ExDiffg}
\end{figure}

For $N \rightarrow \infty$,  the squared variance is $\Sigma_{\tilde{k}}^2 \approx  {R_\perp^2}/{D_\perp}$ and the 
moments simplify (even $n$)

\be
\left<\epsilon_n \right> \approx \frac{b^n}{\left< r_T^n  \right> }   \int_0^1 d \tilde{k}       \left(  \frac{1}{2} -  \tilde{k} \right)^n      
= \frac{b^n}{\left< r_T^n  \right> }  \frac{1}{ 2^n (1 + n)}
\ee
 
\be
 \frac{\left< r_T^2  \right> } {b^2} \approx \frac{1}{12}  +    \frac{2}{ D_\perp} \frac{ R_\perp^2 }{b^2}   \qquad\qquad
 \frac{\left< r_T^4  \right> } {b^4} \approx \frac{1}{80}  + \frac{2}{3}   \frac{R_\perp^2 }{b^2 D_\perp }  +   8     \frac{ R_\perp^4 }{ D_\perp^2 b^4}  
\ee
For small $b$, the lowest moments reduce to 

\be
\left<\epsilon_2 \right>  \approx \frac{D_\perp}{24}  \frac{b^2}{    R_\perp^2  }\qquad\qquad
\left<\epsilon_4\right>  \approx \frac{D_\perp^2}{640}  \frac{b^4}{    R_\perp^4  }
\ee
The numerical results of   $\left< \epsilon_2 \right>$ and   $\left< \epsilon_4 \right>$ with 
a maximum resolution of $N=500$  are displayed in   Fig.~\ref{ExDiffg}.

\begin{figure}[!htb]
\minipage{0.33\textwidth}
\includegraphics[width=53mm]{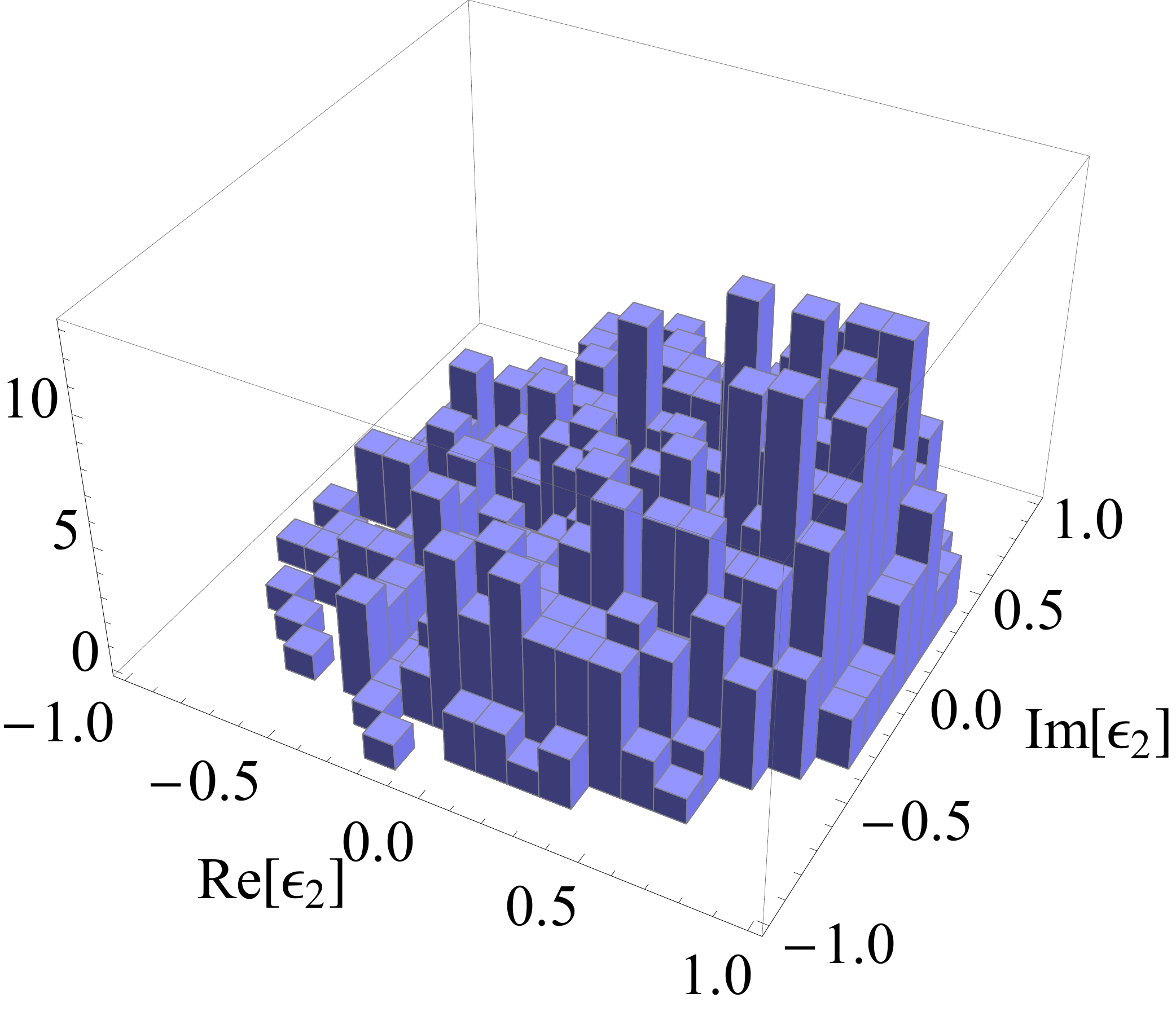}
\endminipage\hfill
\minipage{0.33\textwidth}
\includegraphics[width=53mm]{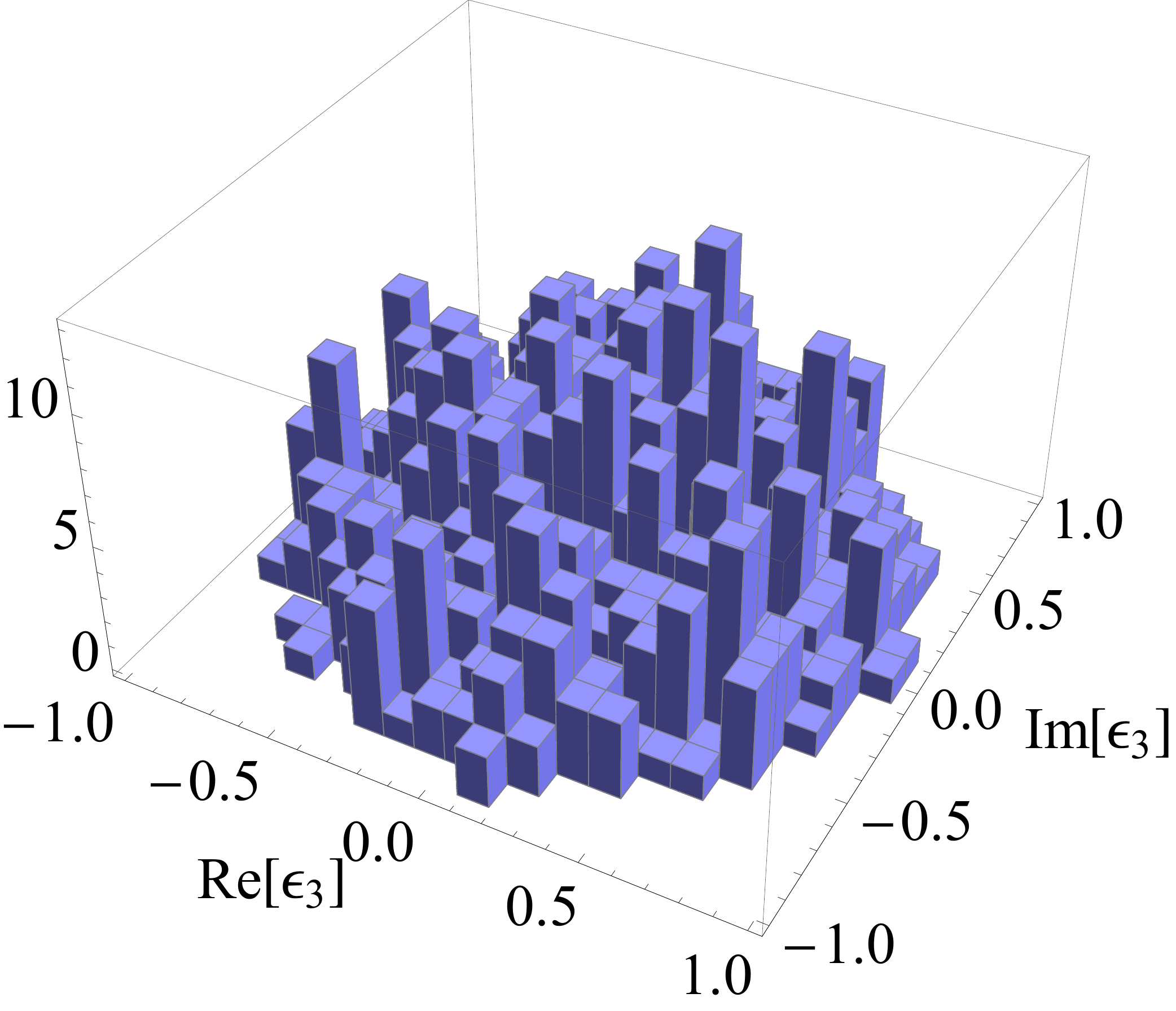}
\endminipage
\minipage{0.33\textwidth}
\includegraphics[width=53mm]{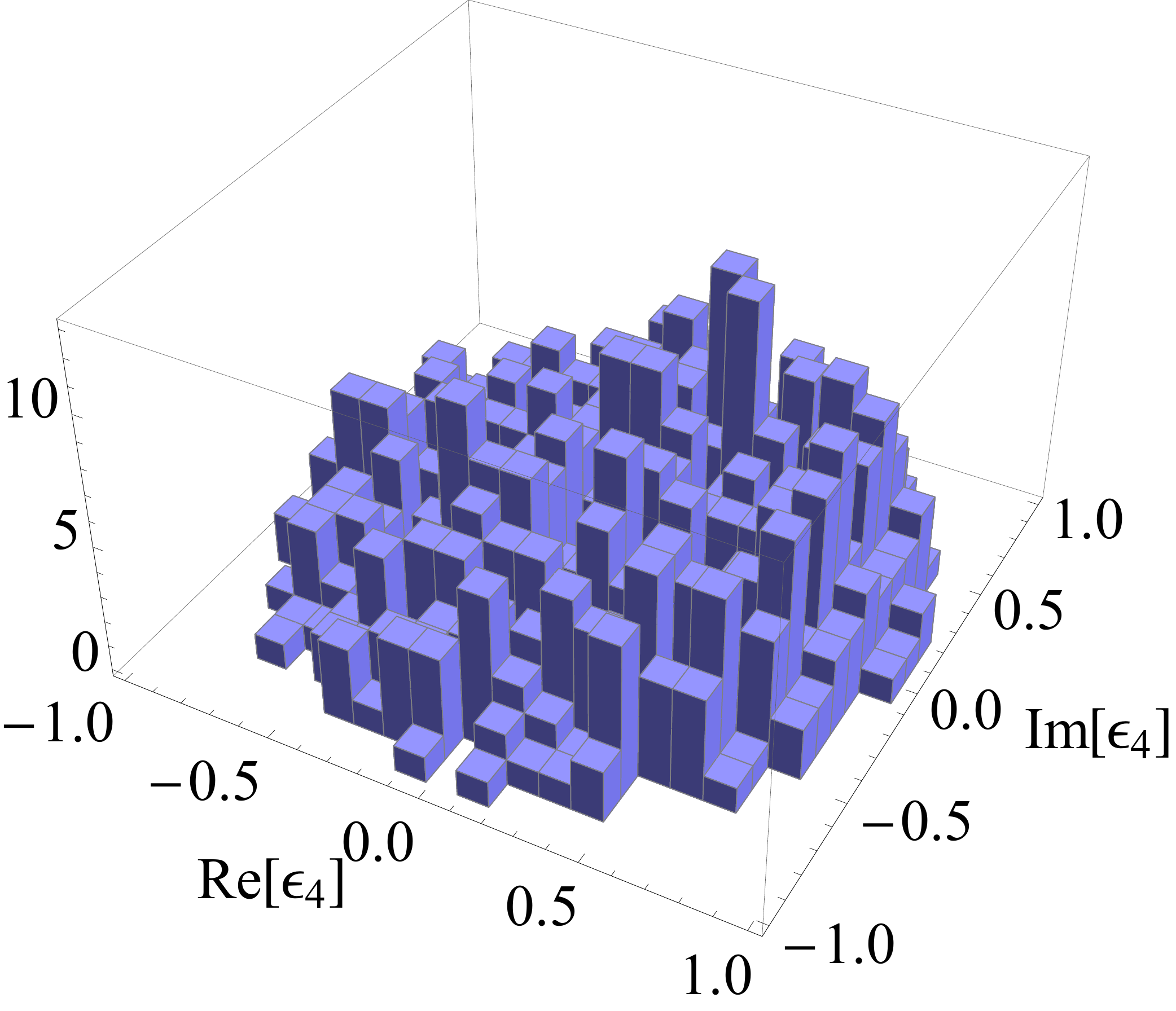}
\endminipage
  \caption{3D Histograms, 1000 random generated strings. N=100 and $N_{ch}=7$.  }\label{Histogram3D}
\end{figure}

\begin{figure}[!htb]
\minipage{0.33\textwidth}
\includegraphics[width=53mm]{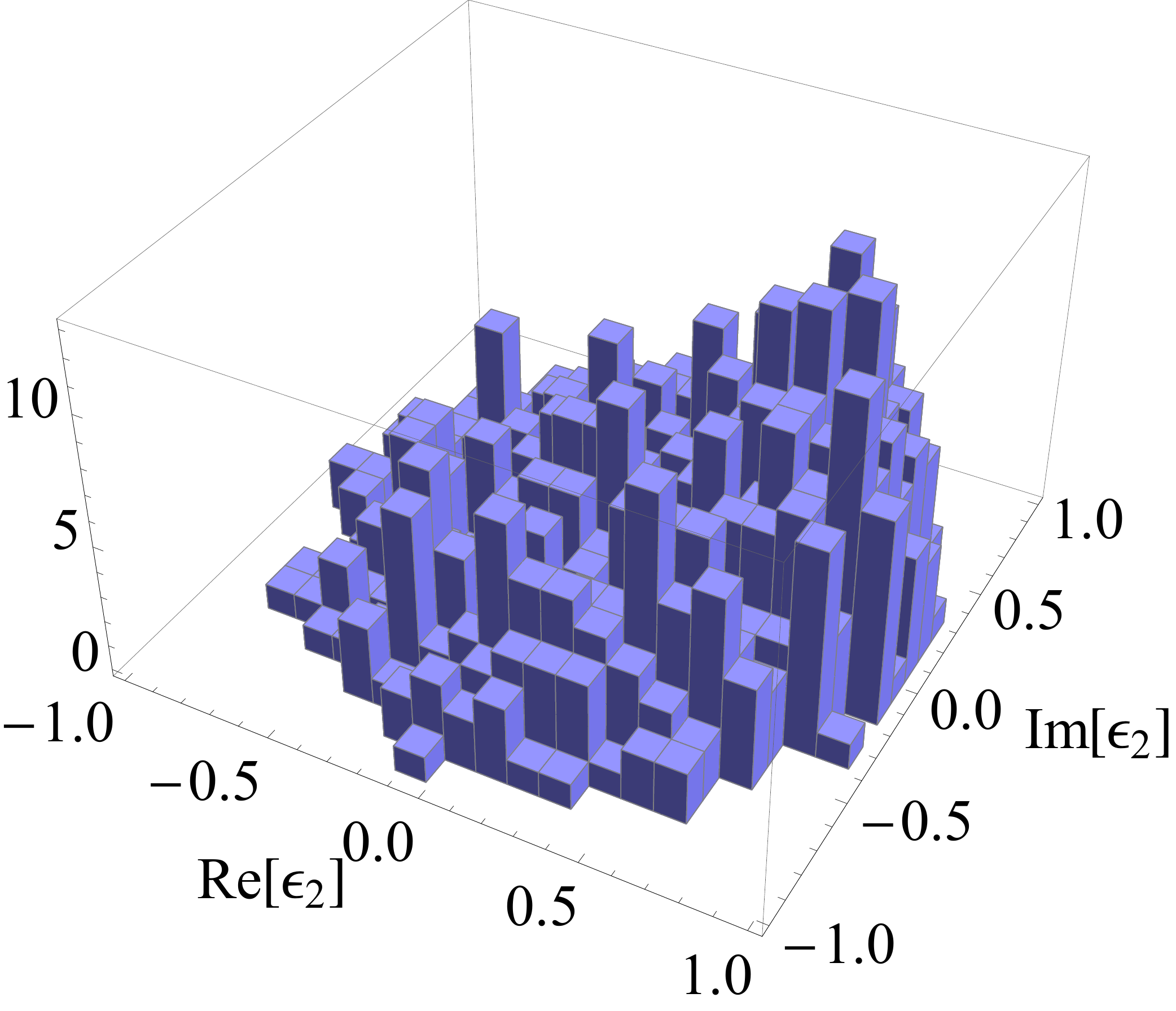}
\endminipage\hfill
\minipage{0.33\textwidth}
\includegraphics[width=53mm]{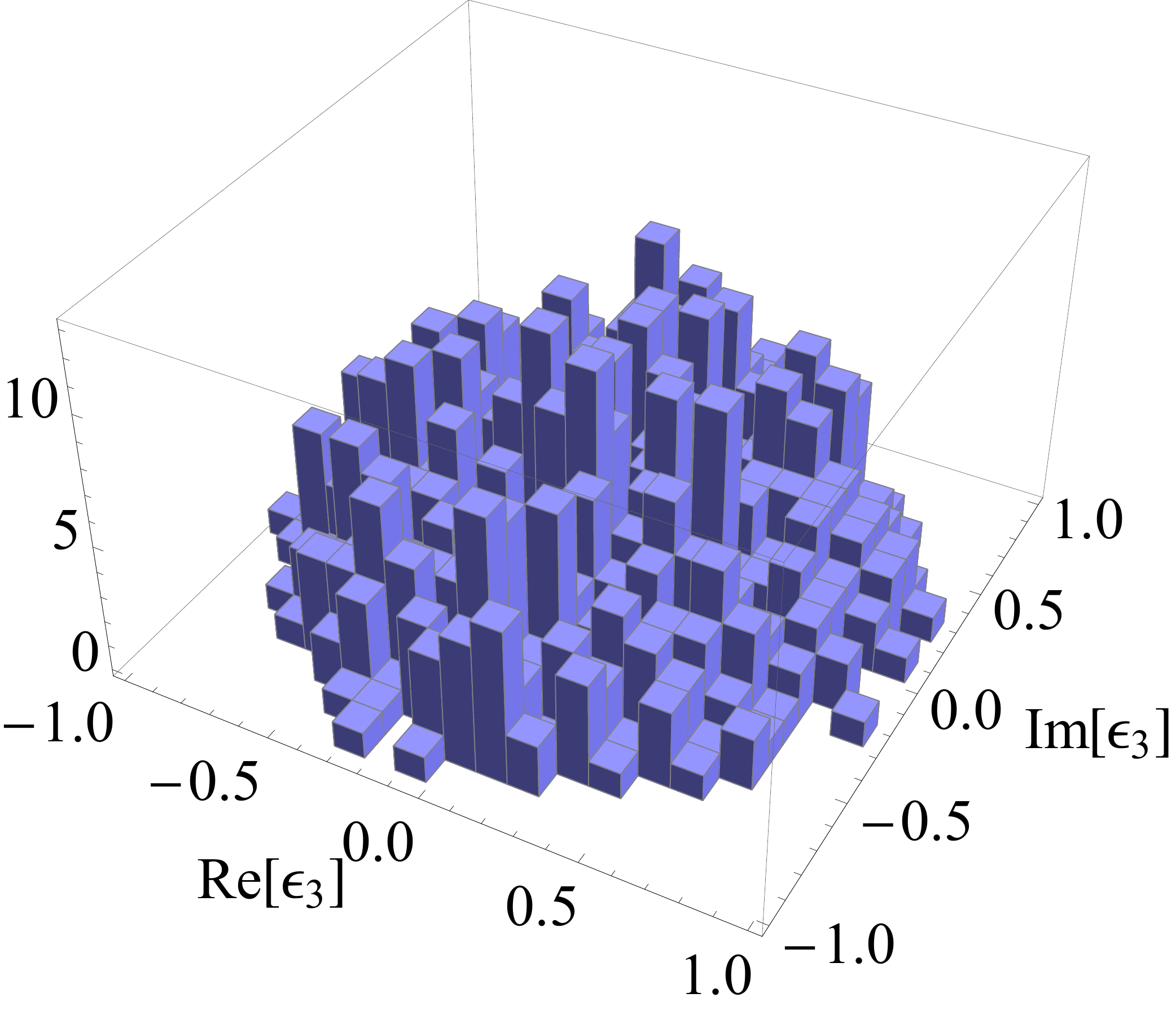}
\endminipage
\minipage{0.33\textwidth}
\includegraphics[width=53mm]{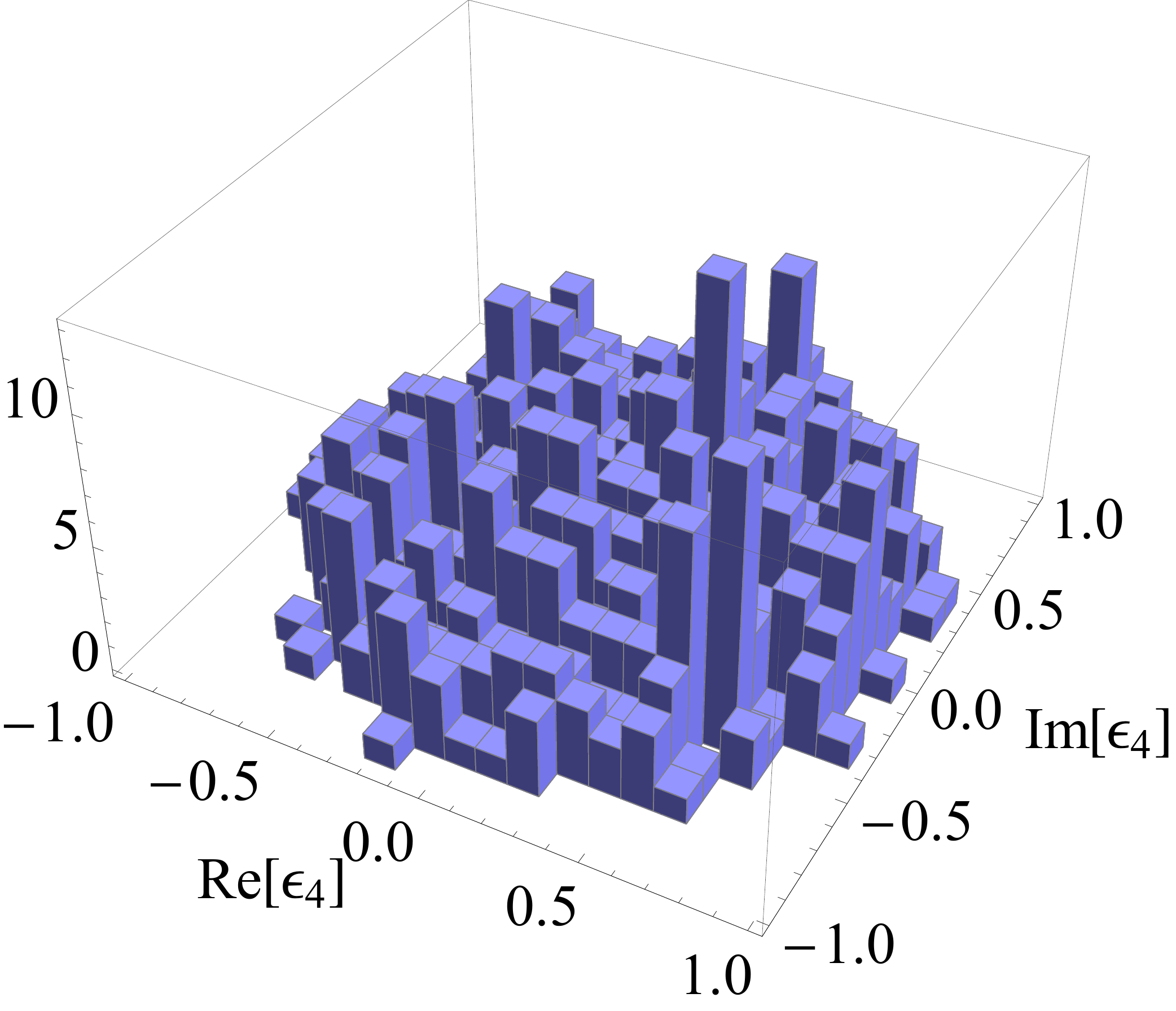}
\endminipage
  \caption{3D Histograms, 1000 random generated strings. N=100 and $N_{ch}=7$ with attractive interaction $g=0.3$.  } \label{Histogram3Dg06}
\end{figure}

To show the transverse cross correlations it is also useful to use the cross moments~\cite{Bzdak:2013rya, Kalaydzhyan:2014tfa}
\bea\la{epsilonnm}
\left(\epsilon_n \{  2 \} \right)^2 &=&  \left< \left|\epsilon_n\right|^2  \right> \nonumber\\
\left(\epsilon_n \{  4 \} \right)^4 &=&   -   \left< \left|\epsilon_n\right|^4  \right>    + 2 \left< \left|\epsilon_n\right|^2  \right>^2 \nonumber\\
\left(\epsilon_n \{  6 \} \right)^6 &=& \frac{1}{4} \left[  \left< \left|\epsilon_n\right|^6  \right>  -  9 \left< \left|\epsilon_n\right|^4 \right>  \left< \left| \epsilon_n\right|^2  \right> + 12 \left< \left|\epsilon_n\right|^2  \right>^3   \right] \nonumber\\
\left(\epsilon_n \{  8 \} \right)^8 &=&   \frac{1}{33} \left[ - \left< \left|\epsilon_n\right|^8  \right> + 16 \left< \left|\epsilon_n\right|^6  \right> \left< \left|\epsilon_n\right|^2  \right>  + 18  \left<  \left|\epsilon_n\right|^4  \right>^2 - 144 \left< \left|\epsilon_n\right|^4  \right> \left< \left|\epsilon_n\right|^2  \right>^2 + 144\left< \left|\epsilon_n\right|^2  \right>^4  \right]
\eea
 For the cross moments (flow), we can only do $N=100$ (randomly generated strings). We use $\bb \sim (0.2, 0.05)$ such that $\bb\leq 0.2$   and $N\bb \geq 5$. 
For a single  string exchange, the multiplicity  range is
$N_{ch} \sim (7, 26)$, while for 5 strings $N_{ch} \sim (35, 130)$ and 10 strings $N_{ch} \sim (70, 260)$. 
To characterize the initial azimuthal deformation of the string
bits in the transverse collision plane, we show in Fig.~\ref{Histogram3D} the pdf distributions of 1000 randomly 
generated single strings  at a resolution of $N=100$  and a multiplicity of $N_{ch} = 7$ 
with no self-interactions $g=0$. The pdf  shown are for
the distributions in $\epsilon_{2,3,4}$ respectively. 
We also show in Fig.~\ref{Histogram3Dg06} the pdf distributions of 1000 randomly generated single strings
at a resolution of $N=100$  with a multiplicity $N_{ch} = 7$  undergoing string bit attractions with $g=0.6$ in the mean-field approximation. Note the strong dipole deformation in the leftmost figure.

\begin{figure}[!htb]
\minipage{0.33\textwidth}
\includegraphics[width=41mm]{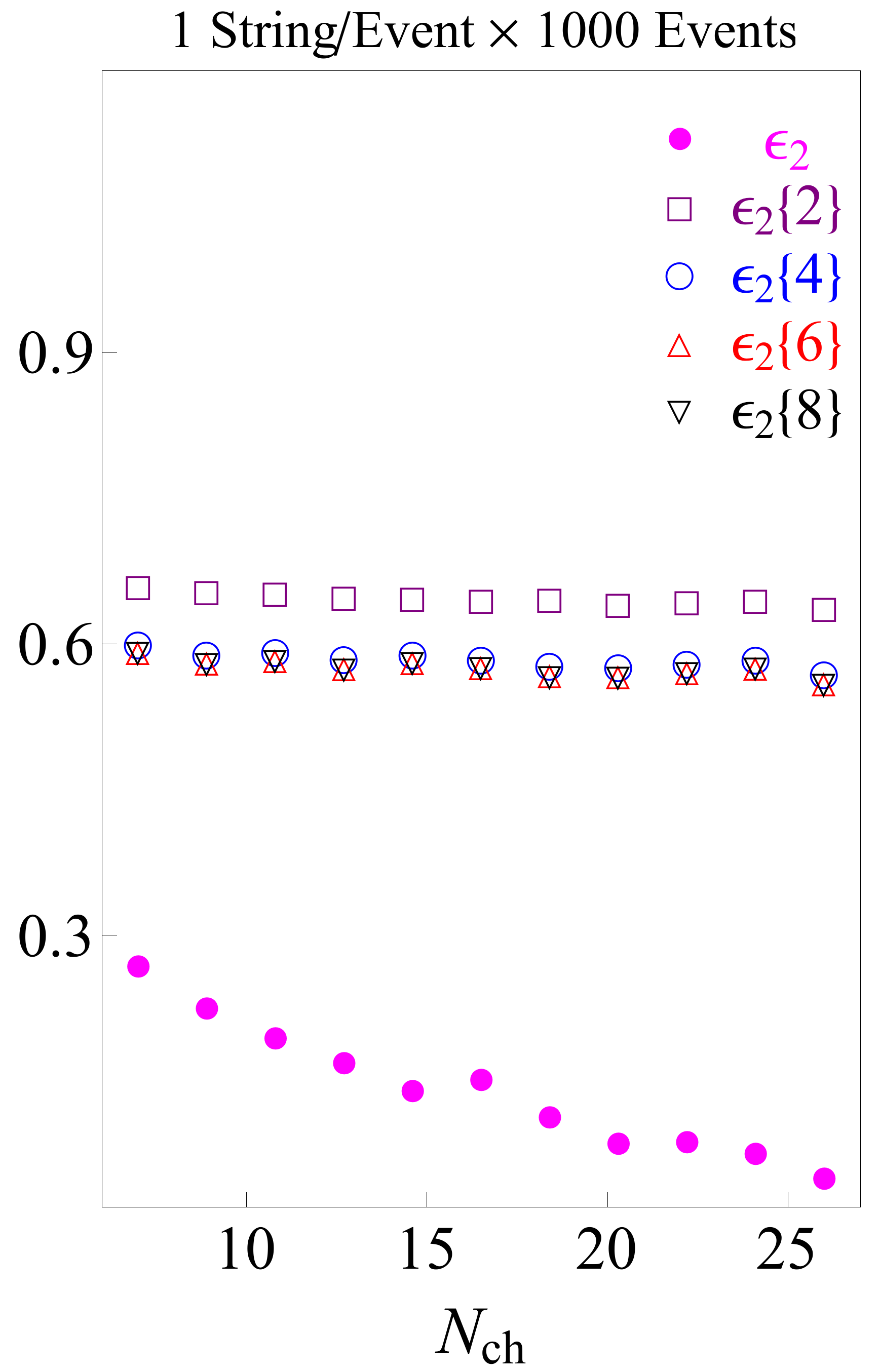}
\endminipage\hfill
\minipage{0.33\textwidth}
\includegraphics[width=41mm]{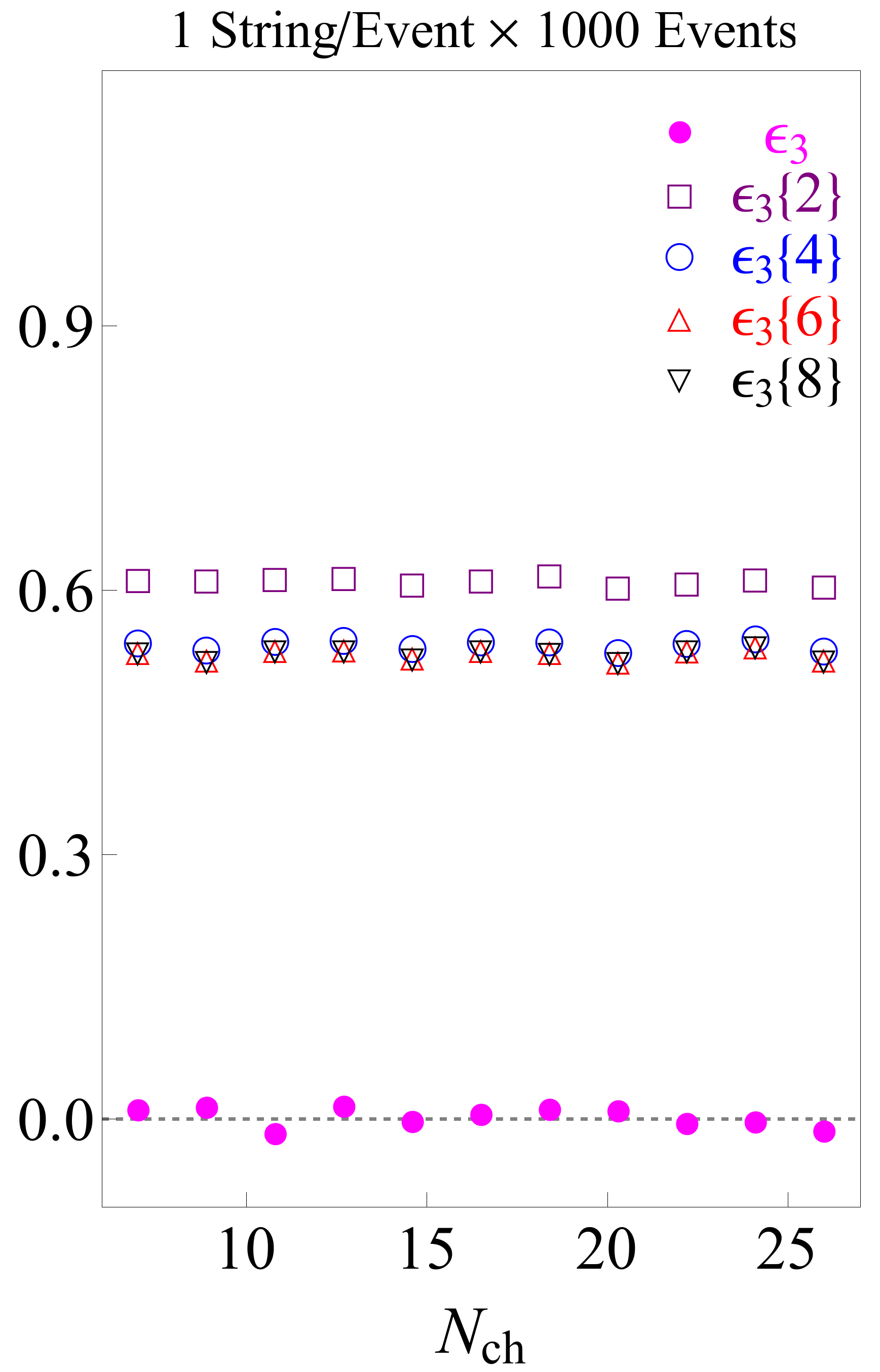}
\endminipage
\minipage{0.33\textwidth}
\includegraphics[width=41mm]{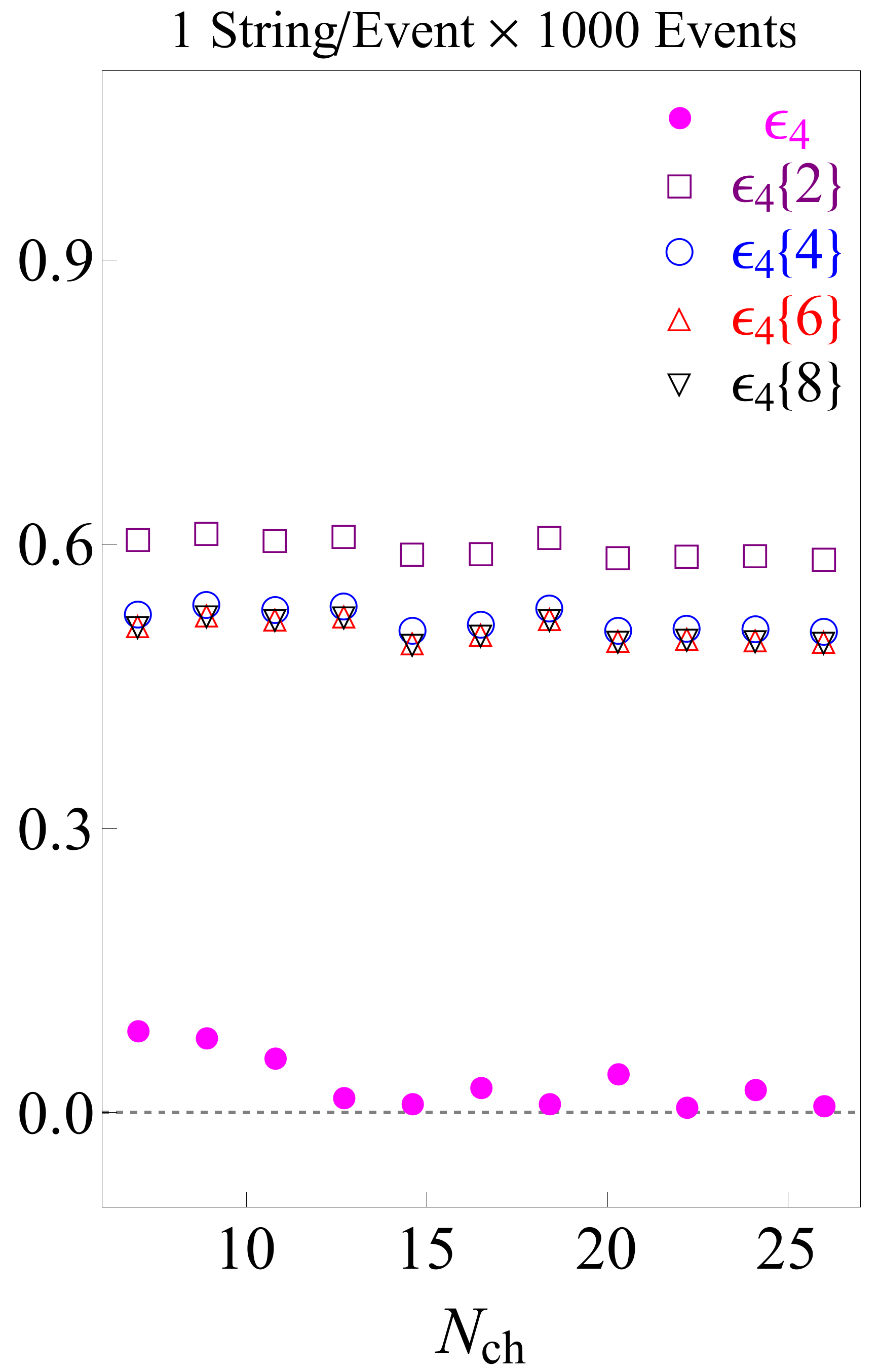}
\endminipage
  \caption{Non-interacting.}\label{EnRandoma}
\end{figure}

\begin{figure}[!htb]
\minipage{0.33\textwidth}
\includegraphics[width=41mm]{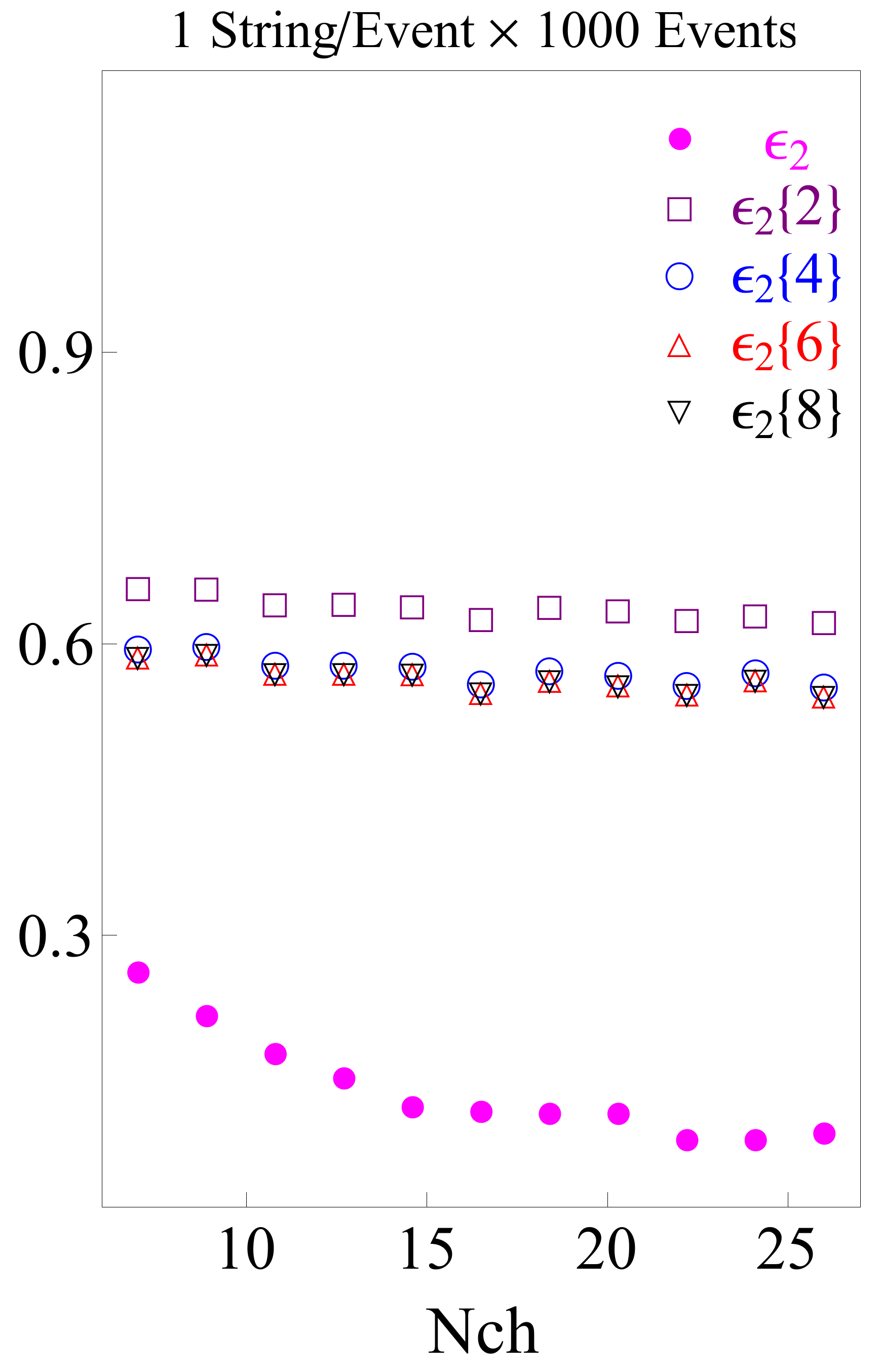}
\endminipage\hfill
\minipage{0.33\textwidth}
\includegraphics[width=41mm]{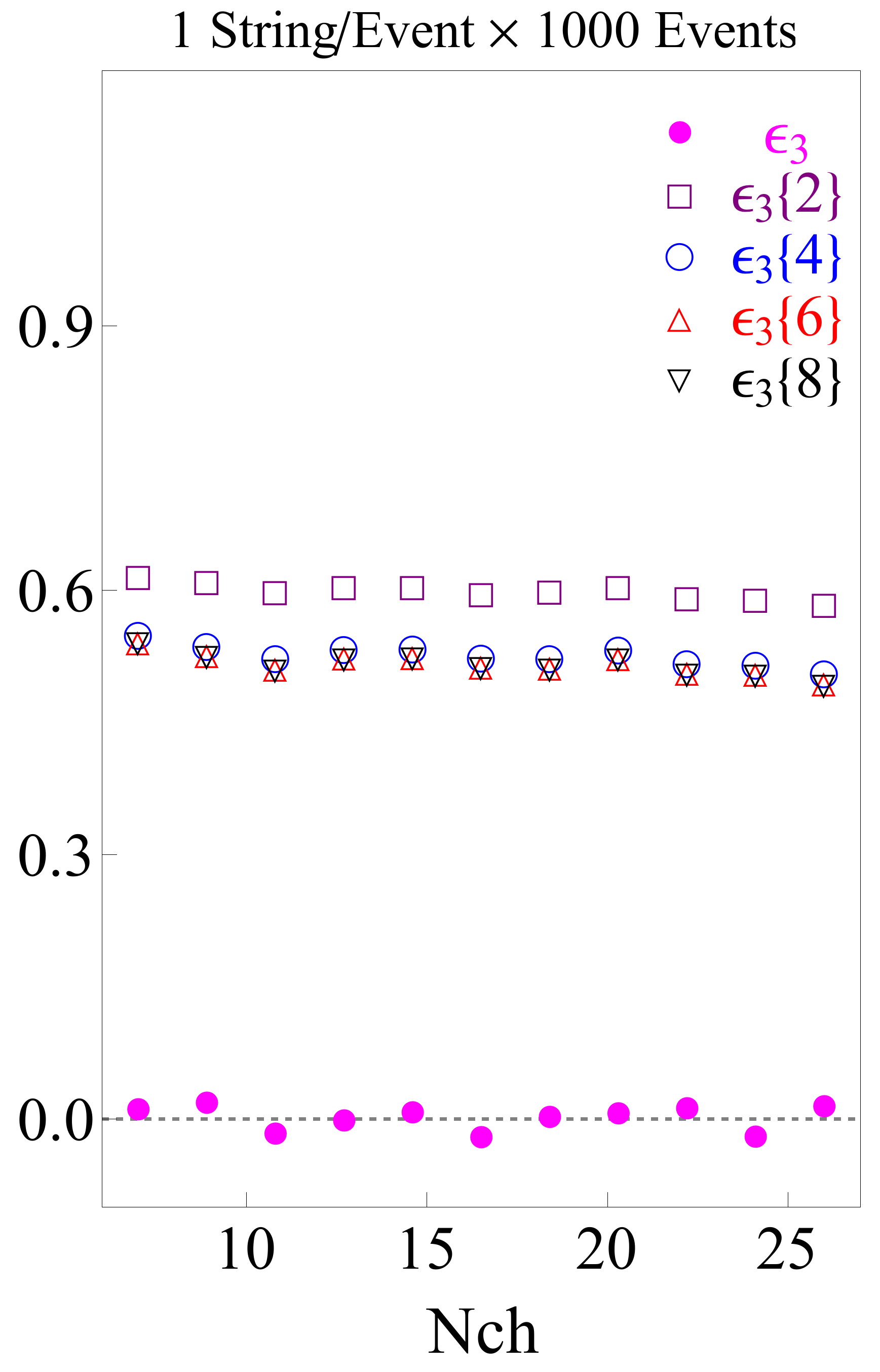}
\endminipage
\minipage{0.33\textwidth}
\includegraphics[width=41mm]{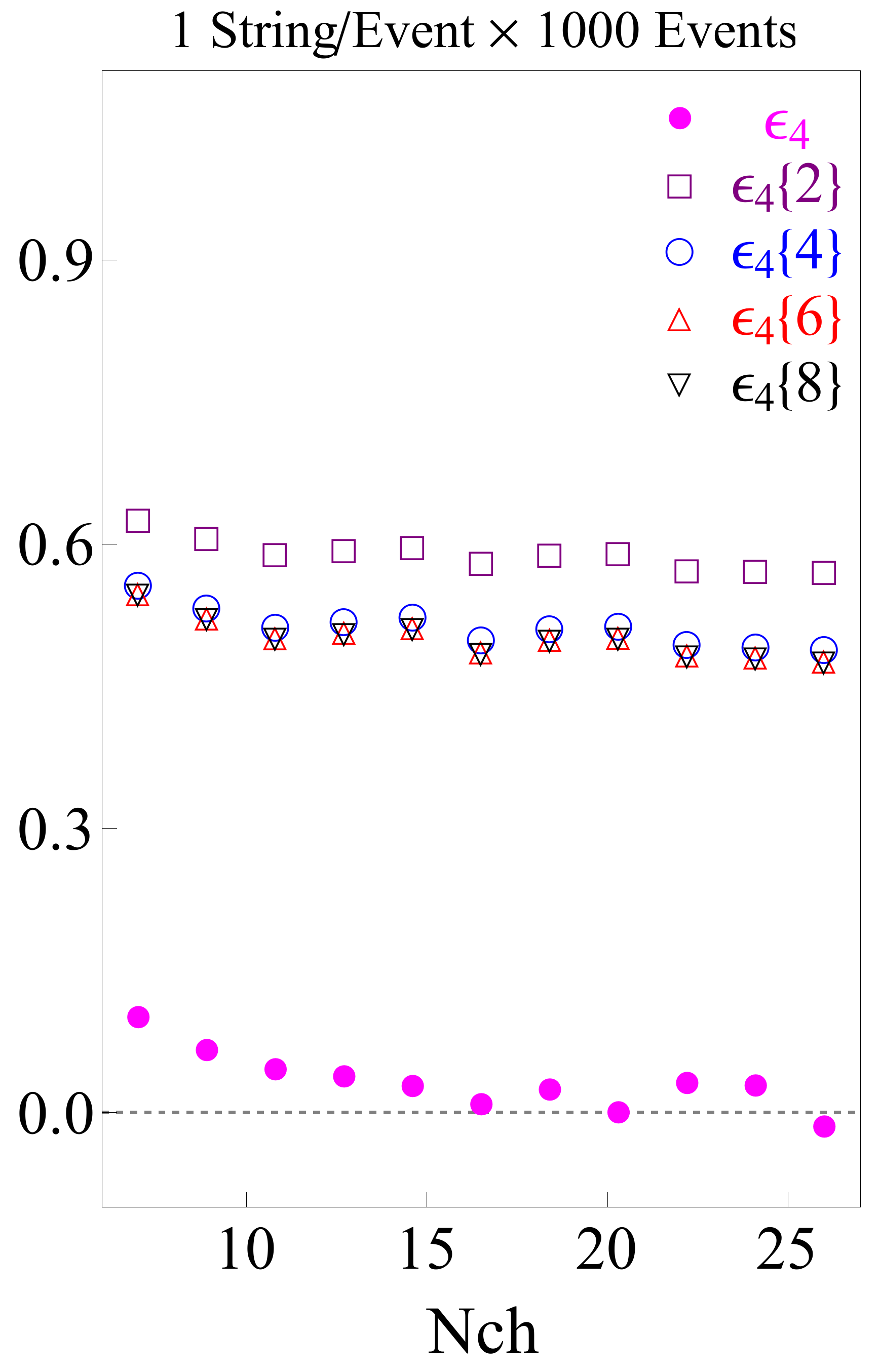}
\endminipage
  \caption{   Attractive interaction $g=0.6$. }  \label{MOMENT6a}
\end{figure}

\begin{figure}[!htb]
\minipage{0.33\textwidth}
\includegraphics[width=41mm]{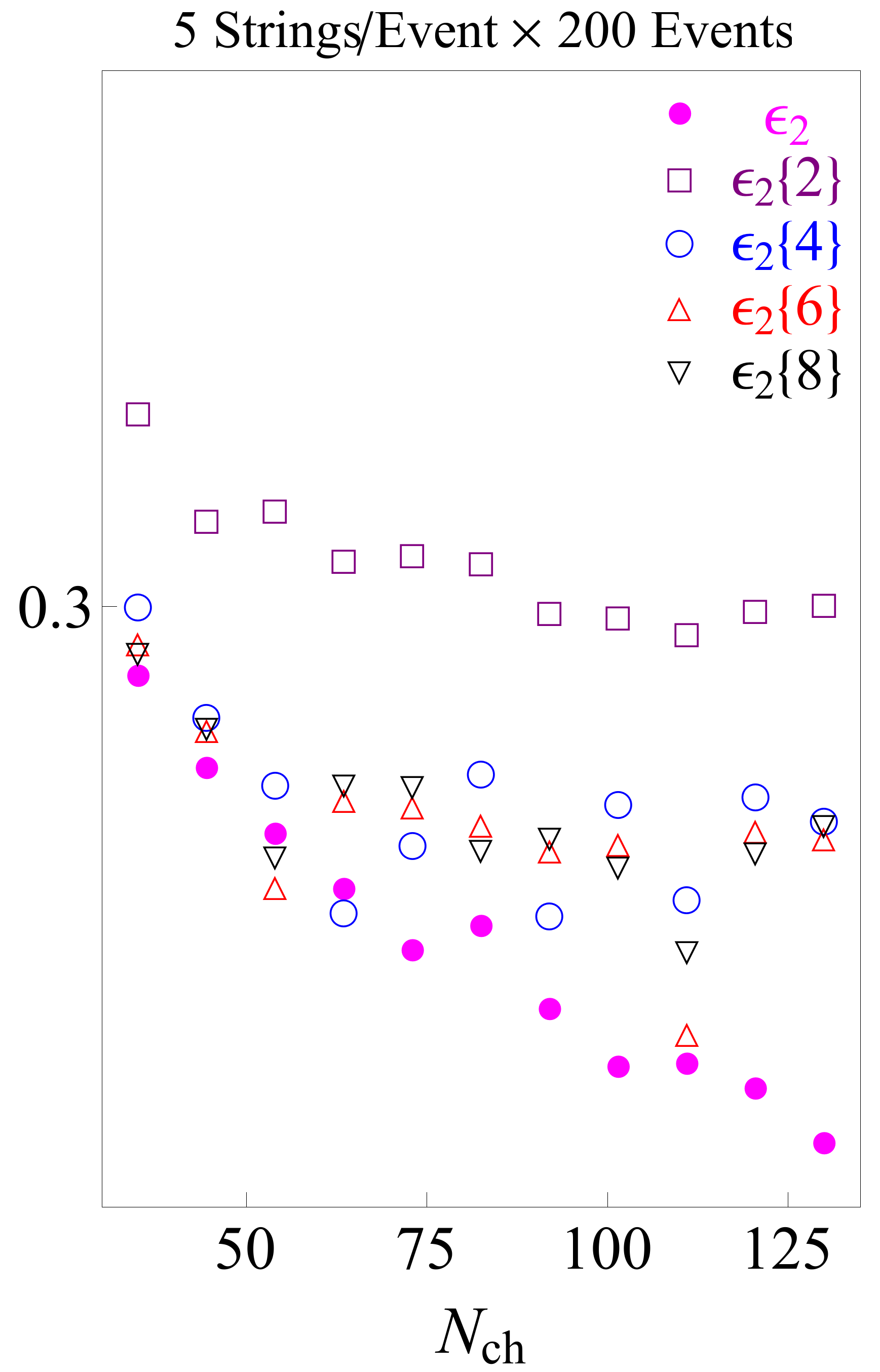}
\endminipage\hfill
\minipage{0.33\textwidth}
\includegraphics[width=41mm]{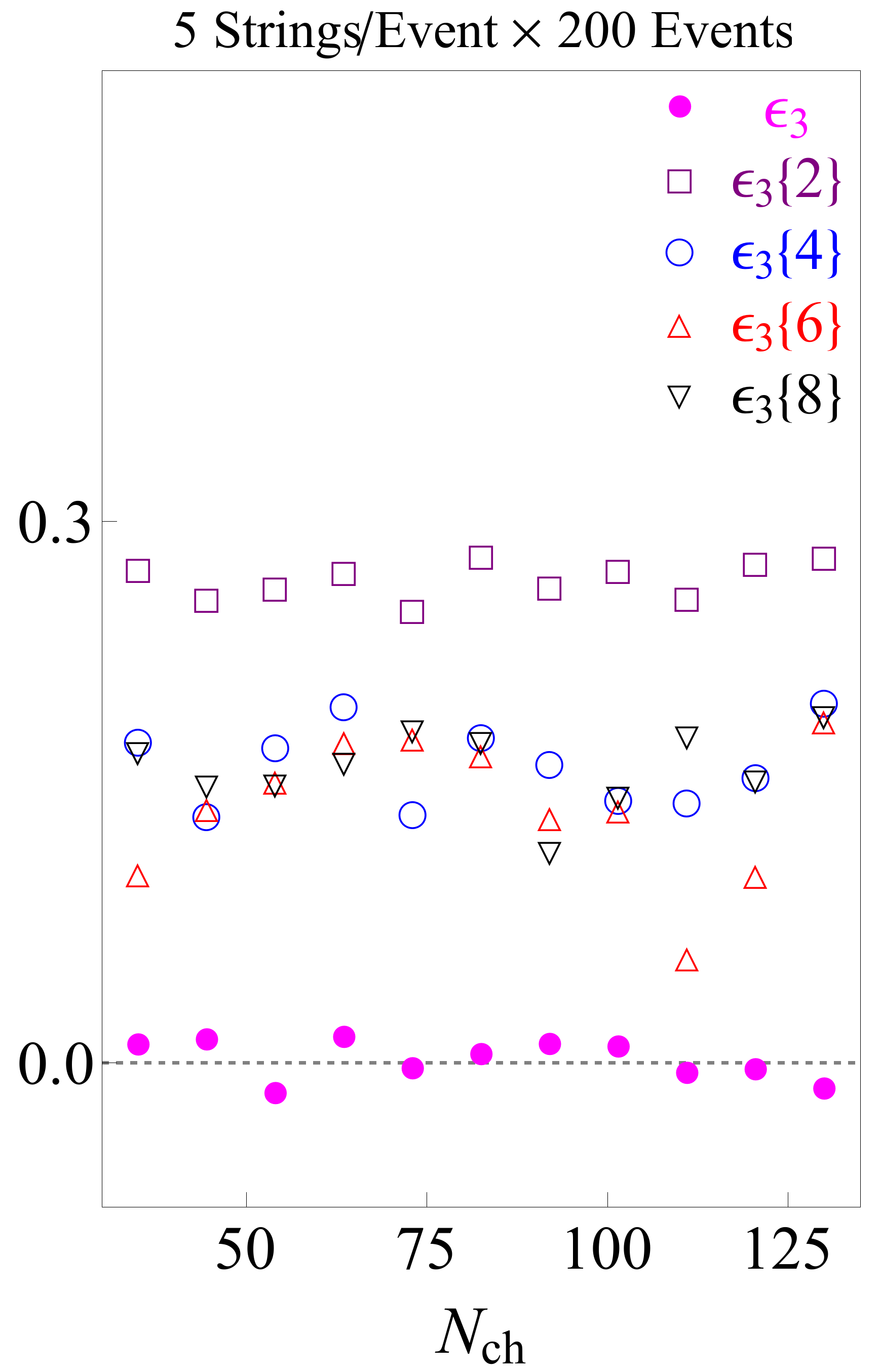}
\endminipage
\minipage{0.33\textwidth}
\includegraphics[width=41mm]{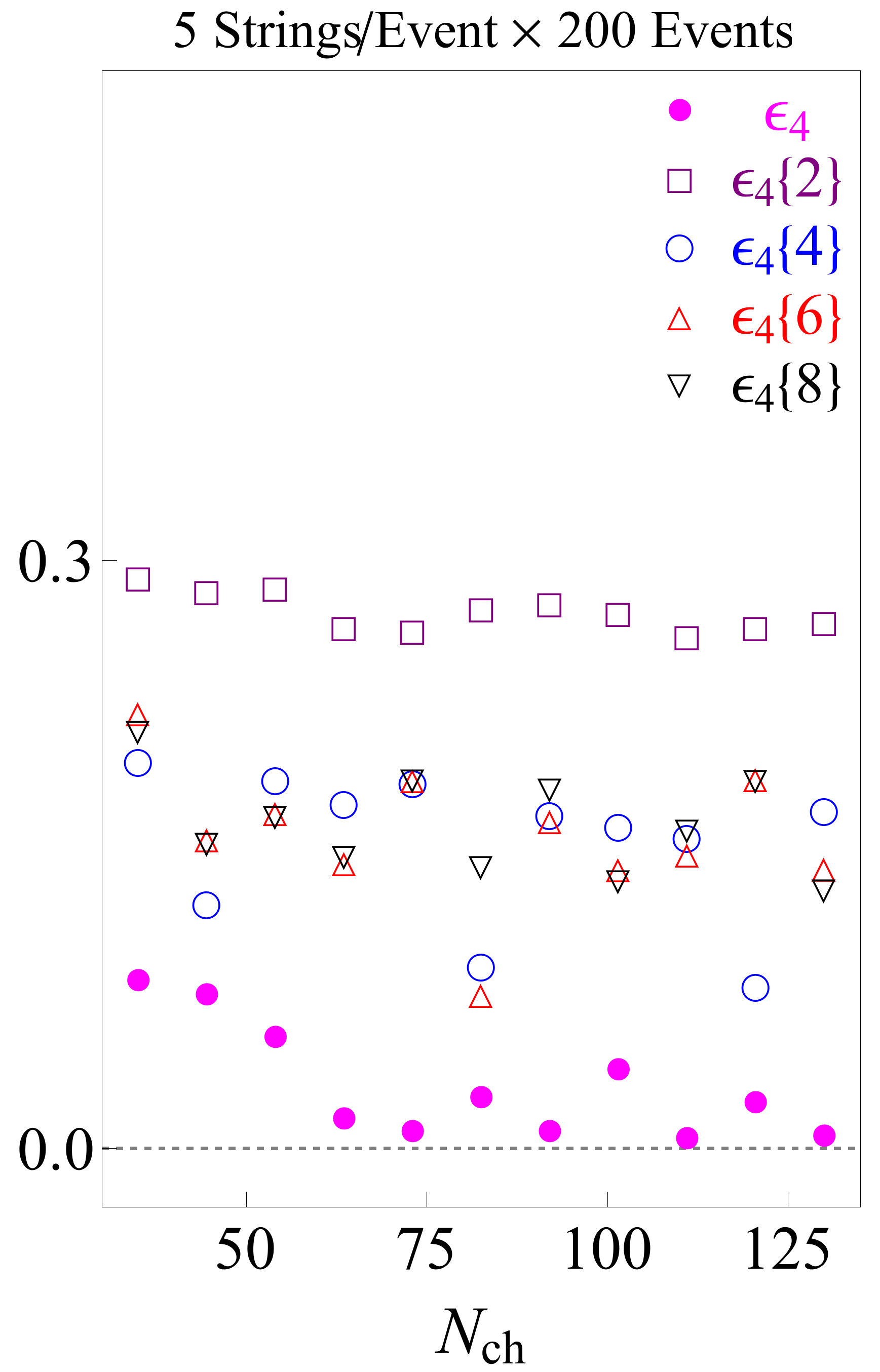}
\endminipage
  \caption{Non-interacting.}\label{EnRandomb}
\end{figure}

\begin{figure}[!htb]
\minipage{0.33\textwidth}
\includegraphics[width=41mm]{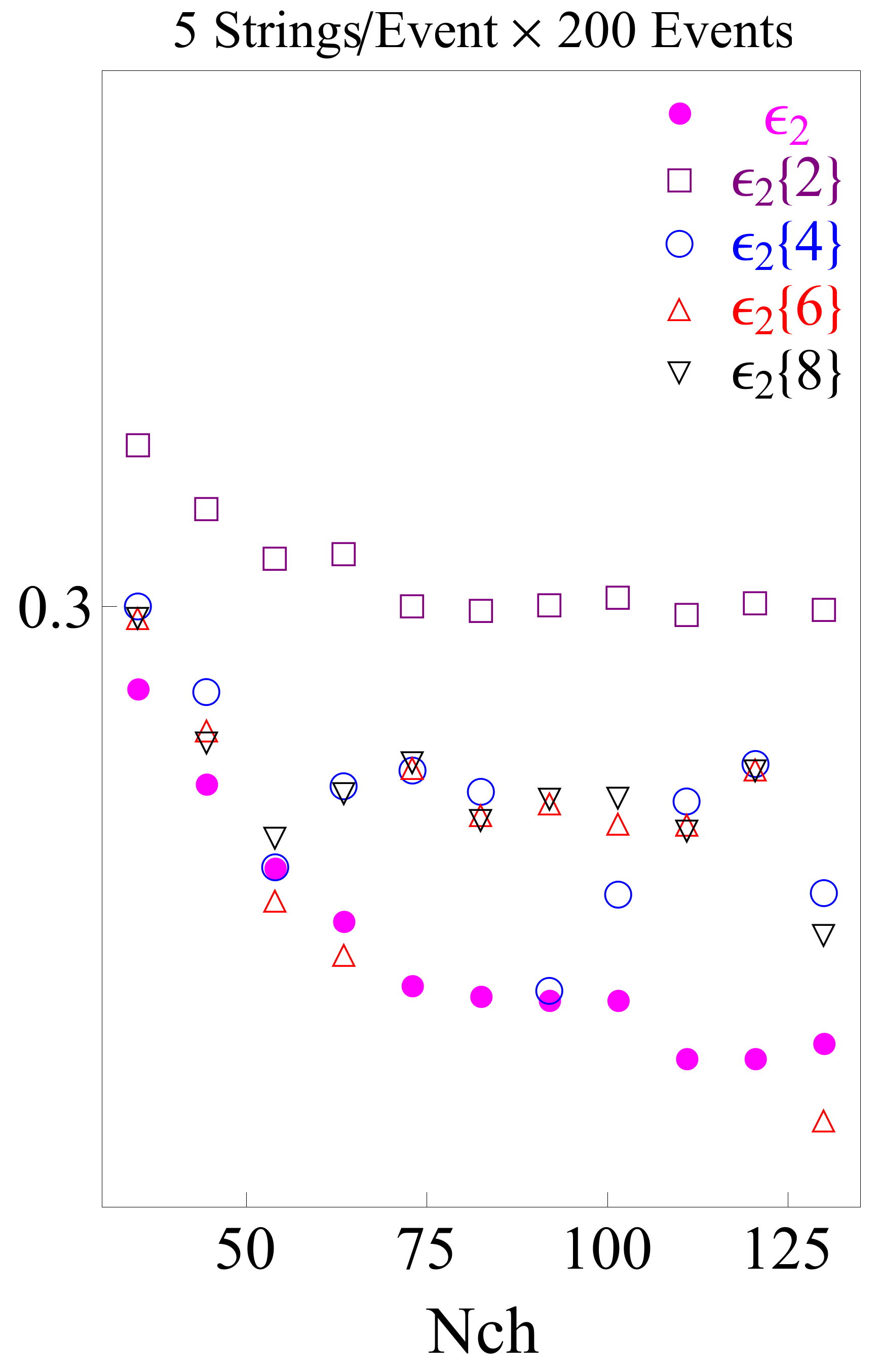}
\endminipage\hfill
\minipage{0.33\textwidth}
\includegraphics[width=41mm]{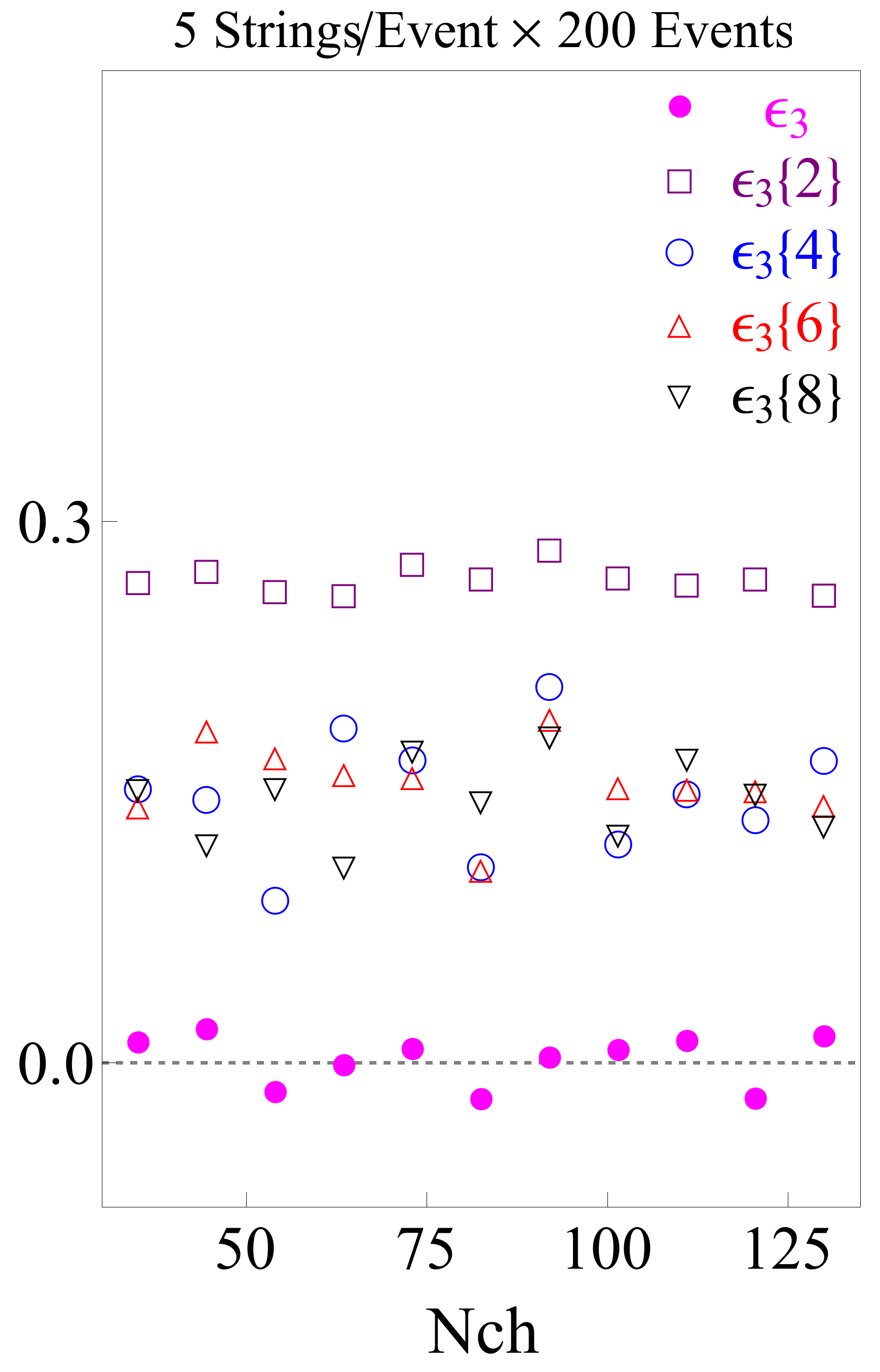}
\endminipage
\minipage{0.33\textwidth}
\includegraphics[width=41mm]{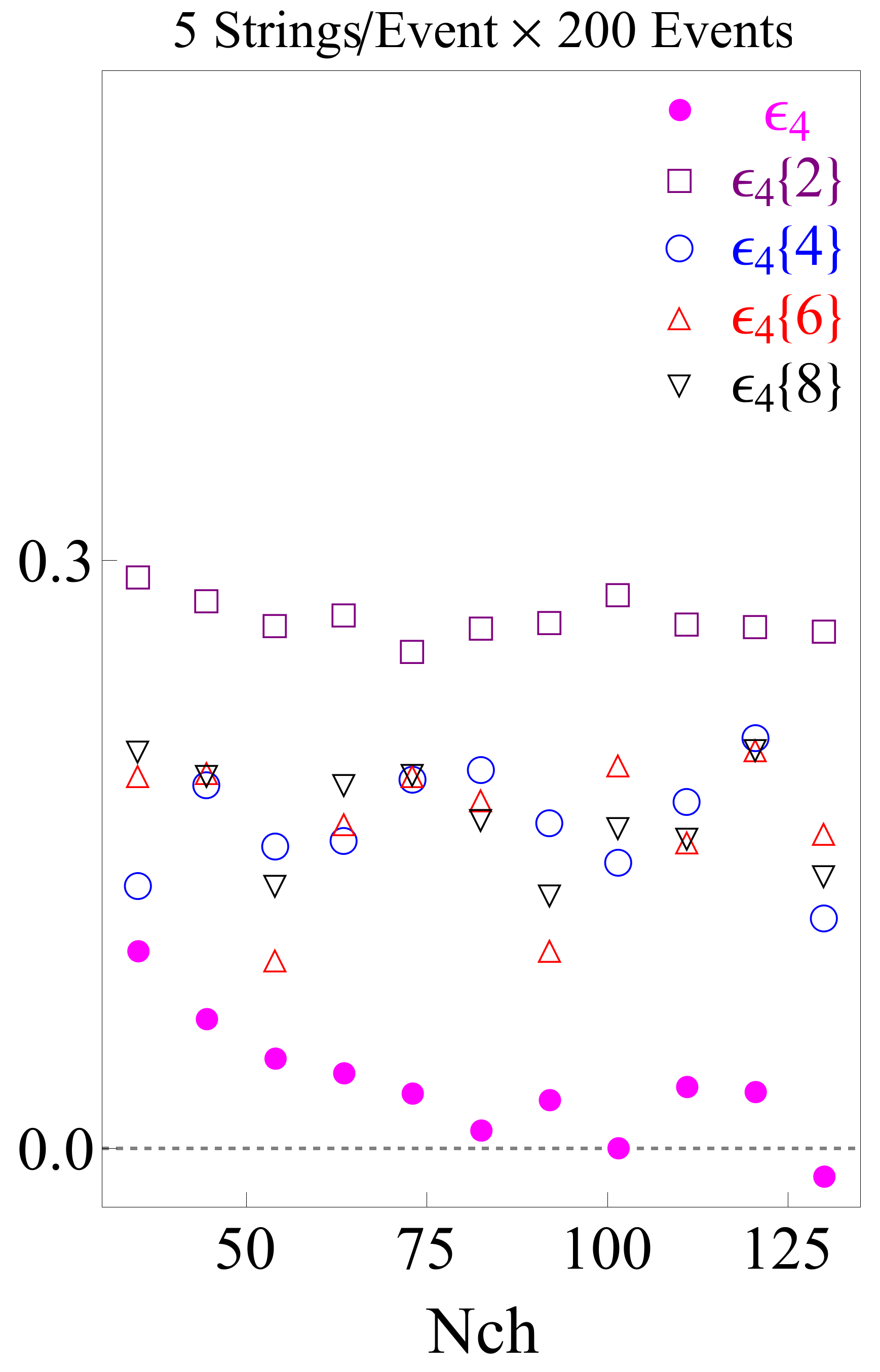}
\endminipage
  \caption{   Attractive interaction $g=0.6$. }  \label{MOMENT6b}
\end{figure}

\begin{figure}[!htb]
\minipage{0.33\textwidth}
\includegraphics[width=41mm]{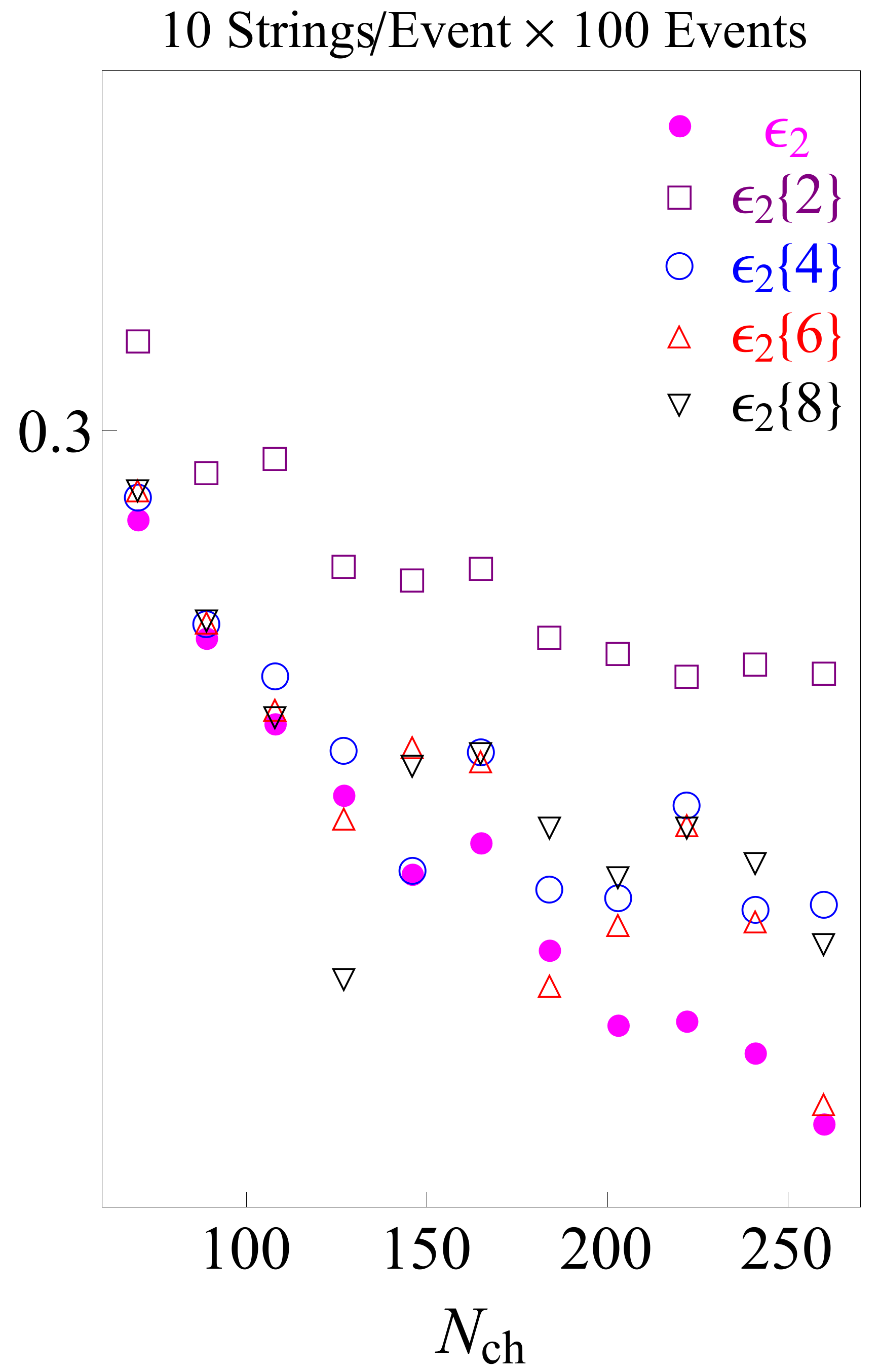}
\endminipage\hfill
\minipage{0.33\textwidth}
\includegraphics[width=41mm]{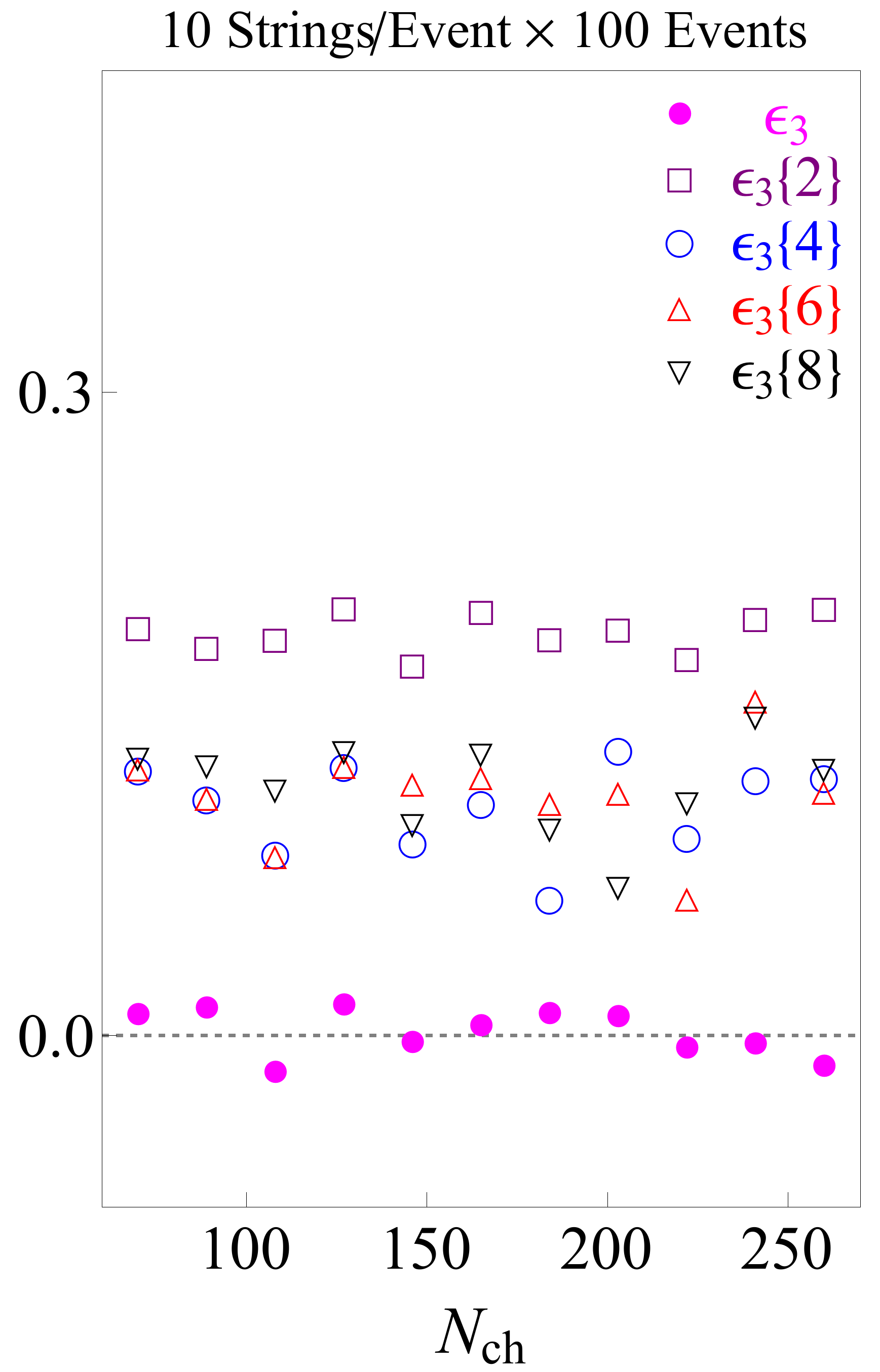}
\endminipage
\minipage{0.33\textwidth}
\includegraphics[width=41mm]{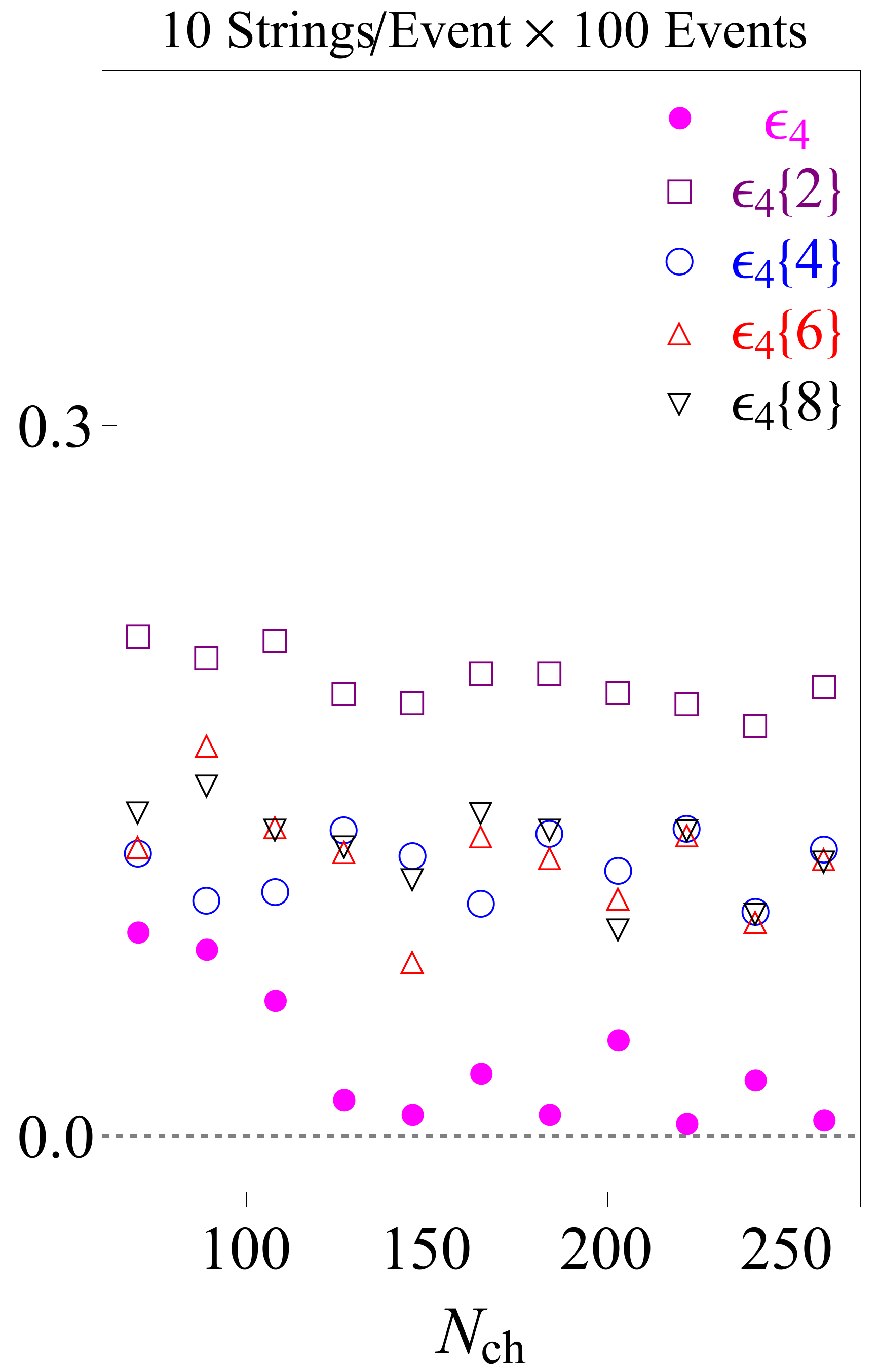}
\endminipage
  \caption{Non-interacting.}\label{EnRandomc}
\end{figure}

\begin{figure}[!htb]
\minipage{0.33\textwidth}
\includegraphics[width=41mm]{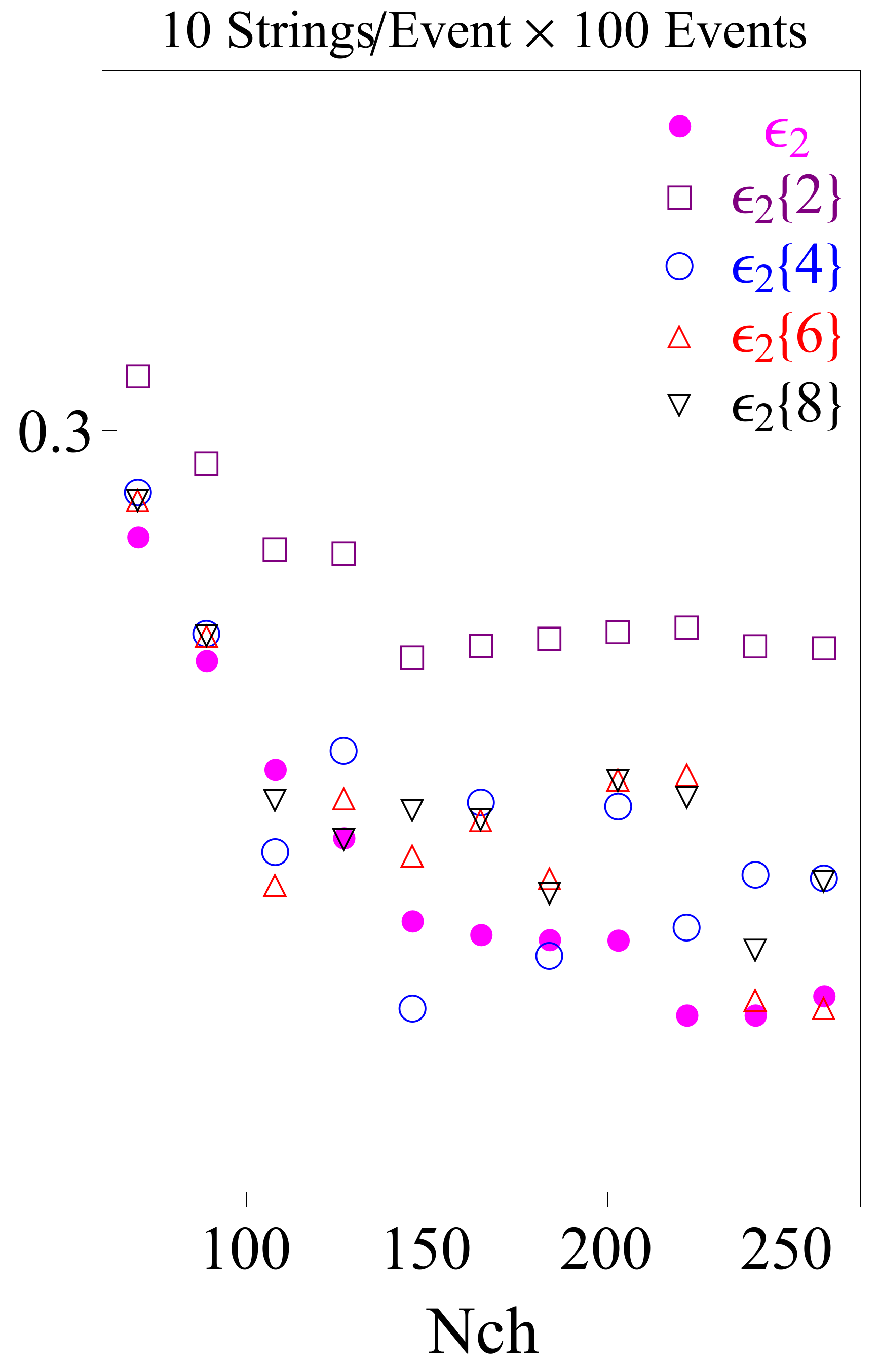}
\endminipage\hfill
\minipage{0.33\textwidth}
\includegraphics[width=41mm]{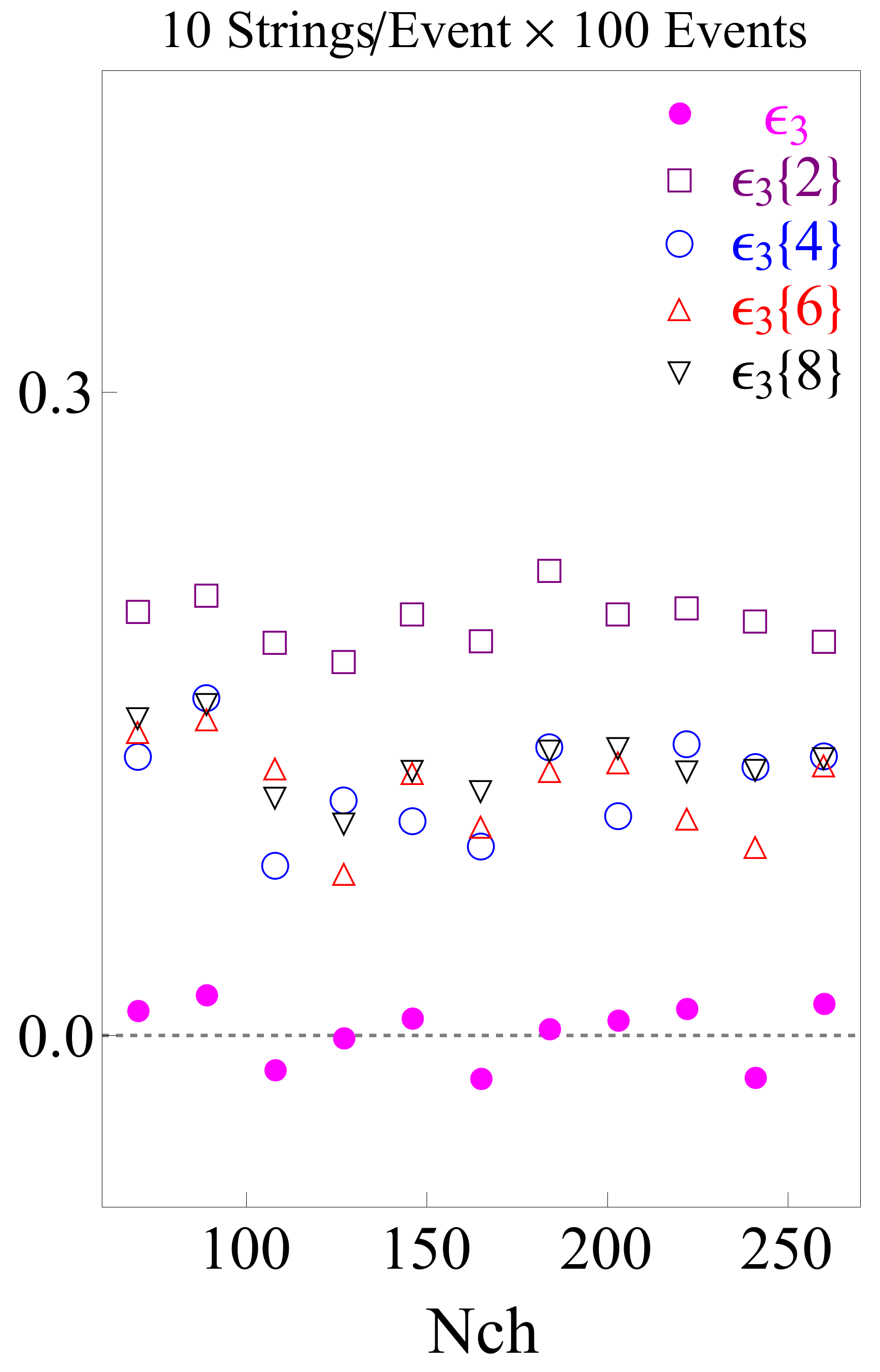}
\endminipage
\minipage{0.33\textwidth}
\includegraphics[width=41mm]{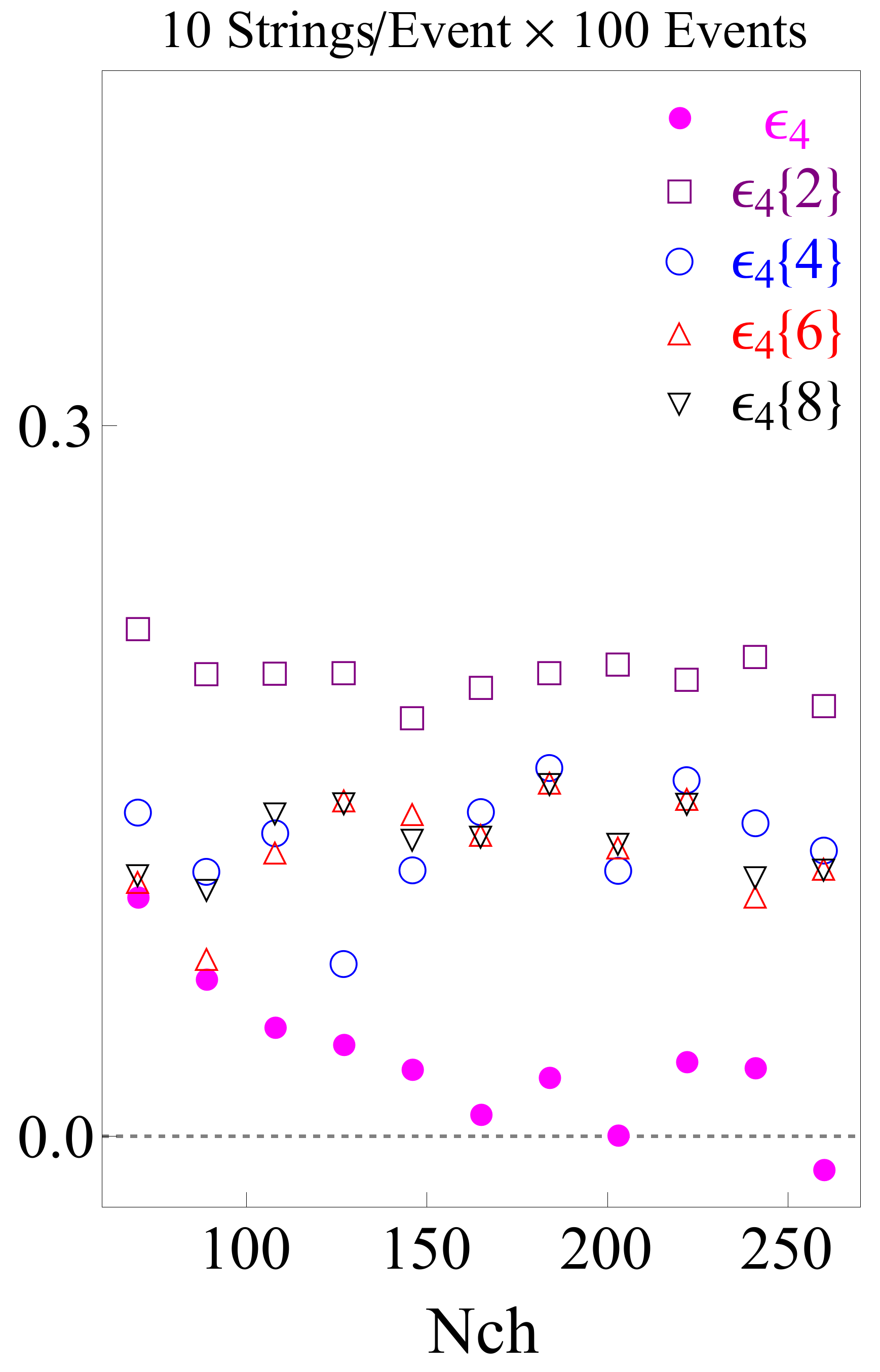}
\endminipage
  \caption{   Attractive interaction $g=0.6$. }  \label{MOMENT6c}
\end{figure}

For completeness we show the behavior of the cross moments with the resolution $N=100$ for a non-interacting and for an attractive
string, in
 Fig.~\ref{EnRandoma} and Fig.~\ref{MOMENT6a} respectively by sampling 1000 times a single string streched 
with $b=5=10l_s$. The attraction is set at $g=0.6$.
   In a typical pp collision at collider energies, we expect to exchange about 10 such long strings~\cite{Stoffers:2012zw,Stoffers:2012ai,Stoffers:2013tla}.  In   Fig.~\ref{EnRandomb} and Fig.~\ref{MOMENT6b} we show the same cross
moments following from the exchange of 5 typical strings streched at $b=5$ sampled 200 times for  non-interacting and attractive case
respectively. The case where 10 string are exchanged is shown in  Fig.~\ref{EnRandomc}  and Fig.~\ref{MOMENT6c} for the
same arrangements of parameters with each 10 string event sampled 100 times.


\section{\label{conclusion} Conclusions}

Holographic strings in walled AdS$_5$   provide a non-perturbative description 
of diffractive scattering, production as well as low-x DIS~\cite{Stoffers:2012zw}. Although a key aspect of  AdS$_5$ is its conformality  which translates to the conformal character of QCD in the UV, the essentials of the walled AdS$_5$ construction for the holographic string with a large rapidity interval can be captured by a string with an effective transverse dimension $2<D_\perp<3$. The holographic Pomeron intercept follows from the zero point motion or Luscher term of the free transverse string with $D_\perp/12$, and the Pomeron slope is fixed by the string tension. 

Low-x physics in the holographic string set up corresponds to a string with higher zero point resolution, whereby the string bits play the
non-perturbative analogue of the wee partons in perturbative QCD. A key aspect of the partonic description is Gribov transverse diffusion
which arises  naturally in the quantum string description as emphasized by Susskind and others~\cite{Susskind:1994hb,Susskind:1994vu,Damour:1999aw,Horowitz:1996nw}. A new aspect of our recent study of the holographic string at low-x consists in the role played by the interactions
between the string bits in $2<D_\perp<3$ and their role in producing a stringy mechanism for saturation~\cite{Qian:2014rda}.

For strings exchanged at smaller impact parameters, the exponential increase in the string excited states dwarf the zero point
fluctuations making the string essentially classical. We have used this observation to construct a micro-canonical description of a
holographic string  by introducing an effective temperature. Close to its Hagedorn temperature, the  string carries large 
entropy and multiplicity and provides a possible and generic mechanism for large multiplicity events  in hadronic collisions in
the Pomeron kinematics.

In flat $D_\perp=3$ dimensions the free string close to its Hagedorn temperature carries large multiplicities and exhibits
large transverse geometrical deformations mostly due to its transverse and classical diffusion. The large outgrowth of the
string bits makes it  ideal for a mean-field analysis of the string self-interactions. We have used the variational analysis to put
a lower bound on the interacting string free energy and use it to detail its geometrical content. Self-interactions cause the
effectively thermal string to contract, a process typical of string-black-hole transmutation in fundamental string theory
\cite{Shuryak:2013sra,Shuryak:2013ke,Bekenstein:1973ur,Bekenstein:1972tm,Bekenstein:1974ax,Polchinski:2001ju,Susskind:1994hb,Susskind:1994vu}.

The geometry of the string bit distributions emerging from streched strings for  small impact parameters 
is rich in structure and transverse deformation.  We have presented a detailed study of its transverse moments and distributions
for single and multiple string exchanges. These prompt and deformed distributions can be used to initialize the prompt parton distributions
in current pp and pA collisions for the recently reported high multiplicity events by the LHC~\cite{CMS:2012qk,Abelev:2012ola,Aad:2012gla}. Our azimuthal and cross-moments provide a specific measure of the prompt asymmetries versus multiplicity.
The holographic string  close to its Hagedorn temperature maybe at the origin of the fire ball mechanism underlying the relevance of a hydrodynamical description in hot but small hadronic volumes~\cite{Shuryak:2013ke,Bozek:2013uha,deSouza:2015ena,Kalaydzhyan:2014zqa,Kalaydzhyan:2014tfa}.

\section{\label{acknowledgements} Acknowledgements}

This work was supported by the U.S. Department of Energy under Contract No.
DE-FG-88ER40388.


\newpage

 \section{Appendix A: Effective Temperature}

It was initially suggested in~\cite{Shuryak:2013ke,Shuryak:2013sra} that at large $\sqrt{s}$ with a rapidity
interval $\chi={\rm ln}(s/s_0)$ and a small impact parameter ${\bf b}$,
the Pomeron as a string exchange senses an Unruh temperature on the worldsheet $1/\beta=\chi/2\pi {\bf b}$. 
The re-summed $1/{\bf b}$ contributions in the Pomeron kinematics led to the concept of a critical Pomeron 
with a Nambu-Goto form for the string amplitude. The energy  and entropy of the critical Pomeron were shown
to be of the form~\cite{Shuryak:2013ke,Shuryak:2013sra}

\bea
E\approx&& \sigma_T \, b\,\left(1-\frac{\beta_H^2}{\beta^2}\right)^{-\frac 12 }\nonumber\\
S\approx&& \sigma_T \, b\,\frac{\beta_H^2}{\beta}\left(1-\frac{\beta_H^2}{\beta^2}\right)^{-\frac 12}
\label{A2}
\eea
Near the Hagedorn temperature $S\approx \beta_H E$ with

\be
E\approx \frac{\sigma_T{\bf b}}{\sqrt{2\beta_H}}\frac{\beta_H}{(\beta-\beta_H)^{\frac 12}}
\label{A3}
\ee
A comparison of (\ref{A2}-\ref{A3})  with (\ref{TS1}-\ref{TS3}) suggests that the effective temperature $1/\bb$ introduced in section III
can be identified with the critical Pomeron parameters as

\be
\bb \approx \frac{1}{{\bf b}}\,\left(1-\frac{\beta_H}{\beta}\right)^{\frac 12}\equiv 
\frac{1}{{\bf b}}\,\left(1-\frac{\beta_H}{2\pi {\bf b}}{\rm ln}(s/s_0)\right)^{\frac 12}
\label{A4}
\ee
which is a measure of how close the kinematical Unruh temperature is to the Hagedorn temperature in units of $l_s$. 
We note that in curved space we have $\beta_H\rightarrow \beta_H(\lambda)=2\pi\sqrt{D_\perp(\lambda)/6}$ in (\ref{A4}).
Although we
have used $1/\bb $ as a phenomenological parameter fixed by the energy/entropy and identified it through the
net charge multiplicity of a dipole-dipole collision, (\ref{A4}) shows that  microscopically it is fixed by 
the collision kinematics. For a fixed rapidity interval, (\ref{A4}) shows that the smaller the collision separation, the closer to the
Hagedorn temperature.

 \section{Appendix B: String versus Black-hole}
 
While our classical string is near its Hagedorn point, its  squared size grows diffusively with its transverse mass.
By letting the string self-interact, we can change both its mass $M_\perp\rightarrow M_\perp (g)$ and thus its size as suggested
by Susskind and others~\cite{Polchinski:2001ju,Susskind:1994hb,Susskind:1994vu, Damour:1999aw,Bergman:1997ki}. In the process, we can adiabatically map the current hot string at its Hagedorn point with
a black-hole in $D_\perp+2$ dimensions. For that we recall that for a Schwarchild black-hole the Bekenstein-Hawking relation
for the entropy is

\be
S_{\rm ch}=\frac{A_\perp}{4G_5}\approx \left(\frac {R_{\rm ch}}{l_P}\right)^{D_\perp}
\label{BH1}
\ee
with the Planck scale $G_5=l_P^{D_\perp}$ and the Schwarchild radius set by the condition

\be
R_{\rm ch}=(G_5M_\perp)^{\frac 1{D_\perp-1}}
\label{BHR}
\ee
(\ref{BH1}) and (\ref{BHR}) map onto  (\ref{TS4}) for a critical string coupling 
$g_s^2M_\perp l_s\approx 1$. To see this, we note that (\ref{BH1}) can be re-written as

\be
S_\perp \approx  \beta_H M_\perp \equiv \beta_H (g_s^2M_\perp)\,\frac 1{g_s^2}\approx  \frac 1{g_s^2}
\label{BH}
\ee
which is equivalent to (\ref{TS4}) when the Schwarchild radius $R_{ch}\rightarrow l_s$ shrinks to 
few string lengths as gravity mediated self-interactions become weak  through
$g_s^2=(l_P/l_s)^{D_\perp}\approx 1/M_\perp l_s$. Specifically, the shrinking 
Schwarchild radius becomes

\be
\frac{R_{ch}}{l_s}=(g_s^2M_\perp l_s)^{\frac 1{D_\perp-1}}\approx 1
\label{BHR1}
\ee
and its Bekenstein-Hawking entropy reduces to

\be
S_{\rm ch}\approx \left(\frac {R_{\rm ch}}{l_P}\right)^{D_\perp}\approx \left(\frac {l_s}{l_P}\right)^{D_\perp}
\approx  \frac 1{g_s^2}
\label{BH2}
\ee
both of which map on the string at the Hagedorn point  provided that the string self-interaction can cause 
a reversal in the growth from a diffusive to a fixed  and smaller size object.

\bibliography{Hightemref}

\begin{thebibliography}{58}
\expandafter\ifx\csname natexlab\endcsname\relax\def\natexlab#1{#1}\fi
\expandafter\ifx\csname bibnamefont\endcsname\relax
  \def\bibnamefont#1{#1}\fi
\expandafter\ifx\csname bibfnamefont\endcsname\relax
  \def\bibfnamefont#1{#1}\fi
\expandafter\ifx\csname citenamefont\endcsname\relax
  \def\citenamefont#1{#1}\fi
\expandafter\ifx\csname url\endcsname\relax
  \def\url#1{\texttt{#1}}\fi
\expandafter\ifx\csname urlprefix\endcsname\relax\def\urlprefix{URL }\fi
\providecommand{\bibinfo}[2]{#2}
\providecommand{\eprint}[2][]{\url{#2}}

\bibitem[{\citenamefont{Chatrchyan et~al.}(2013)}]{CMS:2012qk}
\bibinfo{author}{\bibfnamefont{S.}~\bibnamefont{Chatrchyan}}
  \bibnamefont{et~al.} (\bibinfo{collaboration}{CMS}), \bibinfo{journal}{Phys.
  Lett.} \textbf{\bibinfo{volume}{B718}}, \bibinfo{pages}{795}
  (\bibinfo{year}{2013}), \eprint{1210.5482}.

\bibitem[{\citenamefont{Abelev et~al.}(2013)}]{Abelev:2012ola}
\bibinfo{author}{\bibfnamefont{B.}~\bibnamefont{Abelev}} \bibnamefont{et~al.}
  (\bibinfo{collaboration}{ALICE}), \bibinfo{journal}{Phys.Lett.}
  \textbf{\bibinfo{volume}{B719}}, \bibinfo{pages}{29} (\bibinfo{year}{2013}),
  \eprint{1212.2001}.

\bibitem[{\citenamefont{Aad et~al.}(2013)}]{Aad:2012gla}
\bibinfo{author}{\bibfnamefont{G.}~\bibnamefont{Aad}} \bibnamefont{et~al.}
  (\bibinfo{collaboration}{ATLAS}), \bibinfo{journal}{Phys.Rev.Lett.}
  \textbf{\bibinfo{volume}{110}}, \bibinfo{pages}{182302}
  (\bibinfo{year}{2013}), \eprint{1212.5198}.

\bibitem[{\citenamefont{Shuryak and Zahed}(2013)}]{Shuryak:2013ke}
\bibinfo{author}{\bibfnamefont{E.}~\bibnamefont{Shuryak}} \bibnamefont{and}
  \bibinfo{author}{\bibfnamefont{I.}~\bibnamefont{Zahed}},
  \bibinfo{journal}{Phys.Rev.} \textbf{\bibinfo{volume}{C88}},
  \bibinfo{pages}{044915} (\bibinfo{year}{2013}), \eprint{1301.4470}.

\bibitem[{\citenamefont{Bozek and Broniowski}(2013)}]{Bozek:2013uha}
\bibinfo{author}{\bibfnamefont{P.}~\bibnamefont{Bozek}} \bibnamefont{and}
  \bibinfo{author}{\bibfnamefont{W.}~\bibnamefont{Broniowski}},
  \bibinfo{journal}{Phys.Rev.} \textbf{\bibinfo{volume}{C88}},
  \bibinfo{pages}{014903} (\bibinfo{year}{2013}), \eprint{1304.3044}.

\bibitem[{\citenamefont{de~Souza et~al.}(2015)\citenamefont{de~Souza, Koide,
  and Kodama}}]{deSouza:2015ena}
\bibinfo{author}{\bibfnamefont{R.~D.} \bibnamefont{de~Souza}},
  \bibinfo{author}{\bibfnamefont{T.}~\bibnamefont{Koide}}, \bibnamefont{and}
  \bibinfo{author}{\bibfnamefont{T.}~\bibnamefont{Kodama}}
  (\bibinfo{year}{2015}), \eprint{1506.03863}.

\bibitem[{\citenamefont{Kalaydzhyan and
  Shuryak}(2014{\natexlab{a}})}]{Kalaydzhyan:2014zqa}
\bibinfo{author}{\bibfnamefont{T.}~\bibnamefont{Kalaydzhyan}} \bibnamefont{and}
  \bibinfo{author}{\bibfnamefont{E.}~\bibnamefont{Shuryak}},
  \bibinfo{journal}{Phys. Rev.} \textbf{\bibinfo{volume}{C90}},
  \bibinfo{pages}{014901} (\bibinfo{year}{2014}{\natexlab{a}}),
  \eprint{1404.1888}.

\bibitem[{\citenamefont{Kalaydzhyan and
  Shuryak}(2014{\natexlab{b}})}]{Kalaydzhyan:2014tfa}
\bibinfo{author}{\bibfnamefont{T.}~\bibnamefont{Kalaydzhyan}} \bibnamefont{and}
  \bibinfo{author}{\bibfnamefont{E.}~\bibnamefont{Shuryak}},
  \bibinfo{journal}{Phys. Rev.} \textbf{\bibinfo{volume}{D90}},
  \bibinfo{pages}{025031} (\bibinfo{year}{2014}{\natexlab{b}}),
  \eprint{1402.7363}.

\bibitem[{\citenamefont{Qian and Zahed}(2015)}]{Qian:2014rda}
\bibinfo{author}{\bibfnamefont{Y.}~\bibnamefont{Qian}} \bibnamefont{and}
  \bibinfo{author}{\bibfnamefont{I.}~\bibnamefont{Zahed}},
  \bibinfo{journal}{Phys. Rev.} \textbf{\bibinfo{volume}{D91}},
  \bibinfo{pages}{125032} (\bibinfo{year}{2015}), \eprint{1411.3653}.

\bibitem[{\citenamefont{Grisaru
  et~al.}(1973{\natexlab{a}})\citenamefont{Grisaru, Schnitzer, and
  Tsao}}]{Grisaru:1973vw}
\bibinfo{author}{\bibfnamefont{M.~T.} \bibnamefont{Grisaru}},
  \bibinfo{author}{\bibfnamefont{H.~J.} \bibnamefont{Schnitzer}},
  \bibnamefont{and} \bibinfo{author}{\bibfnamefont{H.-S.} \bibnamefont{Tsao}},
  \bibinfo{journal}{Phys. Rev. Lett.} \textbf{\bibinfo{volume}{30}},
  \bibinfo{pages}{811} (\bibinfo{year}{1973}{\natexlab{a}}).

\bibitem[{\citenamefont{Grisaru
  et~al.}(1973{\natexlab{b}})\citenamefont{Grisaru, Schnitzer, and
  Tsao}}]{Grisaru:1974cf}
\bibinfo{author}{\bibfnamefont{M.~T.} \bibnamefont{Grisaru}},
  \bibinfo{author}{\bibfnamefont{H.~J.} \bibnamefont{Schnitzer}},
  \bibnamefont{and} \bibinfo{author}{\bibfnamefont{H.-S.} \bibnamefont{Tsao}},
  \bibinfo{journal}{Phys. Rev.} \textbf{\bibinfo{volume}{D8}},
  \bibinfo{pages}{4498} (\bibinfo{year}{1973}{\natexlab{b}}).

\bibitem[{\citenamefont{Lipatov}(1976)}]{Lipatov:1976zz}
\bibinfo{author}{\bibfnamefont{L.~N.} \bibnamefont{Lipatov}},
  \bibinfo{journal}{Sov. J. Nucl. Phys.} \textbf{\bibinfo{volume}{23}},
  \bibinfo{pages}{338} (\bibinfo{year}{1976}), \bibinfo{note}{[Yad.
  Fiz.23,642(1976)]}.

\bibitem[{\citenamefont{Kuraev et~al.}(1976)\citenamefont{Kuraev, Lipatov, and
  Fadin}}]{Kuraev:1976ge}
\bibinfo{author}{\bibfnamefont{E.~A.} \bibnamefont{Kuraev}},
  \bibinfo{author}{\bibfnamefont{L.~N.} \bibnamefont{Lipatov}},
  \bibnamefont{and} \bibinfo{author}{\bibfnamefont{V.~S.} \bibnamefont{Fadin}},
  \bibinfo{journal}{Sov. Phys. JETP} \textbf{\bibinfo{volume}{44}},
  \bibinfo{pages}{443} (\bibinfo{year}{1976}), \bibinfo{note}{[Zh. Eksp. Teor.
  Fiz.71,840(1976)]}.

\bibitem[{\citenamefont{Kuraev et~al.}(1977)\citenamefont{Kuraev, Lipatov, and
  Fadin}}]{Kuraev:1977fs}
\bibinfo{author}{\bibfnamefont{E.~A.} \bibnamefont{Kuraev}},
  \bibinfo{author}{\bibfnamefont{L.~N.} \bibnamefont{Lipatov}},
  \bibnamefont{and} \bibinfo{author}{\bibfnamefont{V.~S.} \bibnamefont{Fadin}},
  \bibinfo{journal}{Sov. Phys. JETP} \textbf{\bibinfo{volume}{45}},
  \bibinfo{pages}{199} (\bibinfo{year}{1977}), \bibinfo{note}{[Zh. Eksp. Teor.
  Fiz.72,377(1977)]}.

\bibitem[{\citenamefont{Balitsky and Lipatov}(1978)}]{Balitsky:1978ic}
\bibinfo{author}{\bibfnamefont{I.~I.} \bibnamefont{Balitsky}} \bibnamefont{and}
  \bibinfo{author}{\bibfnamefont{L.~N.} \bibnamefont{Lipatov}},
  \bibinfo{journal}{Sov. J. Nucl. Phys.} \textbf{\bibinfo{volume}{28}},
  \bibinfo{pages}{822} (\bibinfo{year}{1978}), \bibinfo{note}{[Yad.
  Fiz.28,1597(1978)]}.

\bibitem[{\citenamefont{Maldacena}(1999)}]{Maldacena:1997re}
\bibinfo{author}{\bibfnamefont{J.~M.} \bibnamefont{Maldacena}},
  \bibinfo{journal}{Int. J. Theor. Phys.} \textbf{\bibinfo{volume}{38}},
  \bibinfo{pages}{1113} (\bibinfo{year}{1999}), \bibinfo{note}{[Adv. Theor.
  Math. Phys.2,231(1998)]}, \eprint{hep-th/9711200}.

\bibitem[{\citenamefont{Qian and Zahed}(2014)}]{Qian:2014jna}
\bibinfo{author}{\bibfnamefont{Y.}~\bibnamefont{Qian}} \bibnamefont{and}
  \bibinfo{author}{\bibfnamefont{I.}~\bibnamefont{Zahed}}
  (\bibinfo{year}{2014}), \eprint{1410.1092}.

\bibitem[{\citenamefont{Shuryak and Zahed}(2014)}]{Shuryak:2013sra}
\bibinfo{author}{\bibfnamefont{E.}~\bibnamefont{Shuryak}} \bibnamefont{and}
  \bibinfo{author}{\bibfnamefont{I.}~\bibnamefont{Zahed}},
  \bibinfo{journal}{Phys.Rev.} \textbf{\bibinfo{volume}{D89}},
  \bibinfo{pages}{094001} (\bibinfo{year}{2014}), \eprint{1311.0836}.

\bibitem[{\citenamefont{Qian and Zahed}(2012)}]{Zahed:2012sg}
\bibinfo{author}{\bibfnamefont{Y.}~\bibnamefont{Qian}} \bibnamefont{and}
  \bibinfo{author}{\bibfnamefont{I.}~\bibnamefont{Zahed}}
  (\bibinfo{year}{2012}), \eprint{1211.6421}.

\bibitem[{\citenamefont{Stoffers and
  Zahed}(2013{\natexlab{a}})}]{Stoffers:2012mn}
\bibinfo{author}{\bibfnamefont{A.}~\bibnamefont{Stoffers}} \bibnamefont{and}
  \bibinfo{author}{\bibfnamefont{I.}~\bibnamefont{Zahed}},
  \bibinfo{journal}{Phys.Rev.} \textbf{\bibinfo{volume}{D88}},
  \bibinfo{pages}{025038} (\bibinfo{year}{2013}{\natexlab{a}}),
  \eprint{1211.3077}.

\bibitem[{\citenamefont{Stoffers and Zahed}(2012)}]{Stoffers:2012ai}
\bibinfo{author}{\bibfnamefont{A.}~\bibnamefont{Stoffers}} \bibnamefont{and}
  \bibinfo{author}{\bibfnamefont{I.}~\bibnamefont{Zahed}}
  (\bibinfo{year}{2012}), \eprint{1210.3724}.

\bibitem[{\citenamefont{Stoffers and
  Zahed}(2013{\natexlab{b}})}]{Stoffers:2012zw}
\bibinfo{author}{\bibfnamefont{A.}~\bibnamefont{Stoffers}} \bibnamefont{and}
  \bibinfo{author}{\bibfnamefont{I.}~\bibnamefont{Zahed}},
  \bibinfo{journal}{Phys.Rev.} \textbf{\bibinfo{volume}{D87}},
  \bibinfo{pages}{075023} (\bibinfo{year}{2013}{\natexlab{b}}),
  \eprint{1205.3223}.

\bibitem[{\citenamefont{Basar et~al.}(2012)\citenamefont{Basar, Kharzeev, Yee,
  and Zahed}}]{Basar:2012jb}
\bibinfo{author}{\bibfnamefont{G.}~\bibnamefont{Basar}},
  \bibinfo{author}{\bibfnamefont{D.~E.} \bibnamefont{Kharzeev}},
  \bibinfo{author}{\bibfnamefont{H.-U.} \bibnamefont{Yee}}, \bibnamefont{and}
  \bibinfo{author}{\bibfnamefont{I.}~\bibnamefont{Zahed}},
  \bibinfo{journal}{Phys.Rev.} \textbf{\bibinfo{volume}{D85}},
  \bibinfo{pages}{105005} (\bibinfo{year}{2012}), \eprint{1202.0831}.

\bibitem[{\citenamefont{McLerran}(2002)}]{McLerran:2001sr}
\bibinfo{author}{\bibfnamefont{L.~D.} \bibnamefont{McLerran}},
  \bibinfo{journal}{Lect.Notes Phys.} \textbf{\bibinfo{volume}{583}},
  \bibinfo{pages}{291} (\bibinfo{year}{2002}), \eprint{hep-ph/0104285}.

\bibitem[{\citenamefont{Tapia~Takaki}(2010)}]{TapiaTakaki:2010zz}
\bibinfo{author}{\bibfnamefont{J.}~\bibnamefont{Tapia~Takaki}}
  (\bibinfo{collaboration}{ALICE}), \bibinfo{journal}{J.Phys.}
  \textbf{\bibinfo{volume}{G37}}, \bibinfo{pages}{094050}
  (\bibinfo{year}{2010}).

\bibitem[{\citenamefont{Iancu and Venugopalan}(2003)}]{Iancu:2003xm}
\bibinfo{author}{\bibfnamefont{E.}~\bibnamefont{Iancu}} \bibnamefont{and}
  \bibinfo{author}{\bibfnamefont{R.}~\bibnamefont{Venugopalan}}
  (\bibinfo{year}{2003}), \eprint{hep-ph/0303204}.

\bibitem[{\citenamefont{Iancu et~al.}(2002)\citenamefont{Iancu, Leonidov, and
  McLerran}}]{Iancu:2002xk}
\bibinfo{author}{\bibfnamefont{E.}~\bibnamefont{Iancu}},
  \bibinfo{author}{\bibfnamefont{A.}~\bibnamefont{Leonidov}}, \bibnamefont{and}
  \bibinfo{author}{\bibfnamefont{L.}~\bibnamefont{McLerran}}, pp.
  \bibinfo{pages}{73--145} (\bibinfo{year}{2002}), \eprint{hep-ph/0202270}.

\bibitem[{\citenamefont{Navelet and Peschanski}(2002)}]{Navelet:2002zz}
\bibinfo{author}{\bibfnamefont{H.}~\bibnamefont{Navelet}} \bibnamefont{and}
  \bibinfo{author}{\bibfnamefont{R.~B.} \bibnamefont{Peschanski}},
  \bibinfo{journal}{Nucl.Phys.} \textbf{\bibinfo{volume}{B634}},
  \bibinfo{pages}{291} (\bibinfo{year}{2002}), \eprint{hep-ph/0201285}.

\bibitem[{\citenamefont{Iancu}(2001)}]{Iancu:2001yq}
\bibinfo{author}{\bibfnamefont{E.}~\bibnamefont{Iancu}}, pp.
  \bibinfo{pages}{184--191} (\bibinfo{year}{2001}), \eprint{hep-ph/0111400}.

\bibitem[{\citenamefont{Levin}(2001)}]{Levin:2001eq}
\bibinfo{author}{\bibfnamefont{E.}~\bibnamefont{Levin}} (\bibinfo{year}{2001}),
  \eprint{hep-ph/0105205}.

\bibitem[{\citenamefont{McLerran and Venugopalan}(1994)}]{McLerran:1993ka}
\bibinfo{author}{\bibfnamefont{L.~D.} \bibnamefont{McLerran}} \bibnamefont{and}
  \bibinfo{author}{\bibfnamefont{R.}~\bibnamefont{Venugopalan}},
  \bibinfo{journal}{Phys.Rev.} \textbf{\bibinfo{volume}{D49}},
  \bibinfo{pages}{3352} (\bibinfo{year}{1994}), \eprint{hep-ph/9311205}.

\bibitem[{\citenamefont{Gelis et~al.}(2010)\citenamefont{Gelis, Iancu,
  Jalilian-Marian, and Venugopalan}}]{Gelis:2010nm}
\bibinfo{author}{\bibfnamefont{F.}~\bibnamefont{Gelis}},
  \bibinfo{author}{\bibfnamefont{E.}~\bibnamefont{Iancu}},
  \bibinfo{author}{\bibfnamefont{J.}~\bibnamefont{Jalilian-Marian}},
  \bibnamefont{and}
  \bibinfo{author}{\bibfnamefont{R.}~\bibnamefont{Venugopalan}},
  \bibinfo{journal}{Ann.Rev.Nucl.Part.Sci.} \textbf{\bibinfo{volume}{60}},
  \bibinfo{pages}{463} (\bibinfo{year}{2010}), \eprint{1002.0333}.

\bibitem[{\citenamefont{Marquet et~al.}(2005)\citenamefont{Marquet, Mueller,
  Shoshi, and Wong}}]{Marquet:2005hu}
\bibinfo{author}{\bibfnamefont{C.}~\bibnamefont{Marquet}},
  \bibinfo{author}{\bibfnamefont{A.}~\bibnamefont{Mueller}},
  \bibinfo{author}{\bibfnamefont{A.}~\bibnamefont{Shoshi}}, \bibnamefont{and}
  \bibinfo{author}{\bibfnamefont{S.}~\bibnamefont{Wong}},
  \bibinfo{journal}{Nucl.Phys.} \textbf{\bibinfo{volume}{A762}},
  \bibinfo{pages}{252} (\bibinfo{year}{2005}), \eprint{hep-ph/0505229}.

\bibitem[{\citenamefont{Iancu and Mueller}(2004)}]{Iancu:2003uh}
\bibinfo{author}{\bibfnamefont{E.}~\bibnamefont{Iancu}} \bibnamefont{and}
  \bibinfo{author}{\bibfnamefont{A.}~\bibnamefont{Mueller}},
  \bibinfo{journal}{Nucl.Phys.} \textbf{\bibinfo{volume}{A730}},
  \bibinfo{pages}{460} (\bibinfo{year}{2004}), \eprint{hep-ph/0308315}.

\bibitem[{\citenamefont{Iancu}(2009)}]{Iancu:2009nd}
\bibinfo{author}{\bibfnamefont{E.}~\bibnamefont{Iancu}},
  \bibinfo{journal}{Nucl.Phys.Proc.Suppl.} \textbf{\bibinfo{volume}{191}},
  \bibinfo{pages}{281} (\bibinfo{year}{2009}), \eprint{0901.0986}.

\bibitem[{\citenamefont{Gelis}(2013)}]{Gelis:2012ri}
\bibinfo{author}{\bibfnamefont{F.}~\bibnamefont{Gelis}},
  \bibinfo{journal}{Int.J.Mod.Phys.} \textbf{\bibinfo{volume}{A28}},
  \bibinfo{pages}{1330001} (\bibinfo{year}{2013}), \eprint{1211.3327}.

\bibitem[{\citenamefont{Fubini et~al.}(1969)\citenamefont{Fubini, Gordon, and
  Veneziano}}]{Fubini:1969wp}
\bibinfo{author}{\bibfnamefont{S.}~\bibnamefont{Fubini}},
  \bibinfo{author}{\bibfnamefont{D.}~\bibnamefont{Gordon}}, \bibnamefont{and}
  \bibinfo{author}{\bibfnamefont{G.}~\bibnamefont{Veneziano}},
  \bibinfo{journal}{Phys. Lett.} \textbf{\bibinfo{volume}{B29}},
  \bibinfo{pages}{679} (\bibinfo{year}{1969}).

\bibitem[{\citenamefont{Fubini and Veneziano}(1969)}]{Fubini:1969qb}
\bibinfo{author}{\bibfnamefont{S.}~\bibnamefont{Fubini}} \bibnamefont{and}
  \bibinfo{author}{\bibfnamefont{G.}~\bibnamefont{Veneziano}},
  \bibinfo{journal}{Nuovo Cim.} \textbf{\bibinfo{volume}{A64}},
  \bibinfo{pages}{811} (\bibinfo{year}{1969}).

\bibitem[{\citenamefont{Bekenstein}(1973)}]{Bekenstein:1973ur}
\bibinfo{author}{\bibfnamefont{J.~D.} \bibnamefont{Bekenstein}},
  \bibinfo{journal}{Phys.Rev.} \textbf{\bibinfo{volume}{D7}},
  \bibinfo{pages}{2333} (\bibinfo{year}{1973}).

\bibitem[{\citenamefont{Bekenstein}(1972)}]{Bekenstein:1972tm}
\bibinfo{author}{\bibfnamefont{J.}~\bibnamefont{Bekenstein}},
  \bibinfo{journal}{Lett.Nuovo Cim.} \textbf{\bibinfo{volume}{4}},
  \bibinfo{pages}{737} (\bibinfo{year}{1972}).

\bibitem[{\citenamefont{Bekenstein}(1974)}]{Bekenstein:1974ax}
\bibinfo{author}{\bibfnamefont{J.~D.} \bibnamefont{Bekenstein}},
  \bibinfo{journal}{Phys.Rev.} \textbf{\bibinfo{volume}{D9}},
  \bibinfo{pages}{3292} (\bibinfo{year}{1974}).

\bibitem[{\citenamefont{Polchinski and Susskind}(2001)}]{Polchinski:2001ju}
\bibinfo{author}{\bibfnamefont{J.}~\bibnamefont{Polchinski}} \bibnamefont{and}
  \bibinfo{author}{\bibfnamefont{L.}~\bibnamefont{Susskind}}, pp.
  \bibinfo{pages}{105--114} (\bibinfo{year}{2001}), \eprint{hep-th/0112204}.

\bibitem[{\citenamefont{Susskind and Griffin}(1994)}]{Susskind:1994hb}
\bibinfo{author}{\bibfnamefont{L.}~\bibnamefont{Susskind}} \bibnamefont{and}
  \bibinfo{author}{\bibfnamefont{P.}~\bibnamefont{Griffin}}
  (\bibinfo{year}{1994}), \eprint{hep-ph/9410306}.

\bibitem[{\citenamefont{Susskind}(1995)}]{Susskind:1994vu}
\bibinfo{author}{\bibfnamefont{L.}~\bibnamefont{Susskind}},
  \bibinfo{journal}{J.Math.Phys.} \textbf{\bibinfo{volume}{36}},
  \bibinfo{pages}{6377} (\bibinfo{year}{1995}), \eprint{hep-th/9409089}.

\bibitem[{\citenamefont{Damour and Veneziano}(2000)}]{Damour:1999aw}
\bibinfo{author}{\bibfnamefont{T.}~\bibnamefont{Damour}} \bibnamefont{and}
  \bibinfo{author}{\bibfnamefont{G.}~\bibnamefont{Veneziano}},
  \bibinfo{journal}{Nucl.Phys.} \textbf{\bibinfo{volume}{B568}},
  \bibinfo{pages}{93} (\bibinfo{year}{2000}), \eprint{hep-th/9907030}.

\bibitem[{\citenamefont{Horowitz and Polchinski}(1997)}]{Horowitz:1996nw}
\bibinfo{author}{\bibfnamefont{G.~T.} \bibnamefont{Horowitz}} \bibnamefont{and}
  \bibinfo{author}{\bibfnamefont{J.}~\bibnamefont{Polchinski}},
  \bibinfo{journal}{Phys.Rev.} \textbf{\bibinfo{volume}{D55}},
  \bibinfo{pages}{6189} (\bibinfo{year}{1997}), \eprint{hep-th/9612146}.

\bibitem[{\citenamefont{Rho et~al.}(1999)\citenamefont{Rho, Sin, and
  Zahed}}]{Rho:1999jm}
\bibinfo{author}{\bibfnamefont{M.}~\bibnamefont{Rho}},
  \bibinfo{author}{\bibfnamefont{S.-J.} \bibnamefont{Sin}}, \bibnamefont{and}
  \bibinfo{author}{\bibfnamefont{I.}~\bibnamefont{Zahed}},
  \bibinfo{journal}{Phys.Lett.} \textbf{\bibinfo{volume}{B466}},
  \bibinfo{pages}{199} (\bibinfo{year}{1999}), \eprint{hep-th/9907126}.

\bibitem[{\citenamefont{Janik and Peschanski}(2000)}]{Janik:2000aj}
\bibinfo{author}{\bibfnamefont{R.}~\bibnamefont{Janik}} \bibnamefont{and}
  \bibinfo{author}{\bibfnamefont{R.~B.} \bibnamefont{Peschanski}},
  \bibinfo{journal}{Nucl.Phys.} \textbf{\bibinfo{volume}{B586}},
  \bibinfo{pages}{163} (\bibinfo{year}{2000}), \eprint{hep-th/0003059}.

\bibitem[{\citenamefont{Janik}(2001)}]{Janik:2000pp}
\bibinfo{author}{\bibfnamefont{R.~A.} \bibnamefont{Janik}},
  \bibinfo{journal}{Phys.Lett.} \textbf{\bibinfo{volume}{B500}},
  \bibinfo{pages}{118} (\bibinfo{year}{2001}), \eprint{hep-th/0010069}.

\bibitem[{\citenamefont{Stoffers and
  Zahed}(2013{\natexlab{c}})}]{Stoffers:2013tla}
\bibinfo{author}{\bibfnamefont{A.}~\bibnamefont{Stoffers}} \bibnamefont{and}
  \bibinfo{author}{\bibfnamefont{I.}~\bibnamefont{Zahed}},
  \bibinfo{journal}{Acta Phys.Polon.Supp.} \textbf{\bibinfo{volume}{6}},
  \bibinfo{pages}{7} (\bibinfo{year}{2013}{\natexlab{c}}).

\bibitem[{\citenamefont{Karliner et~al.}(1988)\citenamefont{Karliner, Klebanov,
  and Susskind}}]{Karliner:1988hd}
\bibinfo{author}{\bibfnamefont{M.}~\bibnamefont{Karliner}},
  \bibinfo{author}{\bibfnamefont{I.~R.} \bibnamefont{Klebanov}},
  \bibnamefont{and} \bibinfo{author}{\bibfnamefont{L.}~\bibnamefont{Susskind}},
  \bibinfo{journal}{Int.J.Mod.Phys.} \textbf{\bibinfo{volume}{A3}},
  \bibinfo{pages}{1981} (\bibinfo{year}{1988}).

\bibitem[{\citenamefont{Bergman and Thorn}(1997)}]{Bergman:1997ki}
\bibinfo{author}{\bibfnamefont{O.}~\bibnamefont{Bergman}} \bibnamefont{and}
  \bibinfo{author}{\bibfnamefont{C.~B.} \bibnamefont{Thorn}},
  \bibinfo{journal}{Nucl.Phys.} \textbf{\bibinfo{volume}{B502}},
  \bibinfo{pages}{309} (\bibinfo{year}{1997}), \eprint{hep-th/9702068}.

\bibitem[{\citenamefont{Liu and Zahed}(2014{\natexlab{a}})}]{Liu:2014fda}
\bibinfo{author}{\bibfnamefont{Y.}~\bibnamefont{Liu}} \bibnamefont{and}
  \bibinfo{author}{\bibfnamefont{I.}~\bibnamefont{Zahed}}
  (\bibinfo{year}{2014}{\natexlab{a}}), \eprint{1408.3331}.

\bibitem[{\citenamefont{Liu and Zahed}(2014{\natexlab{b}})}]{Liu:2014qrt}
\bibinfo{author}{\bibfnamefont{Y.}~\bibnamefont{Liu}} \bibnamefont{and}
  \bibinfo{author}{\bibfnamefont{I.}~\bibnamefont{Zahed}}
  (\bibinfo{year}{2014}{\natexlab{b}}), \eprint{1407.0384}.

\bibitem[{\citenamefont{Diakonov and Petrov}(1984)}]{Diakonov:1983hh}
\bibinfo{author}{\bibfnamefont{D.}~\bibnamefont{Diakonov}} \bibnamefont{and}
  \bibinfo{author}{\bibfnamefont{V.~{\relax Yu}.} \bibnamefont{Petrov}},
  \bibinfo{journal}{Nucl. Phys.} \textbf{\bibinfo{volume}{B245}},
  \bibinfo{pages}{259} (\bibinfo{year}{1984}).

\bibitem[{\citenamefont{Feynman}(1955)}]{Feynman:1955zz}
\bibinfo{author}{\bibfnamefont{R.~P.} \bibnamefont{Feynman}},
  \bibinfo{journal}{Phys. Rev.} \textbf{\bibinfo{volume}{97}},
  \bibinfo{pages}{660} (\bibinfo{year}{1955}).

\bibitem[{\citenamefont{Donnachie and Landshoff}(1992)}]{Donnachie:1992ny}
\bibinfo{author}{\bibfnamefont{A.}~\bibnamefont{Donnachie}} \bibnamefont{and}
  \bibinfo{author}{\bibfnamefont{P.}~\bibnamefont{Landshoff}},
  \bibinfo{journal}{Phys.Lett.} \textbf{\bibinfo{volume}{B296}},
  \bibinfo{pages}{227} (\bibinfo{year}{1992}), \eprint{hep-ph/9209205}.

\bibitem[{\citenamefont{Bzdak et~al.}(2014)\citenamefont{Bzdak, Bozek, and
  McLerran}}]{Bzdak:2013rya}
\bibinfo{author}{\bibfnamefont{A.}~\bibnamefont{Bzdak}},
  \bibinfo{author}{\bibfnamefont{P.}~\bibnamefont{Bozek}}, \bibnamefont{and}
  \bibinfo{author}{\bibfnamefont{L.}~\bibnamefont{McLerran}},
  \bibinfo{journal}{Nucl.Phys.} \textbf{\bibinfo{volume}{A927}},
  \bibinfo{pages}{15} (\bibinfo{year}{2014}), \eprint{1311.7325}.

\end{thebibliography}

\newpage

\newpage

\end{document}